\documentclass[11pt,a4paper]{article}
\pdfoutput=1
\usepackage{jstyle}
\usepackage{amsmath, amsfonts, amssymb}
\usepackage{graphicx, subfig, hyperref, color, verbatim}
\usepackage[dvipsnames]{xcolor}
\usepackage{float}

\title{Gluon Scattering in AdS from CFT}
\author[a]{Luis F. Alday,}
\author[a]{Connor Behan,}
\author[a]{Pietro Ferrero,}
\author[b]{Xinan Zhou}
\affiliation[a]{Mathematical Institute, University of Oxford, Andrew Wiles Building, Radcliffe Observatory Quarter, Woodstock Road, Oxford, OX2 6GG, U.K.}
\affiliation[b]{Princeton Center for Theoretical Science, Princeton University, Princeton, NJ 08544, USA}
\emailAdd{luis.alday@maths.ox.ac.uk\\  \hskip 42pt connor.behan@maths.ox.ac.uk, \\ \hskip 42pt pietro.ferrero@maths.ox.ac.uk, \\ \hskip 42pt xinanz@princeton.edu}

\abstract{We present a systematic study of holographic correlators in a vast array of SCFTs with non-maximal superconformal symmetry. These theories include 4d $\mathcal{N}=2$ SCFTs from D3-branes near F-theory singularities, 5d Seiberg exceptional theories and 6d E-string theory, as well as 3d and 4d phenomenological models with probe flavor branes. We consider current multiplets and their generalizations with higher weights, dual to massless and massive super gluons in the bulk. At leading order in the  inverse central charge expansion, connected four-point functions of these operators correspond to tree-level gluon scattering amplitudes in AdS. We show that all such tree-level four-point amplitudes in all these theories are fully fixed by symmetries and consistency conditions and explicitly construct them. Our results encode a wealth of SCFT data and exhibit various interesting emergent structures. These include Parisi-Sourlas-like dimensional reductions, hidden conformal symmetry and an AdS version of the color-kinematic duality.}

\begin{document}
\maketitle
\tableofcontents

\newpage

\section{Introduction}
\label{sec:intro}
Recently, there has been significant progress in computing holographic correlators, which just a few years ago was still considered a notoriously difficult problem.\footnote{See \cite{DHoker:1999pj,Arutyunov:2000py,Arutyunov:2002ff,Arutyunov:2003ae,Arutyunov:2002fh} for earlier results in the literature.} Tree-level four-point functions of arbitrary $\frac{1}{2}$-BPS operators have now been fully computed in all maximally superconformal CFTs by using a universal constructive method \cite{Alday:2020lbp,Alday:2020dtb}.\footnote{The three maximally superconformal theories are: IIB supergravity on $AdS_5\times S^5$ (dual to 4d $\mathcal{N}=4$ super Yang-Mills theory) and 11d supergravity on $AdS_4\times S^7$ and  $AdS_7\times S^4$ (dual respectively to the 3d $\mathcal{N}=8$ ABJM theory and the 6d $\mathcal{N}=(2,0)$ theory).} The results exhibit surprising universality and simplicity, and extend the success of earlier bootstrap approaches \cite{Rastelli:2016nze,Rastelli:2017udc,Zhou:2017zaw,Zhou:2018ofp,Rastelli:2017ymc,Rastelli:2019gtj,Goncalves:2019znr} which were most powerful when applied to strongly coupled $\mathcal{N}=4$ SYM. Physically, these  maximally supersymmetric holographic correlators can be interpreted as the scattering amplitudes of an infinite set of scalar fields in AdS, which arise from the Kaluza-Klein reduction of the 10d or 11d supergravity multiplet on a sphere. These scalars are the super primaries of the Kaluza-Klein multiplets, and can be viewed as super partners of (massless and massive) gravitons, {\it i.e.}, super gravitons. Further studies of AdS super graviton amplitudes have revealed many interesting mathematical structures. In $AdS_5\times S^5$ IIB supergravity, super graviton amplitudes exhibit a curious ten dimensional hidden conformal symmetry \cite{Caron-Huot:2018kta}, which allows them to be succinctly packaged into a simple generating function. This property, however, is not shared by super graviton amplitudes in 11d supergravity compactified on $AdS_4\times S^7$ and  $AdS_7\times S^4$. Another independent dimensional reduction structure was recently discovered in \cite{Behan:2021pzk}, which allows all super graviton amplitudes to be expressed in terms of the amplitudes of a simple scalar seed theory in lower dimensions. This emergent structure is universal to all maximally supersymmetric theories, and appears to be closely related to the Parisi-Sourlas supersymmetry \cite{Parisi:1979ka}. 

These fascinating new developments provide strong motivations to explore more ambitious extensions. One exciting direction is to consider theories which are not maximally supersymmetric. There is a variety of SCFTs in diverse dimensions which preserve half of the maximal superconformal symmetry ({\it i.e.}, eight Poincar\'e supercharges) while still admitting a weakly coupled holographic dual. Constrained by less supersymmetry, such theories have richer structures which are otherwise forbidden, such as global symmetries, and therefore are expected to exhibit new interesting features. Moreover, in five dimensions eight-supercharge SCFTs are also in a sense ``maximal'', since the 5d $\mathcal{N}=2$ supersymmetry algebra does not admit a superconformal extension. On the other hand, it is also extremely interesting to ask what happens when {\it gauge theories}, instead of supergravity, are placed in AdS.  Gauge theories in AdS were first explored in the late `80s \cite{Callan:1989em}, where the original motivation was to use the negative curvature of AdS to bring the infrared behavior under analytic control without conflicting with gauge invariance. This problem was later revisited in {\it e.g.}  \cite{Aharony:2012jf,Aharony:2015hix}.\footnote{There is also an extensive list of works studying general quantum field theories in AdS. See \cite{Doyon:2004fv,Aharony:2010ay,Paulos:2016fap,Carmi:2018qzm,Giombi:2020rmc} for a sampling of results from various points of view. } However, much remains {\it terra incognita} due to enormous computational difficulties in AdS. 

In this paper, we develop powerful techniques to explore both directions in an array of theories. More precisely, we will approach the problem by studying tree-level four-point functions of {\it super gluons}. These super gluons are the scalar super primaries of a family of superconformal short multiplets whose super descendants have at most Lorentz spin 1. A distinguished member of the family is the flavor current multiplet, which harbors spin-1 conserved currents associated with continuous global flavor symmetries. These conserved currents are dual to  massless spin-1 gauge fields in AdS, while the spin-1 operators in other short multiplets are dual to massive vector bosons.  The existence of such multiplets demands a certain amount of supersymmetry. The required amount of supersymmetry is only half as much (at most) as in the maximally superconformal cases. However, as we will see, this amount of supersymmetry also makes the calculation tractable. By focusing on correlators of super gluons, we can avoid dealing with the complicated kinematics associated with spinning operators. On the other hand, correlators of spinning operators can be obtained from the super gluon correlators by using superconformal symmetry.

While we expect our techniques to be useful in other setups, in this paper we will consider super gluon scattering amplitudes in two classes of theories. The first class arises from D-branes or M-branes near singularities, which preserve {\it eight} Poincar\'e supercharges. Theories in this class are  full-fledged SCFTs  which include 4d $\mathcal{N}=2$ SCFTs arising from D3-branes near F-theory singularities \cite{Fayyazuddin:1998fb,Aharony:1998xz}, 5d $F_4$ Seiberg exceptional theories \cite{Seiberg:1996bd}  engineered from D4-D8/O8 systems, as well as the 6d $\mathcal{N}=(1,0)$ E-string theory \cite{Ganor:1996mu,Seiberg:1996vs} from M5-branes on an end-of-the-world M9-brane. These theories play important roles in the landscape of SCFTs. For example, the 6d E-string theory is related to various lower-dimensional conformal field theories. The circle compactification of the E-string theory in the presence of $E_8$ Wilson lines gives rise to the 5d Seiberg theories, while compactifying it on Riemann surfaces leads to 4d $\mathcal{N}=1$ theories. These theories share a common feature in their dual holographic descriptions:  the background geometries all have a singular locus of the form $AdS_{d+1}\times S^3$ where $d$ is the spacetime dimension of the SCFT. On this locus there is a vector multiplet with a certain gauge group (or flavor symmetry group from the boundary CFT perspective)\footnote{This is because in AdS/CFT global symmetry currents in the boundary couple to gauge fields in the bulk. We will use the terms {\it flavor}, {\it gauge}, {\it color} interchangeably in this paper depending on the perspective we want to take.}, which upon reduction on $S^3$ gives rise to the aforementioned super gluons. Another important feature of these theories is that the self-coupling of super gluons is stronger than their coupling to super gravitons by powers of $N$, where $N$ is the number of branes probing the singularities and is taken to be large. This hierarchy of couplings implies that the dominant contribution in tree-level processes is given by super gluon exchanges, and graviton exchanges will only appear at higher orders in $1/N$. Essentially, at leading order we are only dealing with supersymmetric gauge theories in AdS.

The second class of theories is a type of phenomenological model of holography in which flavors (fundamental matter) in the boundary supersymmetric gauge theories are included by adding probe D-branes. The two theories we will focus on are 4d $\mathcal{N}=4$ SYM and 3d $\mathcal{N}=6$ ABJM with fundamental matter. In the first case, D7-branes which fill the $AdS_5$ and wrap an $S^3$ inside $S^5$ are added \cite{Karch:2002sh}. In the second case, D6-branes which fill the $AdS_4$ and wrap an $\mathbb{RP}^3$ inside $\mathbb{CP}^3$  are introduced \cite{Hohenegger:2009as,Gaiotto:2009tk,Hikida:2009tp}. The number of flavor branes in both cases needs to be much smaller than the number of D3- or M2-branes so that the theories remain conformal. The flavor branes can then be treated as probes and do not backreact to the geometry. However, the added flavor branes break half of the supersymmetry, to 4d $\mathcal{N}=2$ and 3d $\mathcal{N}=3$ respectively. The low energy degrees of freedom on the probe branes are again vector multiplets, and give rise to super gluons in AdS after Kaluza-Klein reduction.

The main result of this paper is a complete list of all tree-level four-point functions of super gluons with {\it arbitrary} Kaluza-Klein levels in {\it all} the theories mentioned above. Our main tool is the maximally R-symmetry violating (MRV) method developed in \cite{Alday:2020lbp,Alday:2020dtb} together with constraints from the superconformal Ward identities. Using these ingredients, we develop a streamlined algorithm for computing general four-point super gluon amplitudes that can be applied to any spacetime dimension. By only inputting the spectrum and imposing superconformal symmetry, we show that four-point amplitudes in  vast families of theories are completely fixed. The answer for the amplitudes takes the universal form as a sum over exchange contributions, with no additional contact terms
\begin{equation}
\nonumber \mathcal{M}=\mathtt{c}_s M_s+\mathtt{c}_t M_t+\mathtt{c}_u M_u\;.
\end{equation}
Here $\mathtt{c}_s=f^{I_1I_2J}f^{JI_3I_4}$, $\mathtt{c}_t=f^{I_1I_4J}f^{JI_2I_3}$, $\mathtt{c}_u=f^{I_1I_3J}f^{JI_4I_2}$ are color structures associated with exchanging adjoint representations, and $f^{IJK}$ are the structure constants of the gauge group. Remarkably, we find that the only dependence on the gauge group is through these color structures.  The amplitude factors $M_{s,t,u}$ capture nontrivial dynamics, and will be explicitly given for each theory. These super gluon amplitudes encode a wealth of data in the dual SCFTs. In particular, our analysis outputs the {\it complete} three-point functions of all super primaries of these short multiplets (see Section \ref{sec:correlators} and \ref{sec:flavor}), which to our knowledge were not known in most of these theories. Note that the structure above is very reminiscent of that of flat space gluon amplitudes at tree level -- {\it e.g.}, the dependence on gauge groups is exactly the same. In fact, the comparison can be sharpened by taking the flat space limit of our results. In this limit AdS amplitudes simplify drastically and reduce to 
\begin{equation*}
\left. {\cal M} \right|_{\text{flat}} = \bigg(\frac{\mathtt{c}_sN_s}{s}+\frac{\mathtt{c}_tN_t}{t}+\frac{\mathtt{c}_uN_u}{u}\bigg)\times \big(\text{wavefunctions on $S^3$}\big),
\end{equation*}
with simple polynomials $N_{s,t,u}$  homogeneous in Mandelstam variables obeying $N_s+N_t+N_u=0$. The information of different Kaluza-Klein levels of AdS super gluons in the flat space limit is factored out in an $S^3$ wavefunction factor. We will show that this result matches exactly with the flat space four-gluon scattering amplitude after choosing appropriate polarizations. However, to appreciate the full richness of our results we need to go back to finite AdS curvature where more interesting mathematical structures become visible. Some of these structures have analogues in super graviton amplitudes in maximally supersymmetric theories, while  other features are only possible with less supersymmetry and therefore are new.  Let us highlight some of these structures below.

\vspace{0.2cm}
\noindent{\bf Parisi-Sourlas supersymmetry:} As mentioned above, all super gluon amplitudes can be expressed as the sum of exchange amplitudes without additional contact terms. The exchange amplitudes receive contributions from only a finite set of super gluon multiplets due to selection rules. For each multiplet, we find its contribution can be written as a differential operator acting on the sum of two $AdS_{d+1}$ scalar exchange Witten diagrams. However, a closer look at the sum reveals that it is identical to just a single scalar exchange Witten diagram which lives in a lower dimensional $AdS_{d-1}$ space. This curious dimensional reduction phenomenon turns out to be related to a holographically realized Parisi-Sourlas supersymmetry \cite{Zhou:2020ptb}, and is a feature shared by all the AdS supersymmetric gauge theories considered in this paper. Similar dimensional reduction was observed in maximally supersymmetric super graviton amplitudes \cite{Behan:2021pzk}. However, in that case the dimension of the AdS space is reduced by four instead of two. 

\vspace{0.2cm}

\noindent{\bf Hidden conformal symmetry:} For 4d $\mathcal{N}=2$ theories, we also observe an {\it eight} dimensional hidden conformal symmetry. In position space, we can define a {\it reduced} correlator $H_{k_1k_2k_3k_4}$ to fully capture the dynamical information, where $k_i$ labels the Kaluza-Klein levels of the external super gluons. We find that the reduced correlator with lowest Kaluza-Klein level $H_{2222}(x_{ij}^2)$, depending on the $AdS_5$ distances $x_{ij}^2$ only, can be promoted into a generating function $\mathbf{H}$. The promotion is done by replacing the five dimensional distances in  $H_{2222}(x_{ij}^2)$  with the eight dimensional distances $x_{ij}^2\to x_{ij}^2-t_{ij}$ where $t_{ij}$ are the distances on $S^3$. Since the $AdS_5\times S^3$ background is conformally flat, they can also be viewed as the conformally invariant distances in $\mathbb{R}^{7,1}$ after a Weyl transformation. General correlators $H_{k_1k_2k_3k_4}$ can then be obtained by Taylor expanding the generating function $\mathbf{H}$ in $t_{ij}$ and collecting all admissible monomials of $t_{ij}$ in the reduced correlator. The same type of hidden conformal structures also made appearances in IIB supergravity in $AdS_5\times S^5$ \cite{Caron-Huot:2018kta} and $AdS_3\times S^3\times K3$ \cite{Rastelli:2019gtj,Giusto:2020neo}, where the symmetries are ten and six dimensional respectively. However, our new example makes it clear that such hidden structures are not unique to supergravity theories, but can be found in supersymmetric gauge theories as well. It also provides further evidence that these structures originate from the conformal flatness of the background geometry. 

\vspace{0.2cm}

\noindent{\bf AdS color-kinematic duality:} A fascinating property of flat space amplitudes is the color-kinematic duality \cite{Bern:2008qj}. A natural question is whether this structure extends to AdS. In this paper, we provide an affirmative answer by finding an AdS version of the color-kinematic duality. This AdS color-kinematic duality holds for the massless super gluons, {\it i.e.}, the gauge fields in AdS, and is almost identical to the flat space relation. These massless four-point Mellin amplitudes have a form analogous to the flat space four-gluon amplitude
\begin{equation}
\nonumber \mathcal{M}\sim \mathtt{c}_s\mathtt{n}_s\left(\frac{1}{s-d
+2}+\ldots\right)+\mathtt{c}_t\mathtt{n}_t\left(\frac{1}{t-d+2}+\ldots\right)+\mathtt{c}_u\mathtt{n}_u\left(\frac{1}{u-d+2}+\ldots\right)
\end{equation}
The kinematic factors $\mathtt{n}_{s,t,u}$ are linear in the Mellin-Mandelstam variables, and the $\ldots$ denote satellite poles at $d-2+2\mathbb{Z}_+$ which are completely fixed by symmetry. Thanks to the Jacobi identity, $\mathtt{c}_{s,t,u}$ satisfy $\mathtt{c}_s+\mathtt{c}_t+\mathtt{c}_u=0$. Remarkably, $\mathtt{n}_{s,t,u}$ satisfy the same identity $\mathtt{n}_s+\mathtt{n}_t+\mathtt{n}_u=0$, giving an AdS relation that completely parallels the flat space one.  This structure for massless super gluons also admits an interesting modification to massive modes, as we will discuss in the paper.

\vspace{0.2cm}

The rest of the paper is organized as follows. In Section \ref{sec:preliminaries} we give a brief review of the theories which we will consider and provide some background. We also discuss various kinematical aspects and introduce the Mellin representation. In Section \ref{MRVandbootstrap} we outline the strategy of our computation. We introduce the MRV method for non-maximally superconformal theories, and explain how to fix the super gluon amplitudes by imposing symmetry constraints. Sections \ref{sec:correlators} and \ref{sec:flavor} contain the main results of this paper. In Section \ref{sec:correlators} we give all four-point super gluon amplitudes in the three full-fledged SCFTs, and in Section  \ref{sec:flavor} we give the results for two models with flavor branes. In Section \ref{sec:flatspace} we study the flat space limit of AdS super gluon amplitudes and show how all amplitudes in this limit are reproduced from the flat space gluon scattering amplitudes. Using the flat space limit, we also outline an alternative method to derive our results. In Section \ref{sec:hiddenstructures} we discuss in detail various interesting mathematical structures in these super gluon amplitudes: Parisi-Sourlas supersymmetry, hidden conformal symmetry and AdS color-kinematic duality. Finally, we conclude in Section \ref{sec:conclusions} with an outlook of future directions. The paper also includes several appendices, which collect useful formulae and some technical results. Appendix \ref{app:exchangeWittenMellin} contains the Mellin amplitude expressions for exchange Witten diagrams from \cite{Alday:2020dtb}. A few technical comments on 3d superconformal blocks are made in  Appendix \ref{app:3dsuperconformalblocks}. In Appendix \ref{sec:chiralalgebra} we extract the chiral algebra correlators from the holographically computed super gluon amplitudes for 4d $\mathcal{N}=2$ theories, and compare the results with independent field theory calculations. In Appendix \ref{app:contact} we discuss contact terms and higher-derivative corrections in various dimensions.

\section{Preliminaries}
\label{sec:preliminaries}
In this section we discuss a few preliminary ingredients for the rest of the paper. In Section \ref{subsec:theories} we describe the theories to be considered. In Section \ref{subsec:kinematics} we specify the correlators we will consider and discuss their kinematic features. In Section \ref{subsec:Mellinrepresentation} we review the Mellin representation and how superconformal constraints are implemented in Mellin space.

\subsection{Theories}
\label{subsec:theories}
In this paper we will discuss holographic correlators in two types of theories. The first type is a distinguished class of SCFTs  in 4d, 5d and 6d, which arises from branes probing singularities. These theories preserve half of the maximal superconformal symmetry, {\it i.e.}, eight Poincar\'e supercharges, and are listed in Table \ref{8PSC}. Let us describe them in more detail.

\begin{table}[htp]
\begin{center}
\begin{tabular}{|c|c|c|}
\hline
Dimension & SCFT$_d$ & holographic origin \\\hline\hline
$d=4$ & $\mathcal{N}=2$ theories & D3 near F-theory singularities \\\hline
$d=5$ & Seiberg exceptional theories & D4-D8/O8 system \\\hline
$d=6$ & E-string theory & M5 on an end-of-the-world 9-brane \\\hline
\end{tabular}
\end{center}
\caption{A class of holographic theories with eight Poincar\'e supercharges and a supergravity limit. These theories arise in a similar fashion from branes probing singularities.}
\label{8PSC}
\end{table}

\noindent{\it $\mathcal{N}=2$ theories from F-theory singularities}  \cite{Fayyazuddin:1998fb,Aharony:1998xz}: These theories arise from $N$ D3-branes, with $N$ large, near an F-theory 7-brane singularity. The resulting near horizon geometry has a metric similar to $AdS_5\times S^5$, but with one of the angular coordinates of the compact space having a changed periodicity
\begin{equation}
ds^2=d\theta^2+\sin^2\theta\, d\phi^2+\cos^2\theta\, d\Omega^2_3\;.
\end{equation} 
Here $d\Omega^2_3$ is the metric on $S^3$, $0\leq \theta\leq \frac{\pi}{2}$, and $\phi$ is periodic with period $2\pi(1-\nu/2)$. Seven types of singularities give rise to a conformal field theory on the D3-branes and those correspond to  
\begin{equation}\label{alphavalue}
\nu=\frac{1}{3}\,,\,\frac{1}{2}\,,\,\frac{2}{3}\,,\,1\,,\,\frac{4}{3}\,,\,\frac{3}{2}\,,\,\frac{5}{3}\;.
\end{equation}
The 7-brane wraps the $AdS_5$ and is located at $\theta=0$, which is an $S^3$ in the compact space. Therefore the $SO(6)$ isometry is broken to 
\begin{equation}
SO(4)\times SO(2) \simeq SU(2)_R\times SU(2)_L\times U(1)_r
\end{equation}
where the first factor $SU(2)_R$ becomes the $SU(2)$ R-symmetry of $\mathcal{N}=2$, and the other $SU(2)_L$ becomes a flavor symmetry. On the 7-brane, there is a $7+1$ dimensional $\mathcal{N}=1$ SYM theory with gauge group $G_F$. From the point of view of the four-dimensional theory living on the $D3$-brane $G_F$ plays the role of a global symmetry. The above choices for $\nu$ in (\ref{alphavalue}) correspond to the following global symmetries 
\begin{equation} 
G_F=U(1)\,,\, SU(2)\,,\, SU(3)\,,\, SO(8)\,,\, E_6\,,\, E_7\,,\, E_8\;.
\end{equation}
The simplest case corresponds to a $\mathbb{Z}_2$ orientifold singularity with $\nu=1$. In this case the four dimensional theory is a $USp(2N)$ $\mathcal{N}=2$ gauge theory with $SO(8)$ global symmetry and with one antisymmetric and four fundamental hypermultiplets.

\vspace{0.3cm}

\noindent{\it Seiberg exceptional theories} \cite{Seiberg:1996bd}: This class of theories comes from the UV fixed point of a $USp(2N)$ gauge theory coupled to $N_f\leq 7$ hypermultiplets in the fundamental representation, and a single hypermultiplet in the antisymmetric representation. The flavor symmetry is enhanced to $G=E_{N_f+1}\times SU(2)_L$ at the fixed point.\footnote{Note $E_1=SU(2)$, $E_2=SU(2)\times U(1)$, $E_3=SU(3)\times SU(2)$, $E_4=SU(5)$, $E_5=SO(10)$. } These theories can also be constructed from Type IIA string theory (or Type I${}^\prime$ string theory) by a D4-D8/O8 setup. The dual geometry is a warped product of $AdS_6$ and a hemisphere $HS^4$. Therefore we are left with an $SO(4)\simeq SU(2)_R\times SU(2)_L$ isometry. The first factor becomes the $SU(2)$ R-symmetry of the 5d $F_4$ superconformal group, while the $SU(2)_L$ becomes a flavor symmetry. The boundary of the four-hemisphere gives an 8-brane that fills $AdS_6$. Similar to the above F-theory singularity case, there is a 9d $\mathcal{N}=1$ vector multiplet propagating on this $AdS_6\times S^3$ boundary  with a gauge group $E_{N_f+1}$.

\vspace{0.3cm}

\noindent{\it E-string theory} \cite{Ganor:1996mu,Seiberg:1996vs}: The E-string theory can be engineered by placing $N$ M5-branes on top of an ``end-of-the-world'' 9-brane of the Ho\v{r}ava-Witten compactification of M-theory. The dual geometry is $AdS_7\times S^4/\mathbb{Z}_2$, which has a $\mathbb{Z}_2$ fixed locus $AdS_7\times S^3$. The $\mathbb{Z}_2$ quotient breaks the isometry from $SO(5)$ down to $SO(4)\simeq SU(2)_R\times SU(2)_L$. Again, the first factor is identified with the $SU(2)$ R-symmetry of the 6d $\mathcal{N}=(1,0)$ superconformal group, and the other $SU(2)_L$ becomes part of the flavor symmetry. In addition, on the $\mathbb{Z}_2$ locus there is a 10d $\mathcal{N}=1$ SYM multiplet with $E_8$ gauge group, which is a flavor group from the point of view of the six dimensional SCFT.  

\vspace{0.3cm}

In these three families of theories, the Kaluza-Klein fields in AdS come from two sources: supergravity in the full 10d or 11d space, and degrees of freedom living on the singular locus. The spectrum of the supergravity modes in $AdS_5$ and $AdS_7$ can be analyzed following the method of  \cite{Aharony:1998xz} (or in the simpler orbifold cases, the spectrum can be obtained by projecting to the singlets \cite{Fayyazuddin:1998fb,Gimon:1999yu}). The analysis of the $AdS_6$ case  is more involved due to the warped product structure \cite{Passias:2018swc}. The supergravity modes can be organized  into superconformal multiplets, and contain both short and long types. On the other hand, analyzing the spectrum from the singular locus is much simpler. Because the vector super multiplets contain fields with at most spin 1, all the KK modes must belong to $\frac{1}{2}$-BPS multiplets. Their spectrum therefore is completely fixed by their R-symmetry representations \cite{Aharony:1998xz,Gimon:1999yu,Brandhuber:1999np}, and there is no need to analyze the linearized equations of motion. In this paper, we will focus on the latter type of AdS excitations, {\it i.e.}, field modes descending from the singular loci, and compute all four-point functions of their dual operators. 

\vspace{0.5cm}

The second class of models shares a lot similarities with the theories discussed above. These are the theories arising from adding ``flavors'', {\it i.e.}, fundamental quarks, to supersymmetric gauge theories. We list these theories in Table \ref{addflavor}. 

\begin{table}[H]
\begin{center}
\begin{tabular}{|c|c|c|}
\hline
Dimension & SCFT$_d$ & holographic origin\\\hline\hline
$d=3$ & $\mathcal{N}=6$ ABJM with flavors & D6 wrapped over $AdS_4\times \mathbb{RP}^3$ \\\hline
$d=4$ & $\mathcal{N}=4$ SYM with flavors & D7 wrapped over $AdS_5\times S^3$\\\hline
\end{tabular}
\end{center}
\caption{A class of holographic theories obtained by adding flavor branes. Note the number of preserved supersymmetries after adding flavor branes is halved, {\it i.e.}, $\mathcal{N}=3$ for the flavored ABJM and $\mathcal{N}=2$ for the flavored SYM.}
\label{addflavor}
\end{table}

Holographically, these models can be obtained by adding probe flavor branes. Such a construction was first considered for 4d $\mathcal{N}=4$ SYM in \cite{Karch:2002sh}, where $N_f\ll N$  D7-branes are added as probes to $AdS_5\times S^5$.\footnote{This is also known as the ``quenched'' approximation.} The D7-branes fill $AdS_5$ and wrap an equatorial $S^3$ inside the $S^5$. The presence of the probe flavor branes breaks the $\mathcal{N}=4$ superconformal symmetry to $\mathcal{N}=2$. Similarly, D6-branes can also be added to $AdS_4\times \mathbb{CP}^3$, which adds fundamental matter to the $\mathcal{N}=6$ ABJM theory \cite{Hohenegger:2009as,Gaiotto:2009tk,Hikida:2009tp}. The D6-branes fill $AdS_4$ and wrap an $\mathbb{RP}^3$ inside $\mathbb{CP}^3$. Note that the flavored theory now has only $\mathcal{N}=3$ superconformal symmetry, {\it i.e.}, six Poincar\'e supercharges. Nevertheless, we will see that it can be treated in the same fashion. We also note that the flavor branes are only probes and do not backreact to the bulk geometry, which  differs from the situation for theories in Table \ref{8PSC}. Finally, using the same argument as above, we can see that the AdS Kaluza-Klein modes from the low energy degrees of freedom of the flavor branes contain only short multiplets \cite{Kruczenski:2003be,Hikida:2009tp}. It is this sector of operators for which we will compute correlators.

\subsection{Kinematics of four-point functions}\label{subsec:kinematics}

\vspace{0.2cm}

\noindent{\it Operators}

\vspace{0.2cm}

\noindent The primary focus of this paper is the class of holographic SCFTs with eight Poincar\'e supercharges summarized in Table \ref{8PSC} in the previous subsection. The examples include 4d $\mathcal{N}=2$ theories constructed from D3-branes near F-theory singularities \cite{Fayyazuddin:1998fb,Aharony:1998xz}, 5d $F(4)$ theories constructed from D4-D8 systems in massive Type IIA supergravity \cite{Ferrara:1998gv}, and 6d $(1,0)$ theories from M5-branes on an ``end-of-the-world'' M9-brane \cite{Berkooz:1998bx}. However, the same techniques will also apply to the theories with flavor branes summarized in Table \ref{addflavor}, even though the 3d theory has only six supercharges. As already mentioned in Section \ref{subsec:theories}, what these theories have in common is that there are branes that fill the AdS factors and an $S^3$ (or $\mathbb{RP}^3$) in the compact space. The presence of the branes breaks the compact space isometries as follows
\begin{equation}
\nonumber\begin{split}
&AdS_7:\quad\quad SO(5)\to SO(4)=SU(2)_R\times SU(2)_L\;,\\
&AdS_6:\quad\quad SO(5)\to SO(4)=SU(2)_R\times SU(2)_L\;,\\
&AdS_5:\quad\quad SO(6)\to SO(4)\times SO(2)= SU(2)_R\times SU(2)_L\times U(1)\;,\\
&AdS_4:\quad\quad SO(6)\to SO(4)= SU(2)_R\times SU(2)_L
\end{split}
\end{equation}
where $SO(4)=SU(2)_R\times SU(2)_L$ is the isometry of an $S^3$ (or $\mathbb{RP}^3$). One of the $SU(2)$ factors is identified with the $SU(2)_R$ R-symmetry of the superconformal group\footnote{For concreteness, we list the superconformal groups. The 3d $\mathcal{N}=3$ superconformal group is $OSp(3|4)$; the 4d $\mathcal{N}=2$ superconformal group is  $SU(2,2|2)$; in 5d the only superconformal group is $F_4$; in 6d the $\mathcal{N}=(1,0)$ superconformal group is $OSp(8^*|2)$.}, while the other $SU(2)$ factor becomes part of the global symmetry. The $U(1)$ factor for the $AdS_5$ case is identified with the $U(1)_r$ symmetry for the 4d $\mathcal{N}=2$ superconformal symmetry. 

The low-energy theories on these AdS-filling branes are SYM with a certain gauge group $G_F$. An important feature of these theories at large $N$ is that the degrees of freedom localized on the branes interact more strongly with themselves than with the gravitational degrees of freedom in the bulk. This can be seen from the fact that the flavor symmetry central charge is larger than the stress tensor central charge by powers of $N$, and the exchange contributions are inversely proportional to the central charges. Therefore, when considering the leading tree-level processes with external brane degrees of freedom we can focus only on the branes and decouple completely the supergravity modes. 

Note that the vector multiplet on the branes contains fields with Lorentz spin at most 1. This condition guarantees that the Kaluza-Klein reduction leads  to only short multiplets, which allows us to fix the spectrum from R-symmetry charges without analyzing the linearized equations of motion. Such short multiplets are $\frac{1}{2}$-BPS for $4\leq d\leq 6$, and $\frac{1}{3}$-BPS for 3d $\mathcal{N}=3$. This strategy to obtain the spectrum was used in \cite{Aharony:1998xz,Gimon:1999yu,Brandhuber:1999np}, and the analyses led to the following results. The super multiplets are labelled by an integer Kaluza-Klein level $k=2,3,\ldots$, with $k=2$ corresponding to the current multiplet. The super primaries are scalar operators which have spin-$\frac{k}{2}$ under the $SU(2)_R$ R-symmetry group and spin-$\frac{k-2}{2}$ under $SU(2)_L$. In the 4d $\mathcal{N}=2$ case there is also a $U(1)_r$ R-symmetry. However, the super primaries are neutral under this $U(1)_r$. 
The super primaries have protected conformal dimensions
\begin{equation}
\Delta_k=\epsilon k\;,\quad \epsilon=\frac{d-2}{2}
\end{equation}
where $d$ is the spacetime dimension of the boundary theory.  Finally, all components of the multiplet transform in the adjoint representation of the gauge group $G_F$, which is a flavor symmetry from the CFT point of view. We denote the super primaries as
\begin{equation}
\mathcal{O}^{I;\alpha_1,\ldots,\alpha_k;\bar{\alpha}_1,\ldots,\bar{\alpha}_{k-2}}(x) \label{ext-op}
\end{equation}
where $\alpha_i=1,2$, $\bar{\alpha}_i=1,2$ are the $SU(2)_R$, $SU(2)_L$ indices respectively, and $I=1,\ldots, {\rm dim}(G_F)$ is a flavor symmetry index.

\vspace{0.5cm}

\noindent{\it Four-point functions}

\vspace{0.2cm}

\noindent To handle the R-symmetry and $SU(2)_L$ flavor symmetry indices, it is convenient to multiply the super primary operators with auxiliary $SU(2)$ spinors 
\begin{equation}
\mathcal{O}^I_k(x,v,\bar{v})=\mathcal{O}^{I;\alpha_1,\ldots,\alpha_k;\bar{\alpha}_1,\ldots,\bar{\alpha}_{k-2}}(x)v^{\beta_1}\ldots v^{\beta_k}\epsilon_{\alpha_1\beta_1}\ldots\epsilon_{\alpha_k\beta_k}\bar{v}^{\bar{\beta}_1}\ldots \bar{v}^{\bar{\beta}_{k-2}}\epsilon_{\bar{\alpha}_1\bar{\beta}_1}\ldots\epsilon_{\bar{\alpha}_{k-2}\bar{\beta}_{k-2}}\;.
\end{equation}
The contraction with the spinors automatically projects the indices to the spin-$\frac{k}{2}$ and spin-$\frac{k-2}{2}$ representations of $SU(2)_R$ and $SU(2)_L$. The four-point correlation functions 
\begin{equation}
G^{I_1I_2I_3I_4}(x_i;v_i,\bar{v}_i)=\langle \mathcal{O}^{I_1}_{k_1}(x_1,v_1,\bar{v}_1)\mathcal{O}^{I_2}_{k_2}(x_2,v_2,\bar{v}_2)\mathcal{O}^{I_3}_{k_3}(x_3,v_3,\bar{v}_3)\mathcal{O}^{I_4}_{k_4}(x_4,v_4,\bar{v}_4) \rangle\;,
\end{equation}
now become a function not only of the spacetime coordinates $x_i$, but also internal coordinates $v_i$, $\bar{v}_i$. By exploiting the covariance under conformal symmetry, R-symmetry, and $SU(2)_L$ flavor symmetry, we can extract a kinematic factor such that the correlators depend only on the cross ratios
\begin{equation}\label{GandcalG}
\begin{split}
G^{I_1I_2I_3I_4}(x_i;v_i,\bar{v}_i)=&\prod_{i<j}\left(\frac{(v_i\cdot v_j)(\bar{v}_i\cdot \bar{v}_j)}{x_{ij}^{2\epsilon}}\right)^{\gamma^0_{ij}}\left(\frac{(v_1\cdot v_2)(v_3\cdot v_4)}{x_{12}^{2\epsilon}x_{34}^{2\epsilon}}\right)^{\mathcal{E}}\\
&\times \left((\bar{v}_1\cdot\bar{v}_2)(\bar{v}_3\cdot\bar{v}_4)\right)^{\mathcal{E}-2} \mathcal{G}^{I_1I_2I_3I_4}(U,V;\alpha,\beta)\;,
\end{split}
\end{equation} 
where
\begin{equation}
x_{ij}=x_i-x_j\;,\quad (v_i\cdot v_j)=v_i^{\alpha}v_j^{\beta}\epsilon_{\alpha\beta}\;,\quad (\bar{v}_i\cdot \bar{v}_j)=\bar{v}_i^{\bar{\alpha}}\bar{v}_j^{\bar{\beta}}\epsilon_{\bar{\alpha}\bar{\beta}}\;.
\end{equation}
Here we have assumed without loss of generality that $k_1\leq k_2\leq k_3\leq k_4$ and distinguish between two different cases 
\begin{equation}\label{twocases}
k_1+k_4\geq k_2+k_3 \;\; \text{(case I)}\;,\quad\quad\;\; k_1+k_4< k_2+k_3 \;\; \text{(case II)}\;.
\end{equation}
The exponents are given by 
\begin{eqnarray}\label{defgamma0}
&&\gamma_{12}^0=\gamma_{13}^0=0\;,\;\; \gamma_{34}^0=\frac{\kappa_s}{2}\;,\; \; \gamma_{24}^0=\frac{\kappa_u}{2}\;,\\
\nonumber && \gamma_{14}^0=\frac{\kappa_t}{2}\,,\;\; \gamma_{23}^0=0\,,\;\text{(I)}\,,\;\;\;\gamma_{14}^0=0\,,\;\; \gamma_{23}^0=\frac{\kappa_t}{2}\,,\;\text{(II)}
\end{eqnarray}
where
\begin{equation}
\kappa_s\equiv|k_3+k_4-k_1-k_2|\;,\;\; \kappa_t\equiv|k_1+k_4-k_2-k_3|\;,\;\; \kappa_u\equiv|k_2+k_4-k_1-k_3|\;,
\end{equation}
and 
\begin{equation}
\mathcal{E}=\frac{k_1+k_2+k_3-k_4}{2} \;\;\; \text{(case I)}\;,\;\;\quad\quad \mathcal{E}=k_1 \;\;\; \text{(case II)}
\end{equation}
is the extremality. We have also defined the conformal cross ratios 
\begin{equation}
U=\frac{x_{12}^2x_{34}^2}{x_{13}^2x_{24}^2}=z\bar{z}\;,\quad V=\frac{x_{14}^2x_{23}^2}{x_{13}^2x_{24}^2}=(1-z)(1-\bar{z})\;,
\end{equation}
the R-symmetry cross ratio
\begin{equation}
\alpha=\frac{(v_1\cdot v_3)(v_2\cdot v_4)}{(v_1\cdot v_2)(v_3\cdot v_4)}\;,
\end{equation}
and the $SU(2)_L$ flavor symmetry cross ratio
\begin{equation}
\beta=\frac{(\bar{v}_1\cdot \bar{v}_3)(\bar{v}_2\cdot \bar{v}_4)}{(\bar{v}_1\cdot \bar{v}_2)(\bar{v}_3\cdot \bar{v}_4)}\;.
\end{equation}
It is not difficult to see that the function $\mathcal{G}^{I_1I_2I_3I_4}(U,V;\alpha,\beta)$ is a polynomial in $\alpha$ of degree $\mathcal{E}$ and a polynomial in $\beta$ of degree $\mathcal{E}-2$. 

\vspace{0.5cm}

\noindent{\it Superconformal Ward identities}

\vspace{0.2cm}

\noindent The form (\ref{GandcalG}) exploits only the bosonic part of the superconformal group. The fermionic generators impose further constraints in the form of superconformal Ward identities \cite{Dolan:2004mu}
\begin{equation}\label{scfWardida}
(z\partial_z-\epsilon \alpha\partial_\alpha)\mathcal{G}^{I_1I_2I_3I_4}(z,\bar{z};\alpha,\beta)\bigg|_{\alpha=\frac{1}{z}}=0\;.
\end{equation}
Another set of identities can be obtained by replacing $z$ with $\bar{z}$. 

It is useful to compare these superconformal Ward identities to the ones with maximal superconformal symmetry, which take the same form. In the latter case, the R-symmetry group is $SO(n)$ with $n=5,6,8$ and therefore the four-point functions have two independent R-symmetry cross ratios $\alpha$, $\bar{\alpha}$ which are on the same footing. Meanwhile, maximally superconformal theories do not have flavor symmetries and there is no cross ratio $\beta$. The superconformal Ward identities in the maximal case can be viewed as replacing $\beta$ with $\bar{\alpha}$ in (\ref{scfWardida}) and omitting the flavor indices $I_i$
\begin{equation}
(z\partial_z-\epsilon \alpha\partial_\alpha)\mathcal{G}(z,\bar{z};\alpha,\bar{\alpha})\bigg|_{\alpha=\frac{1}{z}}=0\quad\quad (\text{maximally superconformal})\;.
\end{equation}
There are in total four identities by interchanging $z$, $\bar{z}$, and $\alpha$, $\bar{\alpha}$. While the non-maximal case (\ref{scfWardida}) has naively half as many identities, the independent flavor structures give rise to more constraints, making the superconformal Ward Identities as powerful in the end.

\newpage

\noindent{\it Flavor projectors and crossing matrices}

\vspace{0.2cm}

\noindent That external operators transform in the adjoint representation of the flavor group $G_F$ brings extra complexities. To deal with the flavor symmetry indices, we introduce the following projectors
\begin{equation}
{\rm P}^{I_1I_2|I_3I_4}_{a}\;,\quad {\rm P}^{I_1I_4|I_3I_2}_{a}\;,\quad {\rm P}^{I_1I_3|I_2I_4}_{a}\;,
\end{equation}
for s-, t-, and u-channels respectively. The index $a$ runs over all irreducible representations in the tensor product of ${\bf adj}_{G_F}\times {\bf adj}_{G_F}$. The projector ${\rm P}^{I_1I_2|I_3I_4}_{a}$, for example, represents the flavor tensor structure that corresponds to exchanging the flavor symmetry representation $a$ in the $12\to34$ channel. These projectors satisfy the following relations
\begin{eqnarray}
\nonumber &&{\rm P}^{I_1I_2|I_3I_4}_{a}=(-1)^{|{\bf R}_a|}{\rm P}^{I_2I_1|I_3I_4}_{a}\;,\\
\nonumber &&{\rm P}^{I_1I_2|I_3I_4}_{a}={\rm P}^{I_3I_4|I_1I_2}_{a}\;,\\
 && {\rm P}^{I_1I_2|I_3I_4}_{a}{\rm P}^{I_1I_2|I_3I_4}_{b}=\delta_{ab}{\rm dim}({\bf R}_a)\;,\\
\nonumber && {\rm P}^{I_1I_2|I_3I_4}_{a}{\rm P}^{I_4I_3|I_5I_6}_{b}=\delta_{ab}{\rm P}^{I_1I_2|I_5I_6}_{a}\;,
\end{eqnarray}
which follow from basic representation theory. The use of projectors makes it easy to talk about independent flavor structures. 
Note that projectors in different channels $\{{\rm P}^{I_1I_2|I_3I_4}_{a}\}$, $\{{\rm P}^{I_1I_4|I_3I_2}_{a}\}$, $\{{\rm P}^{I_1I_3|I_2I_4}_{a}\}$, with $a$ running over all representations in ${\bf adj}_{G_F}\times {\bf adj}_{G_F}$, separately form a basis. Using this we can project, for example, the superconformal Ward identities (\ref{scfWardida}) into the intermediate s-channel representations, and obtain independent constraints
\begin{equation}\label{scfWardidb}
(z\partial_z-\epsilon \alpha\partial_\alpha)\left({\rm P}^{I_1I_2|I_3I_4}_{a}\mathcal{G}^{I_1I_2I_3I_4}(z,\bar{z};\alpha,\beta)\right)\bigg|_{\alpha=\frac{1}{z}}=0\;,\quad {\bf R}_a\in {\bf adj}_{G_F}\times {\bf adj}_{G_F}\;.
\end{equation}

In applications it is also often convenient to consider the decomposition of t- and u-channel projectors into the s-channel. This is achieved by contracting the external adjoint indices of the projectors to form flavor crossing matrices
\begin{equation}
({\rm F}_t)_a{}^b=\frac{1}{{\rm dim}({\bf R}_a)}{\rm P}^{I_3I_2|I_1I_4}_{a}{\rm P}^{I_1I_2|I_3I_4}_{b}\;,\quad ({\rm F}_u)_a{}^b=\frac{1}{{\rm dim}({\bf R}_a)}{\rm P}^{I_4I_2|I_3I_1}_{a}{\rm P}^{I_1I_2|I_3I_4}_{b}\;.
\end{equation}
These matrices can be computed using the methods of \cite{Cvitanovic:2008zz}, and explicit examples can be found in Table 6 of \cite{Chang:2017cdx}. Such flavor crossing matrices accompany the standard crossing equations for four-point functions 
\begin{eqnarray}
\nonumber\sum_b (F_t)_a{}^b G^a(x_3,x_2,x_1,x_4)&=&G^b (x_1,x_2,x_3,x_4)\;,\\
\sum_b (F_u)_a{}^b G^a(x_4,x_2,x_3,x_1)&=&G^b (x_1,x_2,x_3,x_4)
\end{eqnarray}
where 
\begin{equation}
G^{I_1I_2I_3I_4}(x_1,x_2,x_3,x_4)=\sum_a {\rm P}^{I_1I_2|I_3I_4}_{a} G^a (x_1,x_2,x_3,x_4)\;,
\end{equation}
and we have suppressed R-symmetry and the other $SU(2)$ flavor symmetry.

\subsection{Mellin representation}\label{subsec:Mellinrepresentation}
The amplitude interpretation of holographic correlators is manifest in the Mellin representation \cite{Mack:2009mi,Penedones:2010ue}
\begin{eqnarray}\label{inverseMellin}
&&\mathcal{G}^{I_1I_2I_3I_4}(U,V;\alpha,\beta)=\int_{-i\infty}^{i\infty} \frac{dsdt}{(4\pi i)^2} U^{\frac{s}{2}-a_s}V^{\frac{t}{2}-a_t}\mathcal{M}^{I_1I_2I_3I_4}(s,t;\alpha,\beta)\,\Gamma^{(\epsilon)}_{\{k_i\}}\;,\\
\nonumber &&\Gamma^{(\epsilon)}_{\{k_i\}}=\Gamma[\tfrac{\epsilon(k_1+k_2)-s}{2}]\Gamma[\tfrac{\epsilon(k_3+k_4)-s}{2}]\Gamma[\tfrac{\epsilon(k_1+k_4)-t}{2}]\Gamma[\tfrac{\epsilon(k_2+k_3)-t}{2}]\Gamma[\tfrac{\epsilon(k_1+k_3)-u}{2}]\Gamma[\tfrac{\epsilon(k_2+k_4)-u}{2}]
\end{eqnarray}
where $\mathcal{M}^{I_1I_2I_3I_4}$, the Mellin amplitude, is naturally identified as the scattering amplitude in AdS, and $a_s=\frac{\epsilon}{2}(k_1+k_2)-\epsilon\mathcal{E}$, $a_t=\epsilon\mathcal{E}-\frac{\epsilon}{2}(k_1-k_4)$,  $s+t+u=\epsilon \sum_{i=1}^4 k_i\equiv \epsilon\Sigma$. The advantage of this representation can already be seen at the level of individual Witten diagrams. For example, the Mellin amplitude of a contact diagram with $2L$ derivatives is a degree-$L$ polynomial in Mellin variables.  On other hand, an exchange Witten diagram with exchanged dimension $\Delta$ and spin $\ell$ has an amplitude that has a series of simple poles with polynomial residues
\begin{equation}\label{MellinWitten}
\mathcal{M}^{(s)}_{\Delta,\ell}(s,t)=\sum_{m=0}^\infty \frac{\mathcal{Q}_{m,\ell}(t,u)}{s-\Delta+\ell-2m}+\mathcal{P}_{\ell-1}(s,t)\;.
\end{equation}
Here $\mathcal{Q}_{m,\ell}$ are degree-$\ell$ polynomials, and the  regular piece $\mathcal{P}_{\ell-1}$ is a polynomial of degree $\ell-1$. Note that the regular piece is ambiguous and can be modified by adding contact diagrams with no more than $2(\ell-1)$ derivatives. This modification corresponds to choosing different cubic couplings which are equivalent only on-shell. Mellin amplitudes of exchange diagrams up to spin 2 are given for a specific (very convenient) choice of contact terms in Appendix \ref{app:exchangeWittenMellin}.   

We now briefly review how to implement the superconformal Ward identity (\ref{scfWardidb}) in Mellin space. The main difficulty is that 
(\ref{scfWardidb}) is not symmetric in $z$ and $\bar{z}$, which causes square roots to appear when expressed in terms of $U$ and $V$. These square roots are difficult to interpret in Mellin space. However, as we will see below, by taking independent linear combinations of (\ref{scfWardidb}) and its counterpart with $z\leftrightarrow\bar{z}$, all the $z$, $\bar{z}$ dependence becomes polynomial in $U$ and $V$. The latter is easy to exploit in Mellin space. The method was first developed in \cite{Zhou:2017zaw}, and was further streamlined in \cite{Alday:2020dtb}. Let us define 
\begin{equation}
\mathcal{M}_a(s,t;\alpha,\beta)={\rm P}^{I_1I_2|I_3I_4}_{a}\mathcal{M}^{I_1I_2I_3I_4}(s,t;\alpha,\beta)=\sum_{q=0}^{\mathcal{E}}\alpha^q\mathcal{M}^{(q)}_a(s,t;\beta)\;.
\end{equation}
Starting from the basic identity
\begin{equation}
z\partial_z=U\partial_U-\frac{z}{1-z}V\partial_V\;,
\end{equation}
and using the fact that $U\partial_U$, $V\partial_V$ act multiplicatively in the Mellin representation (\ref{inverseMellin})
\begin{equation}
U\partial_U\to\left(\frac{s}{2}-a_s\right)\times\;,\quad V\partial_V\to\left(\frac{t}{2}-a_t\right)\times\;,
\end{equation}
we find the Ward identity (\ref{scfWardidb}) takes the following form acting on the Mellin amplitude
\begin{equation}\label{MellinWI1}
\sum_{q=0}^{\mathcal{E}}\left((1-z)z^{\mathcal{E}-q}\left(\frac{s}{2}-a_s-q\right)-z^{\mathcal{E}-q+1}\left(\frac{t}{2}-a_t\right)\right)\mathcal{M}^{(q)}_a(s,t;\beta)=0\;.
\end{equation} 
Here we have made the inverse Mellin integrals implicit, with the understanding that the powers of $z$ and $1-z$ multiply the integrals from outside. Replacing $z$ with $\bar{z}$, we obtain another identity
\begin{equation}\label{MellinWI2}
\sum_{q=0}^{\mathcal{E}}\left((1-\bar{z})\bar{z}^{\mathcal{E}-q}\left(\frac{s}{2}-a_s-q\right)-\bar{z}^{\mathcal{E}-q+1}\left(\frac{t}{2}-a_t\right)\right)\mathcal{M}^{(q)}_a(s,t;\beta)=0\;.
\end{equation} 
We now consider two independent linear combinations of (\ref{MellinWI1}) and (\ref{MellinWI2}), and get 
\begin{equation}\label{MellinWIpm}
\sum_{q=0}^{\mathcal{E}}\left((\zeta^{\mathcal{E}-q}_\pm-
\zeta^{(\mathcal{E}-q+1)}_\pm)\left(\frac{s}{2}-a_s-q\right)-\zeta^{(\mathcal{E}-q+1)}_\pm\left(\frac{t}{2}-a_t\right)\right)\mathcal{M}^{(q)}_a(s,t;\beta)=0
\end{equation} 
where we have defined 
\begin{equation}
\zeta^{(n)}_+=z^n+\bar{z}^n\;,\quad \zeta^{(n)}_-=\frac{z^n-\bar{z}^n}{z-\bar{z}}\;.
\end{equation}
Crucially, $\zeta^{(n)}_\pm$ can be expressed in terms of $U$ and $V$ using 
\begin{align}
\begin{split}
\zeta^{(n)}_+&=2^{1-n}\,\sum_{k=0}^{\lfloor n/2 \rfloor}
\binom{n}{2k}\,\left((1+U-V)^2-4U\right)^k\,(1+U-V)^{n-2k}\,,\\
\zeta^{(n)}_-&=2^{1-n}\,\sum_{k=0}^{\lfloor n/2 \rfloor}
\binom{n}{2k+1}\,\left((1+U-V)^2-4U\right)^k\,(1+U-V)^{n-2k-1}\,,
\end{split}
\end{align}
where we remind the reader that polynomials in $U$ and $V$ act as difference operators in Mellin space.  More precisely, each monomial $U^mV^n$ multiplying the Mellin integral (\ref{inverseMellin}) can be absorbed by shifting $s$ and $t$, and thus becomes a difference operator $\widehat{U^mV^n}$ acting as 
\begin{equation}\label{diffUmVn}
\widehat{U^mV^n}\circ \mathcal{M}(s,t)=\frac{\Gamma^{(\epsilon)}_{\{k_i\}}(s-2m,t-2n)}{\Gamma^{(\epsilon)}_{\{k_i\}}(s,t)}\mathcal{M}(s-2m,t-2n)\;.
\end{equation}
Replacing the polynomials in (\ref{MellinWIpm}) with the difference operators gives the superconformal Ward identities in Mellin space.

\section{MRV limit and bootstrap method}\label{MRVandbootstrap}
From holography, we expect the super gluon Mellin amplitudes to have the following general structure at tree level
\begin{equation}\label{Mgeneralstructure}
\mathcal{M}^{I_1I_2I_3I_4}=\mathcal{M}^{I_1I_2I_3I_4}_s+\mathcal{M}^{I_1I_2I_3I_4}_t+\mathcal{M}^{I_1I_2I_3I_4}_u+\mathcal{M}_{\rm con}^{I_1I_2I_3I_4}\;.
\end{equation}
Here $\mathcal{M}^{I_1I_2I_3I_4}_{s,t,u}$ are exchange contributions and $\mathcal{M}_{\rm con}^{I_1I_2I_3I_4}$ comes from contact terms. The contributions $\mathcal{M}^{I_1I_2I_3I_4}_{t,u}$ are related to $\mathcal{M}^{I_1I_2I_3I_4}_s$ by Bose symmetry as usual, but we need to pay additional attention to the flavor symmetry. It is useful to introduce the following flavor structures 
\begin{equation}\label{csctcu}
\mathtt{c}_s=f^{I_1I_2J}f^{JI_3I_4}\;,\quad \mathtt{c}_t=f^{I_1I_4J}f^{JI_2I_3}\;,\quad \mathtt{c}_u=f^{I_1I_3J}f^{JI_4I_2}
\end{equation}
where we have dropped the flavor indices  in $\mathtt{c}_{s,t,u}$ to avoid cluttering the notation, and $f^{IJK}$ are the structure constants of the flavor symmetry algebra. In this paper we normalize the structure constants as
\begin{equation}
f^{IL}{}_{K}f^{JK}{}_L=\psi^2h^\vee\delta^{IJ}
\end{equation}
where $h^{\vee}$ is the dual Coxeter number and $\psi$ is the length squared of the longest root of the flavor group. Since all the exchanged fields are in the adjoint representation of the flavor group, the exchange contributions $\mathcal{M}^{I_1I_2I_3I_4}_{s,t,u}$ are respectively proportional to $\mathtt{c}_{s,t,u}$. These flavor structures can also be expressed in terms of the projectors introduced in Section \ref{subsec:kinematics}  
\begin{equation}
\mathtt{c}_s=\psi^2h^\vee{\rm P}^{I_1I_2|I_3I_4}_{{\bf adj}}\;,\quad \mathtt{c}_t=\psi^2h^\vee{\rm P}^{I_1I_4|I_2I_3}_{{\bf adj}}\;,\quad \mathtt{c}_u=\psi^2h^\vee{\rm P}^{I_1I_3|I_4I_2}_{{\bf adj}}\;.
\end{equation}
Let us denote
\begin{equation}
\mathcal{M}_s^{I_1 I_2 I_3 I_4}=\mathtt{c}_sM_s^{(k_1,k_2,k_3,k_4)},\; \mathcal{M}_t^{I_1 I_2 I_3 I_4}=\mathtt{c}_tM_t^{(k_1,k_2,k_3,k_4)},\;\mathcal{M}_u^{I_1 I_2 I_3 I_4}=\mathtt{c}_uM_u^{(k_1,k_2,k_3,k_4)}
\end{equation}
where we have added the superscript $(k_1,k_2,k_3,k_4)$ to manifest the Kaluza-Klein level associated with each external super gluon. Then Bose symmetry relates the t- and u-channel exchange contributions to the s-channel as the permutations of labels $(1,2,3,4)\to(1,4,2,3)$ and $(1,2,3,4)\to(1,3,4,2)$, and gives 
\begin{eqnarray}\label{Bosesymm}
\nonumber M_t^{(k_1,k_2,k_3,k_4)}&=&(\alpha-1)^{\mathcal{E}}(\beta-1)^{\mathcal{E}-2}\bigg(M_s^{(k_1,k_4,k_2,k_3)}\big|^{\{s,t,u\}\to\{t,u,s\}}_{\{\alpha,\beta\}\to\{\frac{1}{1-\alpha},\frac{1}{1-\beta}\}}\bigg)\;,\\
M_u^{(k_1,k_2,k_3,k_4)}&=&(-\alpha)^{\mathcal{E}}(-\beta)^{\mathcal{E}-2}\bigg(M_s^{(k_1,k_3,k_4,k_2)}\big|^{\{s,t,u\}\to\{u,s,t\}}_{\{\alpha,\beta\}\to\{\frac{\alpha-1}{\alpha},\frac{\beta-1}{\beta}\}}\bigg)\;.
\end{eqnarray}
Similarly, under crossing the full Mellin amplitude transforms as 
\begin{eqnarray}\label{Bosesymmfull}
\nonumber \mathcal{M}^{I_1I_2I_3I_4}_{(k_1,k_2,k_3,k_4)}&=&(\alpha-1)^{\mathcal{E}}(\beta-1)^{\mathcal{E}-2}\bigg(\mathcal{M}^{I_1I_4I_2I_3}_{(k_1,k_4,k_2,k_3)}\big|^{\{s,t,u\}\to\{t,u,s\}}_{\{\alpha,\beta\}\to\{\frac{1}{1-\alpha},\frac{1}{1-\beta}\}}\bigg)\;,\\
\mathcal{M}^{I_1I_2I_3I_4}_{(k_1,k_2,k_3,k_4)}&=&(-\alpha)^{\mathcal{E}}(-\beta)^{\mathcal{E}-2}\bigg(\mathcal{M}^{I_1I_3I_4I_2}_{(k_1,k_3,k_4,k_2)}\big|^{\{s,t,u\}\to\{u,s,t\}}_{\{\alpha,\beta\}\to\{\frac{\alpha-1}{\alpha},\frac{\beta-1}{\beta}\}}\bigg)\;,
\end{eqnarray}
or in terms of the flavor stripped amplitudes
\begin{eqnarray}\label{Bosesymmfcrossing}
\nonumber \mathcal{M}^{a}_{(k_1,k_2,k_3,k_4)}&=&\sum_b (-1)^{|\mathbf{R}_a|}({\rm F}_u)_a{}^b (\alpha-1)^{\mathcal{E}}(\beta-1)^{\mathcal{E}-2}\bigg(\mathcal{M}^{b}_{(k_1,k_4,k_2,k_3)}\big|^{\{s,t,u\}\to\{t,u,s\}}_{\{\alpha,\beta\}\to\{\frac{1}{1-\alpha},\frac{1}{1-\beta}\}}\bigg)\;,\\
\mathcal{M}^{a}_{(k_1,k_2,k_3,k_4)}&=&\sum_b (-1)^{|\mathbf{R}_a|}({\rm F}_t)_a{}^b(-\alpha)^{\mathcal{E}}(-\beta)^{\mathcal{E}-2}\bigg(\mathcal{M}^{b}_{(k_1,k_3,k_4,k_2)}\big|^{\{s,t,u\}\to\{u,s,t\}}_{\{\alpha,\beta\}\to\{\frac{\alpha-1}{\alpha},\frac{\beta-1}{\beta}\}}\bigg)\;
\end{eqnarray}
where we have projected the flavor representations into the s-channel and the flavor crossing matrices ${\rm F}_t$, ${\rm F}_u$ were defined in Section \ref{subsec:kinematics}. Note that since the sum of the exchange contributions satisfies Bose symmetry, (\ref{Bosesymmfull}) and (\ref{Bosesymmfcrossing}) must also separately hold for the contact contributions. 

We now outline our general strategy to compute the Mellin amplitudes (\ref{Mgeneralstructure}). 
\begin{enumerate}
\item We fix the contribution of the bosonic components inside each super multiplet exchange by looking at the {\it maximally R-symmetry violating} (MRV) limit to be defined in the next subsection. In this limit, superconformal symmetry dictates the amplitude to exhibit a {\it zero} in a Mandelstam-Mellin variable, which allows us to determine the contribution of each component up to an overall factor. 
\item We use R-symmetry to go away from the MRV limit and restore the general R-symmetry polarization dependence in the multiplet exchange amplitudes. 
\item We write down an ansatz in terms of all possible multiplet exchange amplitudes with unfixed coefficients, as well as contact terms with arbitrary flavor structures. The contact terms should only arise from quartic vertices with zero derivatives in order for the result to be consistent with the flat space limit.  
\item We use the superconformal Ward identities to uniquely solve the unknown parameters up to an overall factor. In fact, with a suitable prescription to restore R-symmetry in the second step, we find all contact terms in the ansatz vanish. Furthermore, these unknown overall factors can be reduced to just one which is related to the flavor central charge, by considering different mixed correlators. 
\end{enumerate}
The MRV notion was first introduced in \cite{Alday:2020lbp,Alday:2020dtb} for maximally superconformal theories. There it was also shown that the zeros dictated by superconformal symmetry fix the relative coefficients of component fields inside each multiplet. Unlike the situation in \cite{Alday:2020lbp,Alday:2020dtb}, however, here we do not know the super primary three-point functions and therefore cannot use them to fix the overall coefficient of each multiplet. Therefore, we have to resort to the superconformal Ward identities in Mellin space, reviewed in Section \ref{subsec:Mellinrepresentation}. The method of using Mellin space superconformal Ward identities to fix the exchange and contact contributions was first introduced and used in \cite{Zhou:2017zaw,Zhou:2018ofp}. In Section \ref{sec:MRV} we will define and study the MRV amplitudes, and in Section \ref{fullmultipletamplitude} we will show how to obtain multiplet exchange amplitudes with arbitrary R-symmetry polarizations. The last two steps of the above plan will be explained in detail in Section \ref{subsec:bootstrap}.

\subsection{MRV amplitudes from zeros}
\label{sec:MRV}
Let us begin by listing the fields that can appear in the exchange amplitudes. A quick look at the super descendant content of the short multiplets tells us that there are only three components allowed inside each multiplet of $\mathcal{O}^{I}_{p}$.\footnote{The 3d case may appear to contain more scalar super descendants of dimension $\epsilon p+1$ in the exchange amplitudes. However, these extra fields are not exchanged as we will explain in Section \ref{subsec:3dflavor}.\label{footnotecomponent3d}} The exchanged component fields are summarized by the table below.
{\begin{center}
 \begin{tabular}{||c| c | c | c ||} 
 \hline
component field & $s^I_p$ & $A^I_{p,\mu}$ & $r^I_p$ \\ [0.5ex] 
 \hline\hline
Lorentz spin $\ell$ & 0 & 1 &  0\\ 
 \hline
conformal dimension $\Delta$ & $\epsilon p$ & $\epsilon p+1$ & $\epsilon p+2$ \\
 \hline
$SU(2)_R$ spin $j_R$ &  $\frac{p}{2}$ & $\frac{p}{2}-1$ & $\frac{p}{2}-2$  \\ [0.5ex] 
 \hline
 $SU(2)_L$ spin $j_L$ &  $\frac{p-2}{2}$ & $\frac{p-2}{2}$ & $\frac{p-2}{2}$  \\ [0.5ex] 
 \hline
\end{tabular}
\end{center}}

The s-channel multiplet exchange amplitude therefore can be written as the linear combination of the three fields
\begin{equation}\label{multipletS}
\mathcal{S}_p(s,t;\alpha)=\lambda_{s_p}\mathcal{Y}_{p}(\alpha) \mathcal{M}_{\epsilon p,0}(s,t)+\lambda_{A_p}\mathcal{Y}_{p-2}(\alpha) \mathcal{M}_{\epsilon p+1,1}(s,t)+\lambda_{r_p}\mathcal{Y}_{p-4}(\alpha) \mathcal{M}_{\epsilon p+2,0}(s,t)\;.
\end{equation}
Here we have stripped away the color structure $\mathtt{c}_s$. By $\mathcal{M}_{\Delta,\ell}$ we denote the s-channel exchange Mellin amplitudes for bosonic Witten diagrams, $\lambda_{\rm field}$ are coefficients to be fixed,  and $\mathcal{Y}_p$ are the R-symmetry polynomials\footnote{The normalization is chosen in such a way that in the s-channel OPE limit $\alpha \to \infty$, the leading term $\alpha ^{\frac{1}{4} \left(2 p-\kappa _t-\kappa _u\right)}$ appears with unit coefficient.} 
\begin{equation}\label{YpSU2}
\mathcal{Y}_p(\alpha)=\frac{\Gamma \left[\frac{2 p-\kappa _t-\kappa _u+4}{4} \right] \Gamma \left[\frac{2 p+\kappa _t+\kappa _u+4}{4} \right]}{p!}\,P_{\frac{p}{2}-\frac{\kappa_t+\kappa_u}{4}}^{(\frac{\kappa_t}{2},\frac{\kappa_u}{2})}(2\alpha-1)
\end{equation}
where $P^{(a,b)}_m(x)$ is the Jacobi polynomial and we assumed that we are in case I of (\ref{twocases}).\footnote{Considering case I is sufficient to obtain the general result. The final expressions for the Mellin amplitudes are independent of which of the two cases we consider.}  Moreover, we have temporarily suppressed the $SU(2)_L$ dependence, since the whole multiplet transforms in the same representation as we can see from the above table. This $SU(2)_L$ dependence will only appear as a multiplicative polynomial in $\beta$, which we will restore later on.  

Let us now define the u-channel MRV limit of Mellin amplitudes as the R-symmetry slice with $\alpha=0$, which corresponds to setting $v_1=v_3$.\footnote{Note that we have assumed the ordering $k_1\leq k_2\leq k_3\leq k_4$. For arbitrary ordering, we should set $v_1=v_3$ if $k_1+k_3\leq k_2+k_4$, and $v_2=v_4$ if $k_1+k_3> k_2+k_4$. The generalization of the MRV notion to other channels is obvious.} We expect the MRV amplitudes 
\begin{equation}
{\bf MRV}^{I_1I_2I_3I_4}(s,t)=\mathcal{M}^{I_1I_2I_3I_4}(s,t;\alpha=0)
\end{equation}
to have the following two features: 
\begin{enumerate}
\item there are no poles in $u$;
\item there is a zero in $u$ at $u=\epsilon\max\{k_1+k_3,k_2+k_4\}$. 
\end{enumerate}
Note that $v_1=v_3$ implies that the only R-symmetry representation which can be exchanged in the u-channel has spin-$\frac{k_1+k_3}{2}$. The first property then follows from the fact that the R-symmetry polynomials associated with the u-channel exchanges vanish in the MRV limit, because all the cubic couplings associated with exchange Witten diagrams are non-extremal, {\it i.e.}, with $p<k_1+k_3$. Going to the MRV limit suppresses these exchange contributions. The second property is the statement that long operators with twist $\epsilon\max\{k_1+k_3,k_2+k_4\}$ should decouple in the MRV limit.\footnote{In a long multiplet of which the super primary has $SU(2)_R$ spin $j$, the highest $SU(2)_R$ spin of the super descendants is $j+2$. For the whole long multiplet (exchanged in the u-channel) to fit inside the correlator, we therefore must have $j+2\leq\frac{1}{2}\min\{k_1+k_3,k_2+k_4\}$. Meanwhile, for the multiplet to be visible in the u-channel MRV limit, the super primary must satisfy $j+2=\frac{1}{2}\min\{k_1+k_3,k_2+k_4\}$, and the visible component is the super descendant with $SU(2)_R$ spin $j+2$. For 4d $\mathcal{N}=2$, 5d $F_4$, and 6d $(1,0)$, this super descendant is obtained from the super primary by acting with $Q^4$ (see, {\it e.g.}, \cite{Cordova:2016emh}). The super primary with twist $\tau$ decouples in the MRV limit, while a super descendant with twist $\tau+2$ remains. For 3d $\mathcal{N}=3$, the spin $j+2$ operator first appears in $Q^2$ acting on the super primary, and the minimal visible twist is $\tau+1$.
} The zero in $u$ precisely offsets the double pole at $u=\epsilon\max\{k_1+k_3,k_2+k_4\}$ from the Gamma function factor, which would otherwise lead to logarithmic singularities upon evaluating the inverse Mellin transformation. Such logarithmic singularities are associated with anomalous dimensions, which are characteristic of long operators.\footnote{Strictly speaking, this argument applies when the Gamma function poles overlap, which is not always guaranteed when $\epsilon$ is an half integer. However, for these values of $\epsilon$ we can just assume $k_i\in2\mathbb{Z}$ so that the poles overlap and we work out the coefficients. The coefficients are essentially the coefficients of bosonic conformal blocks in the superconformal block, and do not depend on the parity of $k_i$. Therefore the solution to these coefficients obtained with this assumption is in fact general.}  These conditions are in fact satisfied by each multiplet either in the s-channel or t-channel, and turn out to fix the multiplet exchange amplitudes uniquely up to an overall constant as we will see below. 

The presence of zeros in the MRV limit fixes the contact term ambiguity for each bosonic amplitude $\mathcal{M}_{\Delta,\ell}$ in (\ref{multipletS}). This can be seen by examining the u-channel Regge limit which corresponds to $s\to \infty$, keeping $u$ fixed. Since the process involves at most spin-1 fields, the residue at each pole is linear in Mandelstam variables. On the other hand,  in the MRV limit we must produce a zero factor $(u-\epsilon\max\{k_1+k_3,k_2+k_4\})$, which implies the MRV amplitude can only be of the form
\begin{equation}
{\bf MRV}\sim (u-\epsilon\max\{k_1+k_3,k_2+k_4\})\left(\sum_i \frac{\mu_{s,i}}{s-\nu_{s,i}}+\sum_j \frac{\mu_{t,j}}{t-\nu_{t,j}}\right)\;.
\end{equation}
Here $\mu_{s,i}$, $\nu_{s,i}$, {\it etc.}, are numbers. Note that this behavior of MRV amplitudes forbids us from adding additional contact terms -- the only allowed zero-derivative contact terms correspond to constants, but these would be incompatible with the condition of having a zero in $u$. Therefore we can see that the s-channel part of the MRV amplitude behaves as $s^{-1}$ in the u-channel Regge limit. Note that this Regge behavior is not satisfied by a generic exchange Mellin amplitude of a spin-$\ell$ field, which goes like $s^{\ell-1}$. However, we also note that a spin-$\ell$ exchange Witten diagram can absorb, via redefining cubic vertices, contact terms with up to $2(\ell-1)$ derivatives. These contact terms correspond to degree $\ell-1$ polynomials in Mellin space, and can be precisely used to improve the Regge behavior to $s^{-1}$. In practice, this improvement means we eliminate $t$ in favor of $u$ and $m$  from the numerators in (\ref{MellinWitten})  and discard all  regular terms
\begin{equation}\label{MellinPR}
\mathcal{P}^{(s)}_{\Delta,\ell}(s,t)=\sum_{m=0}^\infty \frac{\mathcal{Q}_{m,\ell}(\sum_i\Delta_i-u-(\Delta-\ell+2m),u)}{s-\Delta+\ell-2m}\;.
\end{equation}
These exchange Witten diagrams with improved u-channel Regge behavior were called the Polyakov-Regge blocks in \cite{Mazac:2019shk,Sleight:2019ive}.

Using the Polyakov-Regge blocks in (\ref{multipletS}), we can now impose the u-channel zeros at $\alpha=0$. This condition must hold for each pole, and therefore imposes strong conditions. We find that $\lambda_{A_p}$ and $\lambda_{r_p}$ can be uniquely solved in terms of $\lambda_{s_p}$
\begin{eqnarray}\label{sollambda}
\lambda_{A_p}&=&-\frac{\mathcal{Y}_{p}(0)}{\mathcal{Y}_{p-2}(0)}\frac{\epsilon  (k_1-k_2+p) (k_3-k_4+p)}{4 p (p \epsilon +1)}\lambda_{s_p}\;,\\
\nonumber \lambda_{r_p}&=&\frac{\mathcal{Y}_{p}(0)}{\mathcal{Y}_{p-4}(0)}\frac{\epsilon ^2 (k_1-k_2+p-2) (k_1-k_2+p) (k_3-k_4+p-2) (k_3-k_4+p)}{16 (p-2) (p-1) (p \epsilon +1) (p \epsilon -\epsilon +1)}\lambda_{s_p}\;.
\end{eqnarray}
The MRV limit of $\mathcal{S}_p$ then reduces to 
\begin{equation}
\mathcal{S}_p^{\rm MRV}=-\lambda_{s_p} \mathcal{Y}_p(0)\sum_{m=0}^\infty \frac{2 (p-1) p (k_2 \epsilon +k_4 \epsilon -u)f_{m,0}|_{\Delta_E=\epsilon p}}{(k_1-k_2-p) (-k_3+k_4+p) (m+p \epsilon -\epsilon )(s-\epsilon p-2m)}\;.
\end{equation}
where $f_{m,0}|_{\Delta_E=\epsilon p}$ is defined in (\ref{fmell}) in Appendix \ref{app:exchangeWittenMellin}. 

With the solution \eqref{sollambda} at hand, let us show that our ansatz (\ref{multipletS}) obeys the s-channel Bose symmetry up to contact terms. These contact terms are important, and will be discussed in the next subsection. Bose symmetry dictates that the contribution to a four-point function from each exchanged field in the s-channel must be separately invariant upon applying $1 \leftrightarrow 2$ and $3 \leftrightarrow 4$ to the external bosons. To determine the $1 \leftrightarrow 2$ (or $3 \leftrightarrow 4$) interchage parity of the fields allowed by our selection rules, we note that any $(\frac{p}{2}-n,\frac{p-2}{2})$ representation of $SU(2)_R\times SU(2)_L$ appearing in the product of $(\frac{k_1}{2},\frac{k_1-2}{2})$ and $(\frac{k_2}{2},\frac{k_2-2}{2})$ has parity $(-1)^{n+1}$. Furthermore, under $1\leftrightarrow 2$ the Lorentz spin of the exchanged field contributes another factor $(-1)^{\ell}$ in a spin-$\ell$ exchange Witten diagram (modulo contact terms). Combined with the fact that the adjoint representation of $G_F$ is antisymmetric in the tensor product of two adjoints, this shows that every component in the table above \eqref{multipletS} has positive total parity. It is therefore necessary for each component field to appear with symmetric three-point function coefficients. This can be easily checked to be the case after using the explicit expressions for $\mathcal{Y}_{p-2r}(0)$, since both lines of \eqref{sollambda} depend only on $(k_1 - k_2)^2$ and $(k_3 - k_4)^2$.

\subsection{Full multiplet exchange amplitudes}\label{fullmultipletamplitude}
From the MRV limit, we can restore the full $\alpha$-dependence by using R-symmetry. To accomplish this, we continue to use the Polyakov-Regge blocks in (\ref{multipletS}) and plug in the solution (\ref{sollambda}). However, we now use the R-symmetry polynomials for general $\alpha$ rather than the special value $\alpha=0$. It is more convenient to write each R-symmetry polynomial (\ref{YpSU2}) as an expansion in powers of $(1-\alpha)$
\begin{equation}
\mathcal{Y}_p(\alpha)=\sum_{i = 0}^{\frac{2p-\kappa_t-\kappa_u}{4}}
\frac{\Gamma \left[\frac{2 p+\kappa _t-\kappa _u+4}{4}\right] \Gamma \left[\frac{4 i-2 p+\kappa _t+\kappa _u}{4}\right] \Gamma \left[\frac{4 i+2 p+\kappa _t+\kappa _u+4}{4}\right]}{i!\,p!\, \Gamma \left[\frac{2i+2+\kappa _t}{2}\right] \Gamma \left[\frac{-2 p+\kappa _t+\kappa _u}{4} \right]}(1-\alpha)^i\;.
\end{equation}
We find that at each simple pole $s=\epsilon p+2m$ there is a linear factor of the Mandelstam variable $u$ of the form 
\begin{equation}
u+\rho(i,m)\;,
\end{equation}
which breaks the s-channel Bose symmetry since $t$, the counterpart of $u$, is missing. To restore the Bose symmetry, we  replace $m$ in this factor by its solution from 
\begin{equation}
s+t+u=\epsilon (k_1+k_2+k_3+k_4)\;,\quad s=\epsilon p+2m\;.
\end{equation}
This prescription amounts to choosing a particular contact term in (\ref{multipletS}). The exchange amplitude now becomes
\begin{equation}\label{Sp}
\mathcal{S}_p(s,t;\alpha)=\sum_{m=0}^\infty \sum_{i = 0}^{\frac{2p-\kappa_t-\kappa_u}{4}} \frac{\mathcal{R}_{p,m}^i(t,u)}{s-\epsilon p-2m}(1-\alpha)^i\;,
\end{equation}
\begin{equation}\label{RescalR}
\mathcal{R}_{p,m}^i(t,u)=\lambda_{s_p} K^i_{p}(t,u)B^i_{p,m}E^i_p
\end{equation}
where 
\begin{equation}
K^i_{p}(t,u)=-2i(2i+\kappa_t)u^++2i(\kappa_u-2)t^+-\frac{1}{4}(t^--2i\epsilon)(2p-\kappa_t-\kappa_u)(2p+\kappa_t+\kappa_u-4)\;,
\end{equation}
\begin{equation}
B^i_{p,m}=\frac{(-1)^{1+p-\frac{\kappa_t+\kappa_u}{2}+i}p(p-1)\Gamma[\epsilon(p-1)]\Gamma[\epsilon p]}{p!\,m!\,(m+\epsilon(p-1))!\Gamma[\frac{\epsilon}{2}(k_1+k_2-p)-m]\Gamma[\frac{\epsilon}{2}(k_3+k_4-p)-m]}\;,
\end{equation}
\begin{align}
\begin{split}
E^i_p=&\frac{\epsilon\,
 \Gamma \left[\frac{2 p-\kappa _t-\kappa _u+4}{4} \right]
 \Gamma \left[\frac{2 p+\kappa _t-\kappa _u}{4} \right] }{4\, i!\,
 \Gamma \left[\frac{2i+2+\kappa _t}{2}\right]
 \Gamma \left[\frac{-4 i+2 p-\kappa _t-\kappa _u+4}{4}\right]
 \Gamma \left[\frac{2 p \epsilon -\left(\kappa _t+\kappa _u\right) \epsilon +4}{4}\right]}\\
& \times \frac{\Gamma \left[\frac{4 i+2 p+\kappa _t+\kappa _u-4}{4} \right]}{\Gamma \left[\epsilon \,\frac{2 p-\kappa _t+\kappa _u}{4} \right] 
\Gamma \left[\epsilon\,\frac{2 p+\kappa _t+\kappa _u}{4}\right] 
 \Gamma \left[\epsilon \,\frac{2 p+\kappa _t-\kappa _u}{4} \right]}\;,
\end{split}
\end{align}
and 
\begin{equation}
u^\pm=u\pm \frac{\epsilon}{2}\kappa_u-\frac{\epsilon}{2}\Sigma\;,\quad\quad t^\pm=t\pm \frac{\epsilon}{2}\kappa_t-\frac{\epsilon}{2}\Sigma\;.
\end{equation}
The same prescription for restoring Bose symmetry was used in \cite{Alday:2020lbp,Alday:2020dtb} for the maximally superconformal cases, and was shown to have the additional benefit that the full amplitudes can be written only in terms of these exchange amplitudes with no extra contact terms. We will see the same happens for the case with half maximal superconformal symmetry. 

\subsection{Bootstrapping full correlators}
\label{subsec:bootstrap}
We are now ready to assemble the pieces together and write down an ansatz for the full correlator. The ansatz has the general form of  (\ref{Mgeneralstructure}). The s-channel exchange contribution  reads
\begin{equation}\label{Mssumofmultiplet}
\mathcal{M}^{I_1I_2I_3I_4}_s=\mathtt{c}_s\sum_{p\in \mathcal{I}_s}\mathcal{S}_p(s,t;\alpha)\mathcal{Y}_{p-2}(\beta)\;.
\end{equation}
Here $\mathtt{c}_s$ is the flavor structure associated with the s-channel exchange defined in (\ref{csctcu}), and $\mathcal{S}_p(s,t;\alpha)$ is the exchange amplitude of the multiplet $p$ given in (\ref{Sp}). The function $\mathcal{Y}_{p-2}(\beta)$ is the $SU(2)_L$ flavor symmetry polynomial given by (\ref{YpSU2})\footnote{In (\ref{YpSU2}) the polynomial was defined for the $SU(2)_R$ symmetry. However, kinematically there is no difference between $SU(2)_R$ and $SU(2)_L$, and therefore they have the same symmetry polynomials.}, associated with the spin-$\frac{p-2}{2}$ $SU(2)_L$ representation of multiplet $p$. Note that each multiplet amplitude $\mathcal{S}_p(s,t;\alpha)$ contains an unknown coefficient $\lambda_{s_p}$, see (\ref{RescalR}), which will also be denoted as $\lambda^{(k_1k_2|k_3k_4)}_{p}$ when we want to emphasize the channel and the external weights. However, we should also note that this coefficient has the interpretation of the product of three-point functions of super primary operators
\begin{equation}\label{lambdaCC}
\lambda^{(k_1k_2|k_3k_4)}_{p}=C_{k_1,k_2,p}C_{k_3,k_4,p}\;,
\end{equation}
with our normalizations used for Witten diagrams and $SU(2)_R$, $SU(2)_L$ symmetry polynomials. The summation of $p$ is over a {\it finite} set $\mathcal{I}_s$
\begin{equation}\label{selection-p}
\mathcal{I}_s=\big\{p\;|\;p-\max\{|k_1-k_2|,|k_3-k_4|\}=2\,,\,4\,,\,\ldots\,,\,2\mathcal{E}-2\big\}\;.
\end{equation}
The finite summation range is determined by two constraints: the selection rules of $SU(2)_L$ flavor symmetry and $SU(2)_R$ R-symmetry, and the requirement that extremal couplings must vanish.\footnote{This is because the contact Witten diagrams associated with extremal couplings $C_{k_1,k_2,k_1+k_2}$ are divergent. In order for the effective action to remain finite, these couplings must be absent. Note that there is no contradiction with extremal three-point functions being ``nonvanishing'' on the CFT side, because in CFTs operators in a different basis are usually considered. The natural single-particle operators in the AdS side are  linear combinations of the ``single-trace'' operators and ``multi-trace'' operators in the CFT side \cite{Arutyunov:1999en,Arutyunov:2000ima}. The mixing is such that the extremal correlators are zero. See, {\it e.g.}, Section 2.1 of \cite{Alday:2019nin} and \cite{Aprile:2019rep,Aprile:2020uxk} for detailed discussions.} The t- and u-channels are similar, and can be obtained from the s-channel by Bose symmetry (\ref{Bosesymm}).

For the contact part $\mathcal{M}^{I_1I_2I_3I_4}_{\rm con}$, we will allow all possible flavor and R-symmetry structures but only terms that are independent of the Mandelstam variables
\begin{equation}
\mathcal{M}^{I_1I_2I_3I_4}_{\rm con}=\sum_{{\bf R}_a\in {\bf adj}_{G_F}\times {\bf adj}_{G_F}}{\rm P}^{I_1I_2|I_3I_4}_{a} \sum_{i=0}^{\mathcal{E}}\sum_{j=0}^{\mathcal{E}-2}\delta_{a;i,j}\alpha^i\beta^j\;.
\end{equation}
Here $\delta_{a;i,j}$ are arbitrary coefficients to be fixed. The independence of the Mandelstam variables corresponds to zero-derivative contact interactions. This is required because the Mellin amplitude in the  $s,t,u\to\infty$ limit should coincide with the flat space amplitude \cite{Penedones:2010ue} which has a constant growth behavior. We will make further comments on the flat space behavior of the Mellin amplitudes in Section \ref{sec:flatspace}.

We now impose the superconformal Ward identities on the ansatz $\mathcal{M}^{I_1I_2I_3I_4}$ in Mellin space, as reviewed in Section \ref{subsec:Mellinrepresentation}. We find that all contact interaction parameters $\delta_{a;i,j}$ vanish, and all exchange parameters $\lambda^{(k_ik_j|k_mk_n)}_{p}$ are fixed up to an overall rescaling factor. Furthermore, the ratios of $\lambda^{(k_ik_j|k_mk_n)}_{p}$ are {\it independent} of the flavor group. Let us first make a few comments on these general results. The explicit solutions for different theories will be given in the next two sections.
\begin{itemize}
\item  The independence of the solution on the flavor groups is easy to understand. With all  $\delta_{a;i,j}$ vanishing, the four-point amplitudes have only three flavor structures $\mathtt{c}_s$, $\mathtt{c}_t$, $\mathtt{c}_u$.  As $\mathtt{c}_{s,t,u}$ satisfy the Jacobi identity
\begin{equation}
\mathtt{c}_s+\mathtt{c}_t+\mathtt{c}_u=0\;,
\end{equation}
only two structures are independent, and can be chosen as $\mathtt{c}_s$ and $\mathtt{c}_t$. Therefore, the Mellin superconformal Ward identities are implemented on the combinations
\begin{equation}
M_s^{(k_1,k_2,k_3,k_4)}-M_u^{(k_1,k_2,k_3,k_4)}\;,\quad M_t^{(k_1,k_2,k_3,k_4)}-M_u^{(k_1,k_2,k_3,k_4)}\;,
\end{equation}
which are agnostic about the flavor group $G_F$.
\item The vanishing of the additional contact terms in the ansatz  may seem to be merely a computational outcome of applying the superconformal conformal Ward identities, and is far from obvious. However, in Section \ref{sec:flatspace} we will explain that the absence of these terms is expected by examining the flat space limit of the Mellin amplitudes. Moreover, their absence is also essential for the correlators to exhibit the Parisi-Sourlas supersymmetry as we will discuss in Section \ref{subsec:PSsusy}. We will also perform a complementary check  in Appendix \ref{app:contact}, where we show that there are no contact term solutions to the superconformal Ward identities with less than four derivatives. 
\item Finally, the unfixed overall factors of different correlators are actually not independent. This follows from their interpretation in terms of OPE coefficients (\ref{lambdaCC}). In fact by considering mixed correlators of the type $\langle k k q q\rangle$, we can extract all three-point function coefficients $C_{k_1,k_2,k_3}$ up to a common factor. This factor is fixed by $C_{2,2,2}$ which is related to the flavor central charge, as the $k=2$ multiplet contains the flavor current.  This allow us to ${\it fully}$ fix the answer. 
\end{itemize}

\section{Correlators in theories with eight supercharges}
\label{sec:correlators}
In this section we implement the strategy laid out in Section \ref{MRVandbootstrap} to compute all tree-level four-point super gluon correlators for the theories listed in Table \ref{8PSC}. We find that all correlators can be written as the sum of multiplet exchange amplitudes (\ref{Mssumofmultiplet}), with no extra contact terms. Moreover, thanks to  (\ref{lambdaCC}) we can fix all four-point correlators up to an overall coefficient which is determined by the flavor central charge. This computation can be done by examining $\langle kkqq\rangle$ correlators (and their permutations), from which we can also extract all three-point functions for super primary operators.\footnote{A famous earlier use of this strategy appeared in the derivation of three-point functions for Virasoro minimal models \cite{df84,df85}.} 

Let us expand on this point. We first look at the $\langle 22qq\rangle$ correlators and consider the following ratios
\begin{equation}
\frac{\lambda_{2}^{(22|qq)}}{\lambda_{q}^{(2q|2q)}}=\frac{C_{2,2,2}C_{2,q,q}}{C_{2,q,q}C_{2,q,q}}\;,
\end{equation}
which are fully determined since the overall coefficient drops out. This ratio arises from the $p=2$ multiplet exchange in the s-channel and $p=q$ multiplet exchange in the t-channel, and gives the three-point functions $C_{2,q,q}$ up to the overall factor $C_{2,2,2}$. We then consider the analogous ratio in  $\langle kkqq \rangle$
\begin{equation}
\frac{\lambda_{2}^{(kk|qq)}}{\lambda_{r}^{(kq|kq)}}=\frac{C_{k,k,2}C_{2,q,q}}{C_{k,q,r}C_{k,q,r}}\;,
\end{equation}
from which we can get $C_{k,q,r}$, again up to the overall factor $C_{2,2,2}$. We can then plug these $C_{k,q,r}$ in (\ref{Mssumofmultiplet}) and verify the superconformal Ward identities for general correlators $\langle pqrs \rangle $. This serves as a consistency check of our method and also confirms the absence of contact terms in the general case as well. Finally, the unfixed coefficient $C_{2,2,2}$, or equivalently $\lambda_{2}^{(22|22)}=(C_{2,2,2})^2$, is related to the flavor central charge $C_{\mathcal{J}}$ that appears in the two-point function 
\begin{equation}
\langle \mathcal{J}_\mu^I(x) \mathcal{J}_\nu^J(0) \rangle=\frac{C_{\mathcal{J}}}{V^2_{\hat{S}^{d-1}}}\frac{\delta^{IJ}\mathcal{I}_{\mu\nu}(x)}{x^{2(d-1)}}\;.
\end{equation}
Here $V_{\hat{S}^{d-1}}=2\pi^{\frac{d}{2}}/\Gamma[\frac{d}{2}]$ is the volume of the unit $(d-1)$-sphere, and $\mathcal{I}_{\mu\nu}(x)=\delta_{\mu\nu}-2\frac{x^\mu x^\nu}{x^2}$ is a conformal structure. The central charge $C_{\mathcal{J}}$ was shown in \cite{Chang:2017xmr} to enter  $\lambda_{2}^{(22|22)}$ as 
\begin{equation}\label{lambdaCJ}
\lambda_{2}^{(22|22)}=\frac{2(2\epsilon+1)}{\epsilon}\frac{1}{C_{\mathcal{J}}}\;.
\end{equation}

To reiterate, the general super gluon four-point amplitudes in any spacetime dimensions can be assembled as  
\begin{equation}
\mathcal{M}=\mathtt{c}_sM_s+\mathtt{c}_tM_t+\mathtt{c}_uM_u
\end{equation}
where $M_{s,t,u}$ are related by the crossing relations (\ref{Bosesymm}), and in each channel $\mathtt{c}_sM_s$ is given by the finite sum over multiplets (\ref{Mssumofmultiplet}). The $\lambda$ coefficients for each exchanged multiplet (\ref{RescalR}) are related to the above three-point function coefficients $C_{k_1,k_2,k_3}$ via (\ref{lambdaCC}). We now give the solution for each theory by writing down the three-point functions. 

\subsection{4d $\mathcal{N}=2$: D3-branes near F-theory singularities}
\label{sec:4dN=2}
We start with the 4d $\mathcal{N}=2$ SCFTs arising from D3-branes probing F-theory singularities. We impose the superconformal Ward identities on the ansatz, and find that additional contact terms are absent. As outlined in the above strategy, we focus on the $\langle kkqq \rangle$ correlators. The solution is simply given by
\begin{align}
\begin{split}
\lambda^{(kk|qq)}_{p}=\lambda^{(kq|kq)}_{p}=\lambda^{(22|22)}_{2}\,.
\end{split}
\end{align}
Hence, from (\ref{lambdaCC}) the OPE coefficients are simply given by
\begin{align}
C_{k_1,k_2,k_3}=\,C_{2,2,2}\,. \label{ope-sol-4d}
\end{align}
The coefficient $\lambda^{(22|22)}_{2}$ is given by (\ref{lambdaCJ}), and the central charges were computed in   \cite{Aharony:2007dj}
\begin{equation}
C_{\mathcal{J}}=\frac{12}{2-\nu}N
\end{equation}
where $\nu$ characterizes the 7-brane singularity  type and was defined in (\ref{alphavalue}). By contrast, the stress tensor central charge, which controls the coupling of super gluons to bulk super gravitons, grows as $C_{\mathcal{T}}\sim N^2$. Tree-level super gluon four-point functions exchanging super gravitons are proportional to $C_{\mathcal{T}}^{-1}$, and are therefore subleading to the correlators considered in this paper where only super gluons are exchanged. 

To demonstrate our general result, we give the explicit amplitudes for the $\langle 22kk\rangle$ correlators. The result is 
\begin{align}
\mathcal{M}=\mathtt{c}_s\, {M}_s+\mathtt{c}_t\,{M}_t+\mathtt{c}_u\,{M}_u\,,
\end{align}
where 
\begin{align}
\begin{split}
{M}_s&=\frac{-2 \alpha  (k+2)+k+\alpha  t+\alpha  u-u+2}{(k-2)!\,(s-2) }\,\left(C_{2,2,2}\right)^2,\\
{M}_t&=\frac{(1-\alpha ) (-2 \alpha  k-k+\alpha  s+u-2)}{(k-2)!\, (t-k)}\,\left(C_{2,2,2}\right)^2,\\
{M}_u&=\frac{-\alpha  (2 \alpha  k-3 k-\alpha  s+s+t-2)}{(k-2)!\,(u-k)}\,\left(C_{2,2,2}\right)^2.
\end{split}
\end{align}

\subsection{5d $F_4$: Seiberg exceptional theories}
\label{sec:5d}

We then move onto the Seiberg exceptional theories in 5d. Imposing the Ward identities on $\langle kkqq \rangle$ correlators, we find that there are no further contact terms. The exchange coefficients are given by
\begin{align}
\begin{split}
\lambda^{(kk|qq)}_{p}=&\frac{2^{2(k+q+p)}3^{\tfrac{1}{2}(7-3k-3q-3p)}\,\Gamma[k]\Gamma[q]\Gamma[p]\,\Gamma[3p/4]^4}{\pi\,\Gamma\left[-\tfrac{1}{3}+k\right]\Gamma\left[\tfrac{1}{3}+k\right]\,\Gamma\left[-\tfrac{1}{3}+q\right]\Gamma\left[\tfrac{1}{3}+q\right]\,\Gamma\left[-\tfrac{1}{3}+p\right]\Gamma\left[\tfrac{1}{3}+p\right]\,\Gamma[p/2]^4}\\
& \frac{\Gamma\left[\tfrac{1}{12}(-2+6k+3p)\right]\Gamma\left[\tfrac{1}{12}(2+6k+3p)\right]\Gamma\left[\tfrac{1}{12}(-2+6q+3p)\right]\Gamma\left[\tfrac{1}{12}(2+6q+3p)\right]}{\Gamma\left[\tfrac{1}{4}(2k+p)\right]\Gamma\left[\tfrac{1}{4}(2q+p)\right]}\\
& \frac{\Gamma\left[\tfrac{3}{4}(2k-p)\right]\Gamma\left[\tfrac{3}{4}(2q-p)\right]}{\Gamma\left[k-\tfrac{p}{2}\right]\Gamma\left[q-\tfrac{p}{2}\right]}  \,\lambda^{(kk|qq)}_{2}\,,\\
\lambda^{(kq|kq)}_{p}=&\frac{2^{2(k+q+p)}3^{\tfrac{1}{2}(7-3k-3q-3p)}\,\Gamma[k]\Gamma[q]\Gamma[p]}{\pi\,\Gamma\left[-\tfrac{1}{3}+k\right]\Gamma\left[\tfrac{1}{3}+k\right]\,\Gamma\left[-\tfrac{1}{3}+q\right]\Gamma\left[\tfrac{1}{3}+q\right]\,\Gamma\left[-\tfrac{1}{3}+p\right]\Gamma\left[\tfrac{1}{3}+p\right]}\\
& \frac{\Gamma\left[\tfrac{3}{4}(k+p-q)\right]^2\,\Gamma\left[\tfrac{3}{4}(k-p+q)\right]^2\,\Gamma\left[\tfrac{3}{4}(-k+p+q)\right]^2}{\Gamma\left[\tfrac{1}{2}(k+p-q)\right]^2\,\Gamma\left[\tfrac{1}{2}(k-p+q)\right]^2\,\Gamma\left[\tfrac{1}{2}(-k+p+q)\right]^2}\\
& \frac{\Gamma\left[\tfrac{1}{12}(-2+3k+3q+3p)\right]^2\,\Gamma\left[\tfrac{1}{12}(2+3k+3q+3p)\right]^2}{\Gamma\left[\tfrac{1}{4}(k+q+p)\right]^2}\,\lambda^{(kk|qq)}_{2}\,.
\end{split}
\end{align}
Using the relation (\ref{lambdaCC}) between $\lambda^{(k_1k_2|k_3k_4)}_{p}$ and the super primary three-point functions, the above results allow us to extract the values of the OPE coefficients
\begin{align}
\begin{split}
C_{k_1,k_2,k_3}&=\frac{2^{2\Xi}\,3^{\tfrac{7-6\Xi}{4}}\Gamma\left[-\tfrac{1}{6}+\tfrac{\Xi}{2}\right] \Gamma\left[\tfrac{1}{6}+\tfrac{\Xi}{2}\right]}{\sqrt{\pi}\,\Gamma\left[\tfrac{\Xi}{2}\right]}\,\prod_{i=1}^3\sqrt{\frac{\Gamma[k_i]}{\Gamma\left[-\tfrac{1}{3}+k_i\right]\Gamma\left[\tfrac{1}{3}+k_i\right]}}\,\frac{\Gamma\left[\tfrac{3\alpha_i}{2}\right]}{\Gamma[\alpha_i]}\,C_{2,2,2}\;.
\end{split}
\end{align}
Here we have introduced
\begin{align}
\alpha_1=\frac{1}{2}(k_2+k_3-k_1)\,, \quad
\alpha_2=\frac{1}{2}(k_3+k_1-k_2)\,, \quad
\alpha_3=\frac{1}{2}(k_1+k_2-k_3)\,, 
\end{align}
and $\Xi=\alpha_1+\alpha_2+\alpha_3$. The overall factor $C_{2,2,2}$ is given by (\ref{lambdaCJ}), and the central charges for the flavor group $E_{N_f+1}$ were computed in \cite{Chang:2017mxc}. We have 
\begin{equation}
C_{\mathcal{J}}^{E_{N_f+1}}=\frac{256\sqrt{2}}{3\pi\sqrt{8-N_f}}N^{\frac{3}{2}}+\mathcal{O}(N^{\frac{1}{2}})\;.
\end{equation}
For comparison, $C_{\mathcal{T}}\sim N^{\frac{5}{2}}$ and super graviton exchanges are again suppressed at large $N$.

Let us also give the $\langle 22kk\rangle$ correlator as an example. The poles in the Mellin amplitudes can be resummed into hypergeometric functions, and the result reads
\begin{align}
\mathcal{M}=\mathtt{c}_s\, {M}_s+\mathtt{c}_t\,{M}_t+\mathtt{c}_u\,{M}_u\,,
\end{align}
where 
\begin{align}
\begin{split}
{M}_s=&\frac{64 (2 \alpha  (-3 k+t+u-6)+3 k-2 u+6) {}_3F_2\left(-\tfrac{1}{2},\tfrac{5-3 k}{2},\tfrac{3-s}{2};\,\tfrac{5}{2},\tfrac{5-s}{2};\,1\right)}{3 \pi ^{5/2} (s-3) \Gamma \left[\tfrac{3}{2} (k-1)\right]}\,\left(C_{2,2,2}\right)^2,\\
{M}_t=&\frac{\Gamma \left[\tfrac{3 k}{2}\right] }{\pi ^2\Gamma \left[\tfrac{1}{2}(3k-1)\right]^2}\, \frac{24 (1-\alpha) (k-1) ((6 \alpha +3) k-2 (\alpha  s+u-3)) }{3 k-2 t}\times \\
& _3F_2\left(-\frac{1}{2},-\frac{1}{2},\frac{3 k-2t}{4};\frac{3 k-1}{2},\frac{3 k-2t+4}{4};1\right)
\,\left(C_{2,2,2}\right)^2,\\
{M}_u=&\frac{\Gamma \left[\frac{3 k}{2}\right]}{\pi ^2 \Gamma \left[\tfrac{1}{2}(3k-1)\right]^2}\,\frac{24 \alpha  (k-1) ((6 \alpha -9) k+2 (-\alpha  s+s+t-3))}{3 k-2 u}\times \\
& _3F_2\left(-\frac{1}{2},-\frac{1}{2},\frac{3 k-2 u}{4};\frac{3 k-1}{2},\frac{3 k-2 u+4}{4};1\right)\,\left(C_{2,2,2}\right)^2.
\end{split}
\end{align}
The $k=2$ case was first computed in \cite{Zhou:2018ofp}.

\subsection{6d $\mathcal{N}=(1,0)$: E-string theory}
\label{sec:6d}
Finally, we consider the E-string theory in 6d. Imposing the Ward identities for $\langle kkqq \rangle$ correlators, one finds that it is not possible to introduce extra contact terms, and
\begin{align}
\begin{split}
\lambda^{(kk|qq)}_{p}=&\frac{6\,\Gamma\left[\tfrac{p+1}{2}\right]^4\,\Gamma\left[\tfrac{1+2k-p}{2}\right]\Gamma\left[\tfrac{1+2q-p}{2}\right]\Gamma\left[\tfrac{-1+2k+p}{2}\right]\Gamma\left[\tfrac{-1+2q+p}{2}\right]}{\pi\,\Gamma\left[-\tfrac{1}{2}+k\right]\Gamma\left[\tfrac{1}{2}+k\right]\Gamma\left[-\tfrac{1}{2}+q\right]\Gamma\left[\tfrac{1}{2}+q\right]\Gamma\left[-\tfrac{1}{2}+p\right]\Gamma\left[\tfrac{1}{2}+p\right]}\,\lambda^{(kk|qq)}_{2}\,,\\
\lambda^{(kq|kq)}_{p}=&\frac{6\,\Gamma\left[\tfrac{1-k+q+p}{2}\right]^2\Gamma\left[\tfrac{1+k-q+p}{2}\right]^2\Gamma\left[\tfrac{1+k+q-p}{2}\right]^2\Gamma\left[\tfrac{-1+k+p+q}{2}\right]^2}{\pi\,\Gamma\left[-\tfrac{1}{2}+k\right]\Gamma\left[\tfrac{1}{2}+k\right]\Gamma\left[-\tfrac{1}{2}+q\right]\Gamma\left[\tfrac{1}{2}+q\right]\Gamma\left[-\tfrac{1}{2}+p\right]\Gamma\left[\tfrac{1}{2}+p\right]}\,\lambda^{(kk|qq)}_{2}\,.
\end{split}
\end{align}
This allows us to extract the values of the OPE coefficients
\begin{align}
C_{k_1,k_2,k_3}=\frac{\sqrt{6}\,\Gamma \left[-\tfrac{1}{2}+\Xi \right]}{\sqrt{\pi}}\,\prod_{i=1}^3\frac{\Gamma\left[\tfrac{1}{2}+\alpha_i\right]}{\sqrt{\Gamma\left[-\tfrac{1}{2}+k_i\right]\Gamma\left[\tfrac{1}{2}+k_i\right]}}\,C_{2,2,2}\,. \label{ope-sol-6d}
\end{align}
Finally, the $E_8$ flavor group central charge of the 6d E-string theory is given by \cite{Chang:2017xmr}
\begin{equation}
C_{\mathcal{J}}=60N^2+90N\;,
\end{equation}
which determines $C_{2,2,2}$ via (\ref{lambdaCJ}). On the other hand, note that the stress tensor central charge scales as $C_{\mathcal{T}}\sim N^3$ at large $N$. Hence super graviton exchanges in the four-point functions are again suppressed. 

As an explicit example of our results, we give here the Mellin amplitudes for the $\langle 22kk\rangle$ correlators in E-string theory
\begin{align}
\mathcal{M}=\mathtt{c}_s\, M_s+\mathtt{c}_tM_t+\mathtt{c}_u\,M_u\,,
\end{align}
where 
\begin{align}
\begin{split}
M_s&=-\frac{2 (k (s-4)-3) (2(2 \alpha -1) k-\alpha  (t+u-8)+u-4)}{(2k-3)!\,(s-6)\, (s-4) }\,\left(C_{2,2,2}\right)^2,\\
M_t&=\frac{2 (1-\alpha ) (k (-2 k+t-2)+1) (-2 (2 \alpha +1) k+\alpha  s+u-4)}{(2k-3)!\,(t-2k)\, (t-2 k-2) }\,\left(C_{2,2,2}\right)^2,\\
M_u&=-\frac{2 \alpha  (k (-2 k+u-2)+1) (2(2 \alpha -3) k-\alpha  s+s+t-4)}{(2k-3)!\,(u-2k) \,(u-2 k-2)}\,\left(C_{2,2,2}\right)^2.
\end{split}
\end{align}
The $k=2$ case reproduces the result found in \cite{Zhou:2018ofp}.

\section{Correlators from flavor branes}
\label{sec:flavor}
\subsection{4d SYM with flavors}
\label{subsec:4dflavor}

In our computation we have shown that given the spectrum as an input, all correlators are fixed by superconformal symmetry up to a common overall coefficient $C_{2,2,2}$. The model of $AdS_5\times S^5$ with flavor D7-branes wrapping $S^3\subset S^5$ has the same vector multiplet spectrum and superconformal symmetry as the 4d $\mathcal{N}=2$ SCFTs arising from D3-brane probing F-theory singularities. Therefore the four-point functions of super gluons (or mesons from the dual field theory perspective) are identical, when expressed in terms of  $C_{2,2,2}$, to those given in section \ref{sec:4dN=2}. The strength of the interactions among mesons was analyzed in \cite{Kruczenski:2003be} where the expected behavior at large $N$ was found to be
\begin{equation}
C_{2,2,2} \sim \frac{1}{\sqrt{N}}.
\end{equation}
In particular, this interaction becomes small for large $N$ but at the same time is dominant with respect to the exchange of gravitons. 

\subsection{3d ABJM with flavors}
\label{subsec:3dflavor}
We now study the $AdS_4\times \mathbb{CP}^3$ model with flavor  D6-branes wrapping $\mathbb{RP}^3\subset \mathbb{CP}^3$,  which preserves 3d $\mathcal{N}=3$ superconformal symmetry. We first clarify the question about what component fields are exchanged in a multiplet, which was alluded to in footnote \ref{footnotecomponent3d}. The Kaluza-Klein spectrum from the D6-branes \cite{Hikida:2009tp} turns out to be the same as the spectrum of a 7d $\mathcal{N}=2$ vector multiplet on $AdS_4\times S^3$ \cite{Imamura:2010sa}. The latter shows up when considering M2-branes near orbifold singularities preserving 3d $\mathcal{N}=4$ superconformal symmetry \cite{Benna:2008zy,Imamura:2008nn,Terashima:2008ba}, and describes the degrees of freedom localized on the singular locus.  These Kaluza-Klein modes are $\frac{1}{3}$-BPS in terms of the 3d $\mathcal{N}=3$ superconformal algebra, while they are $\frac{1}{2}$-BPS with respect to 3d $\mathcal{N}=4$. Let us first consider the $\mathcal{N}=4$ case. The multiplet spectrum is summarized as follows \cite{Imamura:2010sa}
{\begin{center}
 \begin{tabular}{||c| c | c | c |c||} 
 \hline
component field & $s^I_p$ & $A^I_{p,\mu}$ & $r^I_p$ &$t^I_p$\\ [0.5ex] 
 \hline\hline
Lorentz spin $\ell$ & 0 & 1 &  0 & 0\\ 
 \hline
conformal dimension $\Delta$ & $\frac{p}{2}$ & $\frac{p}{2}+1$ & $ \frac{p}{2}+2$ &  $ \frac{p}{2}+1$ \\
 \hline
 $SU(2)_L$ spin $j_L$ &  $\frac{p}{2}-1$ & $\frac{p}{2}-1$ & $\frac{p}{2}-1$ & $\frac{p}{2}-1$ \\ [0.5ex] 
 \hline
 $SU(2)_a$ spin $j_a$ &  $\frac{p}{2}$ & $\frac{p}{2}-1$ & $\frac{p}{2}-2$  & $\frac{p}{2}-1$\\ [0.5ex] 
 \hline
 $SU(2)_b$ spin $j_b$ &  0 & 0 & 0 & 1  \\ [0.5ex] 
 \hline
\end{tabular}
\end{center}}
\noindent where we have kept only the bosonic fields, and $SU(2)_a$, $SU(2)_b$ are the two $SU(2)$ R-symmetry groups of 3d $\mathcal{N}=4$. Notice that the extra scalars $t^I_p$ are charged under $SU(2)_b$ and therefore cannot be exchanged in four-point functions of $s^I_p$.  For $\mathcal{N}=3$, the spectrum is the same as above but $SU(2)_{a,b}$ combine diagonally to give the $SU(2)_R$ R-symmetry group of 3d $\mathcal{N}=3$. One might wonder if the $t^I_p$ scalars are now allowed to appear in the exchange. However we will argue that this does not happen. This is because the multiplet exchange amplitude has the same poles and residues as its associated short multiplet superconformal block in Mellin space. The superconformal blocks of this multiplet with external $s^I_p$ satisfy the {\it same} superconformal Ward identities in $\mathcal{N}=3$ and $\mathcal{N}=4$. Since the $\mathcal{N}=4$ case admits a unique solution \cite{Chang:2017xmr,Bobev:2017jhk,Baume:2019aid}, it would require the conformal blocks associated with $t^I_p$ scalars alone to form another solution to the Ward identities, if they were to appear in the exchange. Such solutions do not exist (see Appendix \ref{app:3dsuperconformalblocks} for details and further comments). To summarize, the 3d $\mathcal{N}=3$ case falls within the situation considered in Section \ref{sec:MRV}, and we can use the techniques developed there to compute four-point functions by setting $\epsilon=\frac{1}{2}$. Moreover, since the exchanged fields are the same as 3d $\mathcal{N}=4$, the correlators we compute can also be interpreted as super gluon four-point functions of 7d $\mathcal{N}=2$ SYM on $AdS_4\times S^3$ by identifying the $SU(2)_R$ spinors as the $SU(2)_a$ spinors.

Following the same procedure and requiring that the Ward identities are satisfied, we find that there are no additional contact terms. The solution for the $\langle kkqq\rangle$ correlators is given by
\begin{align}
\begin{split}
\lambda_{p}^{(kk|qq)}&=
\frac{2^{4-k-p-q}\,\pi\,\Gamma[k]\,\Gamma[q]\,\Gamma[p]}
{\Gamma\left[\tfrac{2+2k-p}{4}\right]\,
\Gamma\left[\tfrac{2+2q-p}{4}\right]\,
\Gamma\left[\tfrac{p+2}{4}\right]^4\,
\Gamma\left[\tfrac{p+2q}{4}\right]\,
\Gamma\left[\tfrac{p+2k}{4}\right]}\,\lambda_{2}^{(kk|qq)}\,,\\
\lambda_{p}^{(kq|kq)}&=\frac{2^{4-k-p-q}\,\pi\,\Gamma[k]\,\Gamma[q]\,\Gamma[p]}
{\Gamma\left[\tfrac{2+k+p-q}{4}\right]^2\,
\Gamma\left[\tfrac{2+k+q-p}{4}\right]^2\,
\Gamma\left[\tfrac{2+p+q-k}{4}\right]^2\,
\Gamma\left[\tfrac{k+p+q}{4}\right]^2}\,\lambda_{2}^{(kk|qq)}\,.
\end{split}
\end{align}
It is straightforward to extract the values of the OPE coefficients which read
\begin{align}
C_{k_1,k_2,k_3}=\frac{2^{2-\Xi}\,\sqrt{\pi}}{\Gamma\left[\tfrac{\Xi}{2}\right]}\,\prod_{i=1}^3\frac{\sqrt{\Gamma[k_i]}}{\Gamma\left[\tfrac{1+\alpha_i}{2}\right]}\,C_{2,2,2},\label{3d-3pt}
\end{align}
where $C_{2,2,2}$ can be determined via (\ref{lambdaCJ}) once the flavor central charge is inputted. 

Let us also write down here the $\langle 22kk\rangle$ correlators as a special case of our general result, which read
\begin{align}
\mathcal{M}=\mathtt{c}_s\, {M}_s+\mathtt{c}_t\,{M}_t+\mathtt{c}_u\,{M}_u\,,
\end{align}
where 
\begin{align}
\begin{split}
{M}_s&=\frac{(2 \alpha  (-k+t+u-2)+k-2 u+2) \, _3F_2\left(\frac{1}{2},\frac{3-k}{2},\frac{1-s}{2};\frac{3}{2},\frac{3-s}{2};1\right)}{\pi ^{5/2} (s-1) \Gamma \left[\frac{k-1}{2}\right]}\,\left(C_{2,2,2}\right)^2,\\
{M}_t&=\frac{(1-\alpha ) \Gamma \left[\frac{k}{2}\right] (2 \alpha  (s-k)-k+2 u-2) \, _3F_2\left(\frac{1}{2},\frac{1}{2},\frac{k-2t}{4};\frac{k+1}{2},\frac{k-2t+4}{4};1\right)}{\pi ^2 (2 t-k) \Gamma \left[\frac{k-1}{2}\right] \Gamma \left[\frac{k+1}{2}\right]}\,\left(C_{2,2,2}\right)^2,\\
{M}_u&=\frac{-\alpha\,  \Gamma \left[\frac{k}{2}\right] (2 \alpha  k-3 k-2 (\alpha -1) s+2 t-2) \, _3F_2\left(\frac{1}{2},\frac{1}{2},\frac{k-2u}{4};\frac{k+1}{2},\frac{k-2u+4}{4};1\right)}{\pi ^2 (2 u-k) \Gamma \left[\frac{k-1}{2}\right] \Gamma \left[\frac{k+1}{2}\right]}\,\left(C_{2,2,2}\right)^2.
\end{split}
\end{align}

\section{Flat space limit}
\label{sec:flatspace}

In this section we examine the flat space limit of Mellin amplitudes \cite{Penedones:2010ue}, where $s,\, t, \, u \to \infty$ with $s+t+u=0$. From our results in Section \ref{sec:correlators} and \ref{sec:flavor} we find the following universal behavior 
\begin{equation}\label{Mfslimit}
\mathcal{M}^{I_1I_2I_3I_4}\bigg|_{s,t\to\infty}=\mathcal{N}_{\{k_i\}}  P_{\{k_i\}}(\sigma,\tau)  \mathcal{A}^{I_1I_2I_3I_4}(s,t;\alpha)\;.
\end{equation}
where $\mathcal{N}_{\{k_i\}}$ is an overall normalization and
\begin{equation}
P_{\{k_i\}}(\sigma,\tau) = \sum_{\substack{i+j+k = \mathcal{E} -2 \\ 0 \leq i,j,k \leq \mathcal{E}-2}} \frac{\sigma^i \tau^j}{i!\,j!\,k!\, (i+\tfrac{\kappa_u}{2})!\, (j+\tfrac{\kappa_t}{2})!\, (k+\tfrac{\kappa_s}{2})!}\;,
\end{equation}
where in $ P_{\{k_i\}}$ we have introduced 
\begin{equation}\label{defsigmatau}
\sigma=\alpha\beta\;,\quad \tau=(1-\alpha)(1-\beta)\;,
\end{equation}
from the $SU(2)_L$ and $SU(2)_R$ cross ratios $\alpha$ and $\beta$. Note that in $P_{\{k_i\}}$, $\alpha$ and $\beta$ are now on the same footing since $\sigma$ and $\tau$ are invariant under $\alpha\leftrightarrow\beta$. These new cross ratios can be understood as the $SU(2)_R\times SU(2)_L$ spinors $v_i$, $\bar{v}_i$ regrouping into $SO(4)$ null vectors $t_i$ which satisfy $t_i\cdot t_i=0$. The factor $P_{\{k_i\}}$ is then the Wick contraction of $k_i-2$ null vectors $t_i$ with $i=1,\ldots,4$
\begin{equation}\label{PfactorWick}
P_{\{k_i\}}\propto \prod_{i<j}t_{ij}^{-\gamma_{ij}^0} (t_{12}t_{34})^{2-\mathcal{E}}\, \mathtt{Wick}\big[\underbrace{t_1\ldots t_1}_{k_1-2}\ldots\underbrace{t_4\ldots t_4}_{k_4-2}\big]
\end{equation}
where $t_{ij}=t_i\cdot t_j=(v_i\cdot v_j)(\bar{v}_i\cdot \bar{v}_j)$ and $\gamma_{ij}^0$ were defined in (\ref{defgamma0}). The cross ratios $\sigma$ and $\tau$ can be expressed in terms of $t_{ij}$ as
\begin{equation}
\sigma=\frac{t_{13}t_{24}}{t_{12}t_{34}}\;,\quad \tau=\frac{t_{14}t_{23}}{t_{12}t_{34}}\;.
\end{equation}
The fact  that theories in different dimensions have the same flat space limit is a nontrivial consistency check of our results, as tree-level amplitudes in flat space are blind to the spacetime dimension. Let us now focus on the color-dependent factor $\mathcal{A}^{I_1I_2I_3I_4}$, which reads
\begin{equation}
\label{Afslimit}
\mathcal{A}^{I_1I_2I_3I_4}=\mathtt{c}_s \frac{u(1-\alpha)-t\alpha}{s}+\mathtt{c}_t\frac{(\alpha-1)(u+s\alpha)}{t}+\mathtt{c}_u\frac{\alpha(t+s(1-\alpha))}{u}\;.
\end{equation} 
Let us now compare this high energy limit of Mellin amplitudes with the expression of tree-level gluon amplitudes in flat space. In the usual setting of flat space scattering, we consider gluons in the adjoint representation of the $SU(N_c)$ gauge group, such that $A_\mu = A_\mu^I T^I$ with $I=1,\cdots,N_c^2-1$. Tree-level amplitudes are usually organized in terms of color-ordered amplitudes, see for instance \cite{Bern:2008qj,Elvang:2015rqa}
\begin{equation}
A_4^{\rm flat} = g^2 \sum_{{\cal P}(2,3,4)} {\rm Tr}\left(T^{I_1}T^{I_2}T^{I_3}T^{I_4} \right) A(1,2,3,4) 
\end{equation}
where the partial amplitudes $A(1,2,3,4)$, {\it etc}, are the color-ordered amplitudes. Color-ordered amplitudes satisfy various relations, including the cyclic and reflection properties
\begin{equation}
A(1,2,3,4) = A(2,3,4,1),~~~A(1,2,3,4) = A(4,3,2,1)\;. 
\end{equation}
For our purposes, however, it becomes more transparent to express the amplitude in the color basis introduced earlier
\begin{equation}
\mathtt{c}_s = f^{I_1 I_2 J} f^{JI_3 I_4}, ~~~ \mathtt{c}_t = f^{I_1 I_4 J} f^{JI_2 I_3},~~~\mathtt{c}_u = f^{I_1 I_3 J} f^{JI_4 I_2}\;,
\end{equation}
which can be written in terms of traces as
\begin{equation}
\mathtt{c}_s = {\rm Tr} \left(T^{I_1}T^{I_2}T^{I_3}T^{I_4} \right) -{\rm Tr} \left(T^{I_1}T^{I_2}T^{I_4}T^{I_3} \right) -{\rm Tr} \left(T^{I_1}T^{I_3}T^{I_4}T^{I_2} \right)+{\rm Tr} \left(T^{I_1}T^{I_4}T^{I_3}T^{I_2} \right)   \;, 
\end{equation}
and similarly for $\mathtt{c}_t$ and $\mathtt{c}_u$. Note that $\mathtt{c}_s$ is given by the only combination with the right symmetry properties under the exchange of indices: antisymmetric under the exchange of $I_1 \leftrightarrow I_2$ or  $I_3 \leftrightarrow I_4$. In this basis $A_4^{\rm flat}$ takes the following form (see, {\it e.g.}, \cite{Elvang:2015rqa})\footnote{Note that our definition for the Mandelstam variables are $s=-(\mathtt{k}_1+\mathtt{k}_2)^2$, $t=-(\mathtt{k}_1+\mathtt{k}_4)^2$, $u=-(\mathtt{k}_1+\mathtt{k}_3)^2$, where the role of $t$ and $u$ is reversed compared to the convention used in some flat space amplitude literature.}
\begin{equation}\label{flatspaceg}
A_4^{\rm flat} = \frac{\mathtt{c}_s N_s}{ s} +  \frac{\mathtt{c}_tN_t}{t} +  \frac{\mathtt{c}_uN_u}{u}
\end{equation}
which coincides exactly with $\mathcal{A}^{I_1I_2I_3I_4}$ in (\ref{Afslimit}) upon the identification
\begin{equation}
\label{colorrep}
N_s= u(1-\alpha)-t \alpha,~~~N_t = (\alpha-1)(u+s\alpha),~~~N_u=\alpha(t +s(1-\alpha)).
\end{equation}
Changing back to the trace basis we can obtain the respective expressions for the color-ordered amplitudes
\begin{eqnarray}
A(1,2,3,4)= -\frac{(s+t-s \alpha)^2}{s t},\\
A(1,3,4,2) = -\frac{(s+t-s \alpha)^2}{s u},\\
A(1,4,2,3)= -\frac{(s+t-s \alpha)^2}{t u}.
\end{eqnarray}
These partial amplitudes satisfy both the photon-decoupling identity 
\begin{eqnarray}
A(1,2,3,4)+ A(1,3,4,2) + A(1,4,2,3)=0,
\end{eqnarray}
where we have used $s+t+u=0$, as well as the BCJ relations
\begin{eqnarray}
 t A(1,2,3,4) = u A(1,3,4,2), ~~~s A(1,2,3,4) = u A(1,4,2,3),~~~t A(1,4,2,3) = s A(1,3,4,2). \nonumber
\end{eqnarray}
Let us now return to (\ref{colorrep}) and ask whether these expressions can be precisely reproduced from the flat space gluon amplitudes with appropriately restricted kinematic configurations. The functions $N_{s,t,u}$ for four-gluon scattering with momenta $\mathtt{k}_i$ and polarizations $\epsilon_i$, $i=1,2,3,4$, in arbitrary flat space dimensions can be found in  \cite{Adamo:2018mpq}. The results are given by\footnote{The relevant expressions of \cite{Adamo:2018mpq} are (4.12) to (4.14) where one needs to drop the $x^-$, $y^-$ dependence and also identify $D_{\mu \nu}$ with the flat space metric.}
\begin{eqnarray}
\label{lionel}
\nonumber N_s = &&2\big( (\epsilon_1 \cdot \epsilon_2) (\mathtt{k}_1-\mathtt{k}_2)^\mu +2 (\epsilon_1 \cdot \mathtt{k}_2) \epsilon_2^\mu -2 (\epsilon_2 \cdot \mathtt{k}_1) \epsilon_1^\mu \big)\\
 &&\times \eta_{\mu \nu}  \big( (\epsilon_3 \cdot \epsilon_4) (\mathtt{k}_4-\mathtt{k}_3)^\nu -2 (\epsilon_3 \cdot \mathtt{k}_4) \epsilon_4^\nu +2 (\epsilon_4 \cdot \mathtt{k}_3) \epsilon_3^\nu \big) \nonumber \\
 &&- 2\big((\epsilon_1 \cdot \epsilon_3) (\epsilon_2 \cdot \epsilon_4)-(\epsilon_1 \cdot \epsilon_4) (\epsilon_2 \cdot \epsilon_3) \big) \big( \mathtt{k}_1 \cdot \mathtt{k}_2+\mathtt{k}_3 \cdot \mathtt{k}_4 \big)\;,
\end{eqnarray}
with analogous expressions for $N_t,N_u$ fixed by crossing. We first consider the flat space limit of the amplitude corresponding to the simplest operator ${\cal O}_2^I= {\cal O}^{I;\alpha_1,\alpha_2}(x) \epsilon_{\alpha_1 \beta_1}\epsilon_{\alpha_2 \beta_2} v^{\beta_1} v^{\beta_2}$.  In order to make the comparison, we consider (\ref{lionel}) in a specially chosen kinematics where the momenta $\mathtt{k}_i$ are restricted to $\mathbb{R}^{d-1,1} \sim \partial AdS_{d+1}$ while the polarization vectors $\epsilon_i$ are restricted to a perpendicular $\mathbb{R}^{4}$, such that $\mathtt{k}_i \cdot \epsilon_j = 0$ for all $i,j$. This leads to
\begin{equation}
\label{lionels}
N_s =2(\epsilon_1 \cdot \epsilon_2)  (\epsilon_3 \cdot \epsilon_4) ((\mathtt{k}_1-\mathtt{k}_2)  \cdot(\mathtt{k}_4-\mathtt{k}_3)) - 2 \left((\epsilon_1 \cdot \epsilon_3) (\epsilon_2 \cdot \epsilon_4)-(\epsilon_1 \cdot \epsilon_4)( \epsilon_2 \cdot \epsilon_3) \right) \left( \mathtt{k}_1 \cdot \mathtt{k}_2+\mathtt{k}_3 \cdot \mathtt{k}_4 \right). \nonumber
\end{equation}
For the polarization vectors we can assemble the doublet of spinors $v^{\beta_1},v^{\beta_2}$ into a null vector $\epsilon$, which leads to $\epsilon_i \cdot \epsilon_j = (v_i \cdot v_j)^2$. This is exactly how we introduced the null vectors earlier in this section, except that we now need to identify $\bar{v}$ with $v$ since ${\cal O}_2^I$ contains two $v$.
Extracting the overall factor $(\epsilon_1 \cdot \epsilon_2)  (\epsilon_3 \cdot \epsilon_4)$, we obtain
\begin{eqnarray}
\label{lionels}
N_s =-2(t-u)+2(\alpha^2-(1-\alpha)^2) s = 4(\alpha s+u)
\end{eqnarray}
which, upon using $s+t+u=0$, agrees precisely with our expression (\ref{colorrep}) for $N_s$ up to an unimportant overall factor. The flat space limit of general four-point amplitudes can also be matched. For operators with $k_i>2$, one should modify the flat space in- and out-states by allowing them to have a nontrivial wavefunction $\Psi_i$ on the angular directions $S^{m}$ of the transverse $\mathbb{R}^{m+1}$ ($m=3$ in the above case). This wavefunction is just a scalar spherical harmonic of rank $k_i-2$, which can be conveniently expressed as\footnote{To see this, we note that the scalar spherical harmonics can be written as $Y^{\mathcal{I}}_k({\rm T})=C^{\mathcal{I}}_{a_1\ldots a_k}{\rm T}^{a_1}\ldots {\rm T}^{a_k}$ where $C^{\mathcal{I}}_{a_1\ldots a_k}$ is a symmetric traceless tensor of $SO(m+1)$. To introduce the null vectors, we can contract the spherical harmonics with $C^{\mathcal{I}}_{b_1\ldots b_k}t^{b_1}\ldots t^{b_k}$. The contraction $C^{\mathcal{I}}_{a_1\ldots a_k}C^{\mathcal{I}}_{b_1\ldots b_k}$ is a product of delta functions of the $a_i$ and $b_i$ indices, and gives rise to (\ref{wavefunction}).} 
\begin{equation}\label{wavefunction}
\Psi_i({\rm T},t_i)=({\rm T}\cdot t_i)^{k_i-2}
\end{equation}
where $t_i$ is a $(m+1)$-dimensional null vector with $t_i\cdot t_i=0$ and ${\rm T}\in \mathbb{R}^{m+1}$, ${\rm T}\cdot {\rm T}=1$ parametrizing $S^m$.\footnote{We need to emphasize here that the sphere polarization $t_i$ and the spacetime polarization $\epsilon_i$ are {\it independent}. From the AdS calculation they turn out to be correlated. However, generally it does not need to be the case.} The flat space gluon amplitudes are now dressed with a factor given by the overlap of the four wavefunctions
\begin{equation}
\int_{S^m}[d{\rm T}]\Psi_1({\rm T},t_1)\Psi_2({\rm T},t_2)\Psi_3({\rm T},t_3)\Psi_4({\rm T},t_4)
\end{equation}
which is easy to evaluate since the integral over ${\rm T}$ gives just the Wick contraction \cite{Lee:1998bxa}. Using this result and the fact that the vectors $t_i$ are null, it is clear that the wavefunction overlap gives the factor $P_{\{k_i\}}(\sigma,\tau)$ in (\ref{PfactorWick}). The above discussion can be straightforwardly generalized to the case of gravitons (see also \cite{Chester:2018aca,Chester:2018dga}). In particular, our derivation explains why the same factor $P_{\{k_i\}}$ also shows up in the flat space limit of Mellin amplitudes in maximally superconformal theories \cite{Alday:2020lbp,Alday:2020dtb}
\begin{equation}\label{flatspaceSUGRA}
\mathcal{M}^{\rm SUGRA}\big|_{s,t\to\infty}\propto \frac{(s+t-s\alpha)^2(s+t-s\bar{\alpha})^2}{s\,t\,u}P_{\{k_i\}}(\sigma,\tau)\;.
\end{equation}
Here $\alpha$, $\bar{\alpha}$ are the two cross ratios of the R-symmetry group $SO(m+1)$ with $m=4,5,7$, and we should identify $\beta=\bar{\alpha}$. 

Going back to (\ref{Afslimit}), let us also note that using $\mathtt{c}_s+\mathtt{c}_t+\mathtt{c}_u=0$ and $s+t+u=0$ we can write the amplitude as
\begin{equation}
\mathcal{A}^{I_1I_2I_3I_4} = \frac{(t \mathtt{c}_s-s \mathtt{c}_t)(s+t-s \alpha)^2}{s\,t\,u}.
\end{equation}
The factor $(s+t-s \alpha)^2$ exactly agrees with the holomorphic part of the corresponding prefactor present in super graviton amplitudes in flat space. In particular, it implies that the above amplitude is a solution to the flat space limit of the Ward identities. In fact, substituting in (\ref{flatspaceg})
\begin{equation}
\mathtt{c}_s\to N_s\;,\quad \mathtt{c}_t\to N_t\;,\quad \mathtt{c}_u\to N_u\;,
\end{equation}
in which $\alpha\to\bar{\alpha}$ is already replaced, the flat space limit of the super gluon amplitudes just become the super graviton amplitudes (\ref{flatspaceSUGRA}). Not surprisingly, this is just the double copy relation in flat space \cite{Bern:2010ue}.

Let us conclude this section with a remark about how to use the flat space limit to constrain the correlators. Recall that in the last step of our bootstrap strategy we used the superconformal Ward identities to fix the OPE coefficients $C_{k_1,k_2,k_3}$ in terms of $C_{2,2,2}$. Then in this section we checked that the resulting expression for the amplitude has the correct flat space limit. However, we can also turn the logic around. We can leave the OPE coefficients unfixed and require the precise answer (\ref{Mfslimit}) is reproduced in the flat space limit. It turns out this requirement is powerful enough to fix all coefficients $C_{k_1,k_2,k_3}$ up to an overall factor which we can choose to be $C_{2,2,2}$. Note that  in matching (\ref{Mfslimit}) we do not need to include any contact terms, which are constant in the flat space limit. However, because of this constant growth behavior, contact terms would contradict the limit (\ref{Mfslimit}) and are not allowed. This provides an intuitive explanation for the vanishing of contact terms observed earlier as a consequence of the Ward identities. In this new line of attack, however, that the superconformal Ward identities hold now serves as a nontrivial consistency check.

\section{Hidden structures in holographic correlators}
\label{sec:hiddenstructures}

\subsection{Parisi-Sourlas supersymmetry}
\label{subsec:PSsusy}
The results in Section \ref{sec:correlators} and \ref{sec:flavor} exhibit an interesting emergent dimensional Parisi-Sourlas-like reduction structure. A similar structure was observed in correlators in maximally superconformal theories \cite{Behan:2021pzk}. However, unlike the maximally superconformal situation where the dimension of the AdS space reduces by four, here we find the AdS space dimension reduces only by two. 

To see this structure, we first note that the factor $K^i_p(t,u)$ in the multiplet exchange amplitude residue (\ref{RescalR}) is a polynomial of $t$, $u$ and $i$. This allows us to move the factor outside the inverse Mellin integrals and interpret it as a differential operator
\begin{equation}\label{replacediff}
U\partial_U\leftrightarrow (\tfrac{s}{2}-a_s)\times\;,\quad V\partial_V\leftrightarrow (\tfrac{t}{2}-a_t)\times\;,\quad (\alpha-1)\partial_\alpha \leftrightarrow i\times\;.
\end{equation}
We can now focus on the other two factors $B^i_{p,m}$, $E^i_p$. The remaining dependence on the Mandelstam variable is contained in a sum over simple poles with residues independent of the Mandelstam variables 
\begin{equation}
\sum_{m=0}^\infty \frac{1}{\,m!\,(m+\epsilon(p-1))!\Gamma[\frac{\epsilon(k_1+k_2-p)-2m}{2}]\Gamma[\frac{\epsilon(k_3+k_4-p)-2m}{2}](s-\epsilon p-2m)}\;,
\end{equation}
which turns out to be proportional to the linear combination of two $AdS_{d+1}$ scalar exchange Witten diagrams with conformal dimensions $\epsilon p$ and $\epsilon p+2$
\begin{equation}\label{2scalars}
\mathcal{M}^{(s),d}_{\epsilon p,0}-\frac{(k_1-k_2+p)(k_2-k_1+p)(k_3-k_4+p)(k_4-k_3+p)\epsilon^2}{16p(p-1)(1+p\epsilon)(1+(p-1)\epsilon)}\mathcal{M}^{(s),d}_{\epsilon p+2,0}\;.
\end{equation}
Here we have added the superscript $d$ to stress that the Witten diagrams are defined in $AdS_{d+1}$. On the other hand, we can check by explicit calculations that the above expression is identical to just a single scalar exchange Witten diagram, however in a lower dimensional space $AdS_{d-1}$
\begin{equation}\label{1scalar}
\mathcal{M}^{(s),d-2}_{\epsilon p,0}\;.
\end{equation}
This scalar exchange Witten diagram has exchange dimension $\epsilon p$ and  the same external dimensions $\epsilon k_i$. The equivalence of (\ref{2scalars}) and (\ref{1scalar}) can be explained in terms of a holographically realized Parisi-Sourlas supersymmetry \cite{Zhou:2020ptb}.\footnote{More generally, \cite{Zhou:2020ptb} showed that an $AdS_{d-1}$ exchange Witten diagram with dimension $\Delta$ and spin $\ell$ can be expressed as the linear combination of five exchange Witten diagrams in $AdS_{d+1}$ with shifted dimensions and spins. These Witten diagram relations generalize nontrivially similar dimensional reduction formulae for conformal blocks discovered in \cite{Kaviraj:2019tbg}. The latter was shown in \cite{Kaviraj:2019tbg} to be a kinematic consequence of an underlying Parisi-Sourlas supersymmetry \cite{Parisi:1979ka}. Similar relations also exist for boundary CFTs and CFTs on real projective space \cite{Zhou:2020ptb,Giombi:2020xah}.} However, it is not clear why Parisi-Sourlas supersymmetry should be present in superconformal correlators. Let us turn to the R-symmetry dependence in $B^i_{p,m}$ and $E^i_p$
\begin{equation}\label{RinBE}
\sum_{i}\frac{(-1)^i\Gamma \left[\frac{4 i+2 p+\kappa_u+\kappa_t-4}{4}\right]}{i!\,\Gamma \left[i+\frac{\kappa_t}{2}+1\right]\Gamma \left[\frac{-4 i+2 p-\kappa_u-\kappa_t+4}{4} \right]}(1-\alpha)^i\;.
\end{equation} 
We can define more generally a polynomial
\begin{eqnarray}\label{defYnu}
\mathbb{Y}_{p,\nu}(\alpha)&=&\sum_{i}\frac{(-1)^i\Gamma \left[\frac{4 i+2 p+\kappa_u+\kappa_t+4-4\nu}{4}\right]\Gamma[\frac{2p-\kappa_t-\kappa_u+4}{4}]\Gamma[\frac{2p+\kappa_t-\kappa_u+4}{4}]}{i!\,\Gamma \left[\frac{2i+\kappa_t+2}{2}\right]\Gamma \left[\frac{-4 i+2 p-\kappa_u-\kappa_t+4}{4} \right]\Gamma[1+p-\nu]}(1-\alpha)^i\\
\nonumber &=&\frac{\Gamma[\frac{2p+\kappa_u+\kappa_t+4-4\nu}{4}]\Gamma[\frac{2p+\kappa_t-\kappa_u+4}{4}]}{(\frac{\kappa_t}{2})!\Gamma[1+p-\nu]}{}_2F_1\left(\tfrac{-2p+\kappa_u+\kappa_t}{4},\tfrac{4+2p+\kappa_u+\kappa_t-4\nu}{4},1+\tfrac{\kappa_t}{2};1-\alpha \right)\;,
\end{eqnarray}
which is proportional to (\ref{RinBE}) when $\nu=2$, and gives the $SO(3)$ R-symmetry polynomials (\ref{YpSU2}) when $\nu=0$
\begin{equation}
\mathcal{Y}_p=\mathbb{Y}_{p,0}\;.
\end{equation}
These functions $\mathbb{Y}_{p,\nu}$ satisfy the differential equation
\begin{equation}
\nonumber y\,(1-y)\,\frac{d^2}{dy^2}\mathbb{Y}_{p,\nu}+\big(\tfrac{2+\kappa_u-2\nu}{2}-y(\tfrac{4+\kappa_t+\kappa_u-2\nu}{2})\big)\frac{d}{dy}\mathbb{Y}_{p,\nu}-(\tfrac{\kappa_t+\kappa_u-2p}{4})(\tfrac{\kappa_t+\kappa_u+2p+4-4\nu}{4})\mathbb{Y}_{p,\nu}=0\;.
\end{equation}
In \cite{Behan:2021pzk}, the modified R-symmetry dependence was shown to be related to reducing the dimension of the internal sphere also by four. However, it is less clear here how to geometrically interpret  $\mathbb{Y}_{p,\nu}$, as the factorization into $SU(2)_R\times SU(2)_L$ is a feature of $S^3$ only and the $SU(2)_L$ factor is left intact throughout.  

Using these observations we can now rewrite the correlators in a more compact form. Restoring the kinematic factor extracted in (\ref{GandcalG}), the s-channel multiplet exchange amplitudes become 
\begin{eqnarray}
\mathbf{S}_p^{(s)}=&&\prod_{i<j}\left(\frac{(v_i\cdot v_j)(\bar{v}_i\cdot \bar{v}_j)}{x_{ij}^{2\epsilon}}\right)^{\gamma^0_{ij}}\frac{(v_1\cdot v_2)^{\mathcal{E}}(v_3\cdot v_4)^{\mathcal{E}}}{(x_{12}^{2\epsilon}x_{34}^{2\epsilon})^{\mathcal{E}}} \left((\bar{v}_1\cdot\bar{v}_2)(\bar{v}_3\cdot\bar{v}_4)\right)^{\mathcal{E}-2}\\
\nonumber &&\times f_{\{k_i, p\}}\, \mathbb{K}_p\circ \int_{-i\infty}^{i\infty} \frac{dsdt}{(4\pi i)^2}U^{\frac{s}{2}-a_s}V^{\frac{t}{2}-a_t} \bigg(\mathcal{M}^{(s),d-2}_{\epsilon p,0}\mathbb{Y}_{p,2}(\alpha)\mathbb{Y}_{p-2,0}(\beta)\bigg)\Gamma^{(\epsilon)}_{\{k_i\}}\;.
\end{eqnarray}
Here $\mathbb{K}_p$ is a differential operator obtained with the replacement (\ref{replacediff}) in the factor $K^i_p(t,u)$, and $f_{\{k_i\},p}$ is an overall factor\footnote{Comparing the original expression (\ref{Sp}) with the expression here we would find an extra factor $(-1)^{p-\frac{\kappa_t+\kappa_u}{2}}$. However, by selection rules this factor is just 1.} 
\begin{equation}
f_{\{k_i\},p}=\frac{2}{(p-1)\epsilon (2p-\kappa_t-\kappa_u)(2p+\kappa_t-\kappa_u)}\;.
\end{equation}
Expressions with similar structures for the t- and u-channel multiplet exchange amplitudes can be obtained from $\mathbf{S}_p^{(s)}$ by crossing via permuting the external labels. The full correlator can then be written as 
\begin{equation}
\begin{split}
G_{k_1k_2k_3k_4}=&\mathtt{c}_s \sum_{p_s} C_{k_1,k_2, p_s}C_{k_3,k_4, p_s}\mathbf{S}_{p_s}^{(s)}+\mathtt{c}_t \sum_{p_t} C_{k_1,k_4, p_t}C_{k_2,k_3, p_t}\mathbf{S}_{p_t}^{(t)}\\
&+\mathtt{c}_u \sum_{p_u} C_{k_1,k_3, p_u}C_{k_2,k_4, p_u}\mathbf{S}_{p_u}^{(u)}
\end{split}
\end{equation}
This form of our results suggests that the super gluon correlators in these theories can be obtained from a lower-dimensional scalar ``seed theory'' in  $AdS_{d-1}$, by dressing the ``seed correlators'' with differential operators. Finally, we remind the reader that the above Parisi-Sourlas dimensional reduction structure relies crucially on the fact that the four-point functions can be written in terms of only exchange contributions. The existence of additional contact terms would spoil this structure.

\subsection{Hidden conformal symmetry}\label{hiddenconformal}
In this subsection, we point out another interesting feature, namely,  the correlators in 4d $\mathcal{N}=2$ theories exhibit an eight dimensional hidden conformal symmetry. To see it, we start from the solution to superconformal Ward identities (\ref{scfWardida}) in 4d, which reads \cite{Nirschl:2004pa}
\begin{equation}\label{GasG0andRH}
G_{k_1k_2k_3k_4}=G_{0,k_1k_2k_3k_4}+R\, H_{k_1k_2k_3k_4}\;,
\end{equation}
with 
\begin{equation}
R=(v_1\cdot v_2)^2(v_3\cdot v_4)^2x_{13}^2x_{24}^2(1-z\alpha)(1-\bar{z}\alpha)\;.
\end{equation}
The function $G_{0,k_1k_2k_3k_4}$ is the protected part of the correlator,  and gives the meromorphic correlator upon performing the chiral algebra twisting $\alpha=1/z$ or $\alpha=1/\bar{z}$ \cite{Nirschl:2004pa,bllprv13} (see Appendix \ref{sec:chiralalgebra} for more discussions). Here and below we have left the color indices implicit.  We will refer to $H_{k_1k_2k_3k_4}$ as the reduced correlator, which has shifted conformal dimensions $k_i+1$, and $SU(2)_R\times SU(2)_L$ spins $(\frac{k_i-2}{2},\frac{k_i-2}{2})$. Recall that the full Mellin amplitude is defined as   
\begin{equation}
\begin{split}
G_{k_1k_2k_3k_4}={}&\int_{-i\infty}^{i\infty}\frac{dsdt}{(4\pi i)^2} (x_{12}^2)^{\frac{s-k_1-k_2}{2}}(x_{34}^2)^{\frac{s-k_3-k_4}{2}}(x_{14}^2)^{\frac{t-k_1-k_4}{2}}(x_{23}^2)^{\frac{t-k_2-k_3}{2}}\\
{}&\times (x_{13}^2)^{\frac{u-k_1-k_3}{2}} (x_{24}^2)^{\frac{u-k_2-k_4}{2}}\mathcal{M}_{k_1k_2k_3k_4}\, \Gamma^{(1)}_{\{k_i\}}
\end{split}
\end{equation}
where $s+t+u=\sum_{i=1}^4 k_i$, and $\Gamma^{(1)}_{\{k_i\}}$ was defined in (\ref{inverseMellin}).
We define an analogous reduced Mellin amplitude $\widetilde{\mathcal{M}}_{k_1k_2k_3k_4}$ as 
\begin{equation}\label{Mtildedef}
\begin{split}
H_{k_1k_2k_3k_4}={}&\int_{-i\infty}^{i\infty}\frac{dsdt}{(4\pi i)^2} (x_{12}^2)^{\frac{s-k_1-k_2}{2}}(x_{34}^2)^{\frac{s-k_3-k_4}{2}}(x_{14}^2)^{\frac{t-k_1-k_4}{2}}(x_{23}^2)^{\frac{t-k_2-k_3}{2}}\\
{}&\times (x_{13}^2)^{\frac{\tilde{u}-k_1-k_3}{2}} (x_{24}^2)^{\frac{\tilde{u}-k_2-k_4}{2}}\widetilde{\mathcal{M}}_{k_1k_2k_3k_4}\, \widetilde{\Gamma}^{(1)}_{\{k_i\}}
\end{split}
\end{equation}
where $s+t+\tilde{u}=\sum_{i=1}^4 k_i-2$, and 
\begin{equation}
\widetilde{\Gamma}^{(1)}_{\{k_i\}}=\Gamma[\tfrac{k_1+k_2-s}{2}]\Gamma[\tfrac{k_3+k_4-s}{2}]\Gamma[\tfrac{k_1+k_4-t}{2}]\Gamma[\tfrac{k_2+k_3-t}{2}]\Gamma[\tfrac{k_1+k_3-\tilde{u}}{2}]\Gamma[\tfrac{k_2+k_4-\tilde{u}}{2}]\;.
\end{equation}
The shift in $\tilde{u}$ is important because Bose symmetry in the reduced Mellin amplitude acts by permutation of $s$, $t$, $\tilde{u}$. Similar to the $\mathcal{N}=4$ case \cite{Rastelli:2017udc}, the protected piece $G_{0,k_1k_2k_3k_4}$ does not contribute to the Mellin amplitude. So it follows from the solution (\ref{GasG0andRH}) that we have the following relation between $\widetilde{\mathcal{M}}_{k_1k_2k_3k_4}$ and $\mathcal{M}_{k_1k_2k_3k_4}$ 
\begin{equation}\label{MeqRhatMtilde}
\mathcal{M}_{k_1k_2k_3k_4}=\widehat{R}\circ \widetilde{\mathcal{M}}_{k_1k_2k_3k_4}
\end{equation}
where $\widehat{R}$ is a difference operator with each $U^mV^n$ in $R$ (after extracting the kinematic factor $(v_1\cdot v_2)^2(v_3\cdot v_4)^2x_{13}^2x_{24}^2$) interpreted as a difference operator
\begin{equation}
\widehat{U^mV^n}'\circ \widetilde{\mathcal{M}}(s,t)=\frac{\widetilde{\Gamma}^{(1)}_{\{k_i\}}(s-2m,t-2n)}{\Gamma^{(1)}_{\{k_i\}}(s,t)}\widetilde{\mathcal{M}}(s-2m,t-2n)\;.
\end{equation}
Notice these monomials act differently than in (\ref{diffUmVn}), and their associated difference operators are distinguished by a prime here. This is because the reduced Mellin amplitudes require a different Gamma function factor in their definition. Using this operator, we find that
\begin{equation}
\widetilde{\mathcal{M}}_{2222}=-4(C_{2,2,2})^2\left(\frac{\mathtt{c}_s}{(s-2)(\tilde{u}-2)}-\frac{\mathtt{c}_t}{(t-2)(\tilde{u}-2)}\right)\;.
\end{equation} 
Here we have used the identity $\mathtt{c}_s+\mathtt{c}_t+\mathtt{c}_u=0$ to eliminate $\mathtt{c}_u$ in order to avoid  ambiguities in writing the answer. But we can also rewrite it in the more symmetric form 
\begin{equation}
\begin{split}
\widetilde{\mathcal{M}}_{2222}={}&-\frac{4}{3}(C_{2,2,2})^2\bigg(\frac{\mathtt{c}_s}{(s-2)(\tilde{u}-2)}-\frac{\mathtt{c}_s}{(s-2)(t-2)}+\frac{\mathtt{c}_t}{(t-2)(s-2)}\\
{}&\quad\quad\quad\quad\quad\;\;-\frac{\mathtt{c}_t}{(t-2)(\tilde{u}-2)}+\frac{\mathtt{c}_u}{(\tilde{u}-2)(t-2)}-\frac{\mathtt{c}_u}{(\tilde{u}-2)(s-2)}\bigg)\;.
\end{split}
\end{equation}
This form of summing over pairs of simultaneous poles turns out to be a generic feature of the reduced Mellin amplitudes. 

The reduced correlators have a remarkable feature -- they are organized by an eight dimensional hidden conformal symmetry. This hidden symmetry allows us to promote $H_{2222}$ into a generating function by replacing AdS distances with higher dimensional distances, and obtain any $H_{k_1k_2k_3k_4}$ by Taylor expansion. More precisely, we find that the $SU(2)_R\times SU(2)_L$ spinors $v_i$, $\bar{v}_i$ regroup into $SO(4)$ null vectors $t_i$ as they do in the flat space limit (see Section \ref{sec:flatspace}). The reduced correlator $H_{k_1k_2k_3k_4}$ can be viewed as a correlator where each operator transforms in the rank-$(k_i-2)$ symmetric traceless representation of $SO(4)$. The $SO(4)$ indices of the operators are contracted with the null vectors $t_i$. The null vectors $t_i$ appear in $H_{k_1k_2k_3k_4}$ only as polynomials of $t_{ij}=t_i\cdot t_j$, and under independent scalings $t_i\to \zeta^i t_i$ the reduced correlators scale as $H_{k_1k_2k_3k_4}\to \prod_{i=1}^4 \zeta_i^{k_i-2} H_{k_1k_2k_3k_4}$. For the lowest Kaluza-Klein level $k_i=2$, $H_{2222}$ is a singlet under $SO(4)$ and therefore depends on $x_{ij}^2$ only. We can define a generating function from the $k_i=2$ reduced correlator as
\begin{equation}
\mathbf{H}(x_i,t_i)=H_{2222}(x_{ij}^2-t_{ij})\;.
\end{equation}
To obtain $H_{k_1k_2k_3k_4}$, we only need to Taylor expand $\mathbf{H}$ in powers of $t_{ij}$, and collect all the monomials of $t_{ij}$ that can appear in this correlator. There are only finitely many such monomials in each $H_{k_1k_2k_3k_4}$ because of the above scalings. The replacement of $x_{ij}^2$ by $x_{ij}^2-t_{ij}$ indicates an eight dimensional conformal symmetry. Since the two factors of $AdS_5\times S^3$ have the same radius, this background can be conformally mapped to $\mathbb{R}^{7,1}$ where $x_{ij}^2-t_{ij}$ is the conformally invariant distance. 
Furthermore, we can check the dimension by looking at the flat space limit of the reduced Mellin amplitude 
\begin{equation}
\widetilde{\mathcal{M}}_{2222}\to -4(C_{2,2,2})^2\left(\frac{\mathtt{c}_s}{su}-\frac{\mathtt{c}_t}{tu}\right)\;. \label{lowest-flatspace}
\end{equation}
This amplitude is annihilated by the flat space special conformal transformation generators  
\begin{equation}
K_\mu=\sum_{i=1}^3\left(\frac{p_{i\mu}}{2}\frac{\partial}{\partial p_i^\nu}\frac{\partial}{\partial p_{i,\nu}}-p_i^\nu\frac{\partial}{\partial p_i^\nu}\frac{\partial}{\partial p_i^\mu}-\frac{d-2}{2}\frac{\partial}{\partial p_i^\mu}\right)\;,
\end{equation}
only when the spacetime dimension is $d=8$. This number of dimensions agrees with the fact that we are essentially studying scattering processes constrained inside the eight dimensional world volume of the 7-branes. Using the generating function, it is easy to find that the reduced Mellin amplitudes for general $k_1,k_2,k_3,k_4$ are
\begin{equation}
\begin{split}\label{Mreducedgeneral}
\widetilde{\mathcal{M}}_{k_1k_2k_3k_4}=&-4(C_{2,2,2})^2\prod_{a<b}t_{ab}^{\gamma_{ab}^0} (t_{12}t_{34})^{\mathcal{E}-2}\!\!\!\!\!\!\!\! \sum_{\substack{i+j+k = \mathcal{E} -2 \\ 0 \leq i,j,k \leq \mathcal{E}-2}} \!\frac{\sigma^i \tau^j}{i!\,j!\,k!\, (i+\tfrac{\kappa_u}{2})!\, (j+\tfrac{\kappa_t}{2})!\, (k+\tfrac{\kappa_s}{2})!}\\
&\;\times\left(\frac{\mathtt{c}_s}{(s-s_M+2k)(\tilde{u}-u_M+2i)}-\frac{\mathtt{c}_t}{(t-t_M+2j)(\tilde{u}-u_M+2i)}\right)
\end{split}
\end{equation}
where $\sigma$, $\tau$ were defined in (\ref{defsigmatau}), and
\begin{eqnarray}
\nonumber s_M&=&\min\{k_1+k_2,k_3+k_4\}-2\;,\\
t_M&=&\min\{k_1+k_4,k_2+k_3\}-2\;,\\
\nonumber u_M&=&\min\{k_1+k_3,k_2+k_4\}-2\;.
\end{eqnarray}
Upon acting on $\widetilde{\mathcal{M}}_{k_1k_2k_3k_4}$ with the difference operator $\widehat{R}$ defined in (\ref{MeqRhatMtilde}), we reproduce the results found in Section \ref{sec:4dN=2}. From the flat space limit of (\ref{Mreducedgeneral}) and $\widehat{R}$, it is straightforward to see that the flat space limit of $\mathcal{M}_{k_1k_2k_3k_4}$ is given by (\ref{Mfslimit}).

The above eight dimensional hidden conformal symmetry parallels  the ten and six dimensional hidden conformal structures for IIB supergravity on $AdS_5\times S^5$ \cite{Caron-Huot:2018kta} and $AdS_3\times S^3\times K3$ \cite{Rastelli:2019gtj}  (see also \cite{Giusto:2020neo}). In the latter two examples, the solutions to the superconformal Ward identity also take the form (\ref{GasG0andRH}), and it is possible to define reduced correlators $H_{k_1k_2k_3k_4}$ which have similar shifted conformal and R-symmetry weights. The lowest-weight reduced correlator $H_{2222}$ is an R-symmetry singlet, and upon the replacement $x_{ij}^2\to x_{ij}^2-t_{ij}$ gives the generating function. The flat space limit of the reduced Mellin amplitudes are also conformally invariant in ten and six dimensions respectively. 

However, an important difference is that the 4d $\mathcal{N}=2$ case arises from a supersymmetric {\it gauge theory} in AdS, while the other two examples are {\it supergravity} theories. Previously it was not clear whether the existence of such higher dimensional conformal symmetries relies on special properties of IIB supergravity. The 4d $\mathcal{N}=2$ example we presented here proves that is not the case. At the moment, a precise understanding of these structures is still lacking. However, finding the same phenomenon in drastically different theories provides strong indication that such hidden conformal structures are due to the conformal flatness of the background.

\subsection{Color-kinematic duality}
The holographic correlators in Sections \ref{sec:correlators} and \ref{sec:flavor} also enjoy another interesting property that can be more easily seen after using the prescription of Section \ref{fullmultipletamplitude} to eliminate the contact terms. This property mimics the celebrated color-kinematic duality \cite{Bern:2008qj} for the lowest Kaluza-Klein modes with $k_i=2$, while higher Kaluza-Klein modes satisfy a modification of it. Note the $k_i=2$ multiplet is where the flavor current belongs, dual to a {\it massless} gauge field in AdS. These four-point amplitudes have the universal factorized structure in each channel  
\begin{equation}\label{Mkeq2}
\mathcal{M}_{2222}=\mathtt{c}_s \mathtt{n}_s \left(\frac{\mathtt{const}}{s-2\epsilon}+\ldots\right)+\mathtt{c}_t \mathtt{n}_t \left(\frac{\mathtt{const}}{t-2\epsilon}+\ldots\right)+\mathtt{c}_u \mathtt{n}_u \left(\frac{\mathtt{const}}{u-2\epsilon}+\ldots\right)
\end{equation}
where $\mathtt{c}_{s,t,u}$ are the color structures defined in (\ref{csctcu}). The normalization $\mathtt{const}$ depends on the theory but is the same for all three channels. The $\ldots$ denote satellite poles at $2\epsilon+2\mathbb{Z}_+$ which complete the series into the sum of two $AdS_{d+1}$ scalar exchange diagrams (\ref{2scalars}), or equivalently, a single scalar exchange diagram in $AdS_{d-1}$ (\ref{1scalar}). The factors $\mathtt{n}_{s,t,u}$ are simple polynomials given by 
\begin{equation}
\mathtt{n}_s=u-4\epsilon+\alpha s\;,\quad \mathtt{n}_t=-(\alpha-1)((\alpha-1)(u-4\epsilon)+\alpha t)\;,\quad \mathtt{n}_u=-\alpha(\alpha(s-4\epsilon)+u)\;,
\end{equation} 
and are related to each other by crossing symmetry. Note that this factorization into polynomial factors $\mathtt{n}_{s,t,u}$ and polar parts corresponding to an $AdS_{d-1}$ scalar exchange diagram is a highly nontrivial feature of our results. The form of (\ref{Mkeq2}) is very similar to the formula for four-point gluon scattering amplitudes in flat space. The spacetime polarizations in flat space are replaced by R-symmetry polarizations. In analogy with the $\frac{1}{s}$ scalar propagator in flat space, we have the $AdS_{d-1}$ scalar propagator which contains a series of simple poles $(\frac{\mathtt{const}}{s-2\epsilon}+\ldots)$. Thanks to the Jacobi identity, we have
\begin{equation}
\mathtt{c}_s+\mathtt{c}_t+\mathtt{c}_u=0\;, \label{jacobi-c}
\end{equation} 
On the other hand, it is also straightforward to check that the kinematic factors satisfy
\begin{equation}
\mathtt{n}_s+\mathtt{n}_t+\mathtt{n}_u=0\;. \label{jacobi-n}
\end{equation}
This relation gives an AdS version of the color-kinematic duality, which is literally the same as that of the flat space gluons! Finding AdS extensions of the flat space color-kinematic duality was also discussed recently in momentum space \cite{Armstrong:2020woi,Albayrak:2020fyp}. However, our realization of this duality in Mellin space appears to be the simplest and bears most resemblance to the flat space relation. One might be tempted by the flat space analogy and replace $\mathtt{c}_{s,t,u}$ by $\mathtt{n}_{s,t,u}$ to obtain new amplitudes. However, we do not recognize the amplitudes following from this naive prescription. It would be very interesting also to find a way to properly ``square'' the super gluon amplitudes so that super gravitons amplitudes are obtained. 

It turns out that a modified version of the color-kinematic duality also holds for correlators with arbitrary weights. For simplicity,  we shall discuss here the case of $\langle kkkk\rangle$ correlators,  although analogous results are also found in general correlators.  We extend the definition of $\mathtt{n}_{s,t,u}$ to each exchanged super multiplet (labelled by $p$) as the coefficient of the pole due to the super primary ($m=0$), dividing by the corresponding $SU(2)_L$ polynomial in order to get rid of the dependence on $\beta$. Explicitly, for each $p$ we have
\begin{align}\label{nskkkk}
\mathtt{n}_s=\sum_{i=0}^{\frac{p}{2}}\mathcal{R}^i_{p,m=0}(t,u)\,(1-\alpha)^i\,,
\end{align}
with $\mathtt{n}_t$ and $\mathtt{n}_u$ related to $\mathtt{n}_s$ by crossing as in \eqref{Bosesymm} as follows
\begin{align}
\begin{split}
\mathtt{n}_t=(\alpha-1)^k\,\left(\left.\mathtt{n}_s\right|_{\alpha\to \frac{1}{1-\alpha}}^{\{t,u\}\to\{u,s\}}\right)\,,\qquad
\mathtt{n}_u=(-\alpha)^k\,\left(\left.\mathtt{n}_s\right|_{\alpha\to \frac{\alpha-1}{\alpha}}^{\{t,u\}\to\{s ,t\}}\right)\,.\\
\end{split}
\end{align}
In this definition we have exploited the fact that the poles of the multiplet exchange amplitude can be completely factored out as an $AdS_{d-1}$ scalar exchange amplitude with the dimension of the super primary (see Section \ref{subsec:PSsusy}). Therefore, stripping away the $SU(2)_L$ polynomials each multiplet exchange amplitude can be expressed as $\mathtt{n}_{s,t,u}$ multiplying a series of simple poles in $s$, $t$ or $u$, just as in (\ref{Mkeq2}). The simple poles have constant coefficients, and the coefficient of the first pole is normalized to one. 
These kinematic factors $\mathtt{n}_{s,t,u}$ can be expressed in terms of $SU(2)_R$ R-symmetry polynomials and read
\begin{align}
\begin{split}
\mathtt{n}_s=&\frac{\lambda_{s_p}\,(-1)^p\,\Gamma [ \epsilon\,p ]}{16 \,\epsilon\,(p-3)\, (p-1)^2 \,  \Gamma \left[\frac{p \epsilon }{2}\right]^4 \Gamma \left[\frac{\epsilon}{2} (2 k-p) \right]^2}\,\\
&\Big[
(p-3) (p-1) \left( -4 \epsilon k  - \epsilon p+t+u+2 \epsilon\right)\,\mathcal{Y}_p(\alpha )\\
& +8 (p-3) (p-1) (t-u)\,\mathcal{Y}_{p-2}(\alpha )+(p-2)^2 (-4 \epsilon k + \epsilon p +t+u)\,\mathcal{Y}_{p-4}(\alpha )
\Big]\,,
\end{split}
\end{align}
where for $\langle kkkk \rangle$ we have
\begin{align}\label{Rsymm_kkkk}
\mathcal{Y}_p(\alpha)=\frac{1}{p!}\,\Gamma\left[\tfrac{p+2}{2}\right]\,\, _2F_1\left(\frac{p}{2}+1,-\frac{p}{2};1;1-\alpha \right)\,.
\end{align}
We find that a generalized version of the relation (\ref{jacobi-n}) holds,  which takes the form
\begin{align}\label{generalizedCK}
\mathtt{p}_s(\alpha)\,\mathtt{n}_s+\mathtt{p}_t(\alpha)\,\mathtt{n}_t+\mathtt{p}_u(\alpha)\,\mathtt{n}_u\,=\,0\,.
\end{align}
Here $\mathtt{p}_{s,t,u}(\alpha)$ are polynomials in $\alpha$,  related by crossing symmetry.\footnote{One might wonder how nontrivial such relations are. However, as a linear function of $s$ and $t$, (\ref{generalizedCK}) gives three equations for two unknowns which are the ratios $\mathtt{p}_{s,t,u}$, and is in general overly-determined. Therefore, the existence of solutions is not guaranteed.}  Explicitly, they are given by
\begin{align}
\begin{split}
\mathtt{p}_s(\alpha)=&\left((1-\alpha ) \alpha \right)^{\frac{p-2}{2}}\,\Big[-2 (\alpha -1) (p-2) \, _2F_1\left(1-\frac{p}{2},\frac{p}{2};1;\frac{\alpha }{\alpha -1}\right) \, _2F_1\left(2-\frac{p}{2},\frac{p}{2};1;\frac{1}{\alpha }\right)\\
&+ (2 (\alpha -2) (p-2)-4 k)\, _2F_1\left(1-\frac{p}{2},\frac{p}{2};1;\frac{\alpha }{\alpha -1}\right) \, _2F_1\left(1-\frac{p}{2},\frac{p}{2};1;\frac{1}{\alpha }\right)\\
&+2 (p-2) \, _2F_1\left(2-\frac{p}{2},\frac{p}{2};1;\frac{\alpha }{\alpha -1}\right) \, _2F_1\left(1-\frac{p}{2},\frac{p}{2};1;\frac{1}{\alpha }\right)\Big]\,,
\end{split}
\end{align}
with
\begin{align}
\mathtt{p}_t(\alpha)=(\alpha-1)^{-1-k+3p/2}\,\mathtt{p}_s\left(\frac{1}{1-\alpha}\right)\,, \qquad
\mathtt{p}_u(\alpha)=(-\alpha)^{-1-k+3p/2}\,\mathtt{p}_s\left(\frac{\alpha-1}{\alpha}\right)\,.
\end{align}
Note that when $p=2$ (corresponding to the exchange of massless particles in AdS), the relation \eqref{generalizedCK} reduces to
\begin{align}
\mathtt{n}_s+(\alpha-1)^{k-2}\,\mathtt{n}_t+(-\alpha)^{k-2}\mathtt{n}_u\,=\,0\,,
\end{align}
which is a minimal modification of the color-kinematic duality relation \eqref{jacobi-n}.  A remarkable fact is that the $\mathtt{p}_{s,t,u}(\alpha)$ do not depend on $\epsilon$,  hence they are the same in any spacetime dimension.

It is also possible to extend the relation (\ref{generalizedCK}) to general $\langle k_1\,k_2\,k_3\,k_4\rangle$ correlators. We notice that in each channel there are always an equal number of $\mathcal{E}-1$ multiplets being exchanged. The generalized version of (\ref{generalizedCK}) holds for any triplet of exchanged multiplets which have the same $i$-th lowest dimension in each channel, where $\mathtt{n}_{s,t,u}$ are defined similarly as in (\ref{nskkkk}).  The degrees of polynomials $\mathtt{p}_{s,t,u}(\alpha)$ are determined only by the extremality $\mathcal{E}$ and the label $i$. For example, for next-to-next-to-extremal correlators which have $\mathcal{E}=2$ and include $\langle 2222\rangle$ as a special example, $\mathtt{p}_{s,t,u}(\alpha)$ are just numbers. However,  $\mathtt{p}_{s,t,u}(\alpha)$ do depend on $\epsilon$ in the general case, and explicit formulae appear to be more cumbersome.

\section{Outlook}
In this paper we have developed powerful techniques to compute holographic correlators in non-maximally supersymmetric conformal field theories. These correspond to tree-level (super) gluon amplitudes on AdS. We applied these techniques to compute all tree-level four-point functions of super gluons with arbitrary Kaluza-Klein levels in a variety of SCFTs in three, four, five and six dimensions. Our strategy consists of two basic steps. In the first step we consider the contribution of each intermediate multiplet in the MRV configuration. The special analytic properties for the multiplet exchange in this limit fix the contribution of each member of the multiplet up to an overall factor. As a result 
\begin{equation}
{\cal M}^{(k_1,k_2,k_3,k_4)} =  \mathtt{c}_s \sum_{p} C_{k_1,k_2,p} C_{p,k_3,k_4} {\cal S}_p + \text{crossed}+ {\cal M}^{(k_1,k_2,k_3,k_4)}_{\mathrm{contact}} 
\end{equation}
where the contribution of each multiplet ${\cal S}_p$ is fully fixed, and the full Mellin amplitude is written in terms of the OPE coefficients $C_{k_1,k_2,p}$  plus a possible contact term, of degree zero in the Mandelstam variables. Here $\mathtt{c}_s$ is the corresponding color structure for the s-channel exchange. While for maximally supersymmetric theories these OPE coefficients were known, this is not the case for the problem at hand. The second step of our strategy is then to impose the superconformal Ward identities for the full correlator (and those related by crossing symmetry). This fixes  all OPE coefficients in terms of $C_{2,2,2}$, which is in turn fixed in terms of the current central charge, and implies that contact terms are actually absent. Our final results have the following structure
\begin{equation}
{\cal M} = \mathtt{c}_s M_s + \mathtt{c}_t M_t+ \mathtt{c}_u M_u
\end{equation}
where $M_s,M_t,M_u$ are exchange contributions in the corresponding channels. The full flavor dependence, or color dependence from the point of view of the dual amplitudes, is encoded in the color factors $\mathtt{c}_s,\mathtt{c}_t,\mathtt{c}_u$. This color dependence is identical to that of tree-level gluon amplitudes  in flat space, but the structure of poles in AdS is more complicated
\begin{equation}
M_s^{\rm flat}= \frac{{\rm Res}(t,u)}{s} \to M_s^{\rm AdS}= \sum_m \frac{{\rm Res}^{(m)}(t,u)}{s-s_0-2m} 
\end{equation}
where generically the sum over poles does not truncate. 

Our techniques and explicit results open the gates for progress in two general directions. On one hand they give us the ability to explore gluon amplitudes in AdS in a rigorous setting, involving full-fledged SCFTs. On the other hand, they also give us a powerful new tool to compute protected and unprotected quantities in a variety of non-maximally supersymmetric conformal field theories, usually much less tractable than their maximally supersymmetric cousins. There are several directions worth exploring.
\begin{itemize}
\item A natural question is which structures present in flat space generalize to AdS and which new structures arise in AdS, without a flat space analogue. In this paper we have seem glimpses of  very rich structures behind AdS amplitudes in the form of a Parisi-Sourlas supersymmetry, hidden conformal symmetry and even a color-kinematic duality. A fascinating question is whether there is an algorithm to ``square'' gluon amplitudes in AdS such as to obtain graviton amplitudes in AdS. 

\item Much of the beautiful structures in gluon amplitudes in flat space require considering higher-point amplitudes. It would be very interesting to extend our methods to higher-point correlators. Note that very much as in flat space, we expect gluon amplitudes in AdS to be simpler than graviton amplitudes in AdS, although the CFTs involved are naively less tractable. As in flat space one should study color-ordered amplitudes. 

\item In this paper we considered the leading contribution to the connected holographic correlators in a $1/N$ expansion. To this order, graviton exchanges can be disregarded and the computation is equivalent to that of a supersymmetric gauge theory on AdS. Similar techniques to the ones used in this paper should also allow us to compute the contribution arising from graviton exchanges (the simplest cases with $k_i=2$ were computed in \cite{Zhou:2018ofp}).  Note that the results of Appendix \ref{app:contact}, in particular the absence of linear order solutions to the Ward identities, suggest that contact terms will also be absent in this case. A related problem is to consider correlators with external gravitons, or mixed correlators with both gravitons and gluons. 

\item It would be interesting to consider higher loop corrections to our computation, proportional to higher inverse powers of $C_{\mathcal{J}}$, following the prescription of \cite{Aharony:2016dwx}. This should allow us to explore the geometry of the world volume of the flavor branes. On the other hand, contributions proportional to higher inverse powers of $C_{\mathcal{T}}$ should allow us to explore the geometry of the entire spacetime. 

\item The infinite families of correlators given in this paper contain a wealth of data for various half maximally supersymmetric conformal field theories. This includes protected data, such as OPE coefficients of $\frac{1}{2}$-BPS operators, but also unprotected data, such as anomalous dimensions of intermediate operators. The vast majority of the OPE coefficients computed in this paper were, to our knowledge, unknown. But it should be possible to match such results to localization computations, and test various conjectures (for instance in relation to the chiral algebra). Furthermore, this data can also be fed/compared to the numerical bootstrap treatment of these theories, see for instance \cite{Chang:2017xmr,Chang:2017cdx}.  
\end{itemize}

\label{sec:conclusions}

\acknowledgments
We would like to thank Shai Chester, Lionel Mason, Silviu Pufu and Yifan Wang for helpful conversations. The work of L.F.A., C.B. and P.F. is supported by funding from the European Research Council (ERC) under the European Union's Horizon 2020 research and innovation programme (grant agreement No 787185). The work of X.Z. is supported in part by Simons Foundation Grant No. 488653.

\appendix

\section{Mellin amplitudes of exchange Witten diagrams}
\label{app:exchangeWittenMellin}
For the reader's convenience, we reproduce here the $AdS_{d+1}$ exchange Mellin amplitudes from \cite{Alday:2020dtb}. The exchanged field has conformal dimension $\Delta_E$ and Lorentz spin $\ell_E$ up to 2. The Mellin amplitudes take the form
\begin{equation}
\mathcal{M}_{\Delta_E,\ell_E}(s,t)=\sum_{m}\frac{f_{m,\ell_E}\, Q_{m,\ell_E}(t,u)}{s-\Delta_E+\ell_E-2m}\;,
\end{equation}
and the residues can be obtained by solving the Casimir equation in Mellin space. The residues consist of a factor 
\begin{equation}\label{fmell}
 f_{m,\ell_E}=\frac{(-1)\,2^{1-2\ell_E}\Gamma[\Delta_E+\ell_E]\big(\tfrac{2-\ell_E-\Delta^{1,2}_E}{2}\big)_m\big(\tfrac{2-\ell_E-\Delta^{3,4}_E}{2}\big)_m}{m!(\tfrac{2\Delta_E-d+2}{2})_m \Gamma[\tfrac{\Delta^{1,2}_E+\ell_E}{2}]\Gamma[\tfrac{\Delta^{3,4}_E+\ell_E}{2}]\Gamma[\tfrac{\Delta^{1,E}_2+\ell_E}{2}]\Gamma[\tfrac{\Delta^{2,E}_1+\ell_E}{2}]\Gamma[\tfrac{\Delta^{3,E}_4+\ell_E}{2}]\Gamma[\tfrac{\Delta^{4,E}_3+\ell_E}{2}]}
\end{equation}
with $\Delta^{i,j}_k\equiv \Delta_i+\Delta_j-\Delta_k$, and a degree-$\ell_E$ polynomial $Q_{m,\ell_E}(t,u)$ in $t$ and $u$
\begin{eqnarray}
\nonumber Q_{m,0}&=&1\;,\\
\nonumber Q_{m,1}&=& \frac{(\delta_u^2-\delta_t^2)(t+u+d-2-\Sigma_\Delta)}{4(\Delta_E-d+1)}+(\Delta_E-1)(t-u)\;,\\
Q_{m,2}&=& \frac{(d-1)T_1}{16d(\Delta_E-d)}-\frac{(d-1)T_2}{16d(\Delta_E-d+1)}+\frac{T_3}{16d}\\
\nonumber &+&\frac{(\delta_u^2-\delta_t^2)}{2}(t-u)(u+t+d-2-\Sigma_\Delta)\\
\nonumber &-&\frac{2(1-\Delta_E+\Delta_E^2)-\delta_t^2-\delta_u^2}{2d}(u+t+d-2-\Sigma_\Delta)^2\\
\nonumber &-&\Delta_E(1-\Delta_E)(t-u)^2\;,
\end{eqnarray}
where the combinations $T_i$ are given by
\begin{eqnarray}
\nonumber T_1&=&(\delta_u^2-\delta_t^2)(t+u+d-2-\Sigma_\Delta)\big(u(\delta_u^2-\delta_t^2-8d)\\
\nonumber &+&t (\delta_u^2-\delta_t^2+8d)-(\delta_u^2-\delta_t^2)(\Sigma_\Delta-d+2)\big)\;,\\
T_2&=&((\delta_u-\delta_t)^2-4)((\delta_u+\delta_t)^2-4)(t+u+d-3-\Sigma_\Delta)\\
\nonumber &\times& (t+u+d-1-\Sigma_\Delta)\;,\\
\nonumber T_3&=&((\delta_u-\delta_t)^2-4)((\delta_u+\delta_t)^2-4)\\
\nonumber &+&8\Delta_E(\Delta_E-1)(2(\Delta_E^2-d(\Delta_E+3)+d^2+1)-\delta_t^2-\delta_u^2)\;.
\end{eqnarray}
In the above we have also defined
\begin{eqnarray}
\nonumber\delta_t&\equiv&\Delta_1+\Delta_4-\Delta_2-\Delta_3\;,\\\delta_u&\equiv&\Delta_2+\Delta_4-\Delta_1-\Delta_3\;,\\\nonumber\Sigma_\Delta&\equiv&\Delta_1+\Delta_2+\Delta_3+\Delta_4\;.
\end{eqnarray}

\section{Comments on 3d superconformal blocks}
\label{app:3dsuperconformalblocks}
We have seen in Section \ref{subsec:3dflavor} that $\frac{1}{3}$-BPS multiplets of $\mathfrak{osp}(3|4)$ and $\frac{1}{2}$-BPS multiplets of $\mathfrak{osp}(4|4)$, which are both generated by the action of four supercharges, contain the same bosonic conformal primaries. This can be further checked at the level of characters using the algorithm in \cite{Cordova:2016emh}. However, this fact does not immediately make it clear whether all such conformal primaries can appear in the OPE between two $\frac{1}{3}$-BPS super primaries of a 3d $\mathcal{N} = 3$ theory.
The operators in question are the scalars $t_p^I$ discussed in Section \ref{subsec:3dflavor}, with $SU(2)_R$ spins $\frac{p}{2}$, $\frac{p-2}{2}$ and $\frac{p-4}{2}$, obtained by acting on the super primary with $Q^{\intercal}_i C Q_j$ where $C$ is the charge conjugation matrix.  In deriving \eqref{3d-3pt}, we have assumed that these operators cannot appear, and we used multiplet exchange amplitudes which only contain the other three components $s^I_p$, $A^I_{p,\mu}$ and $r^I_p$. This assumption turns out to be consistent with superconformal symmetry, as we found unique solutions to the superconformal Ward identities for four-point correlators. However, this more stringent selection rule can also be established without reference to any particular model by looking at superconformal blocks, as was suggested in Section \ref{subsec:3dflavor}. In this appendix, we present the details of this check. As a byproduct, we also observe a bonus $\mathbb{Z}_2$ selection rule for 3d $\mathcal{N} = 4$ superconformal blocks, which however does not hold in general for 3d $\mathcal{N} = 3$.

Our check is based on the fact that a conformal block in Mellin space only differs from the corresponding Witten diagram by an entire function. Let us first consider in any spacetime dimension $d\geq 3$ the linear combination
\begin{align}
\mathcal{G}_p(U, V; \alpha) = \mathcal{Y}_p(\alpha) g_{\epsilon p, 0}(U, V) + c_1 \mathcal{Y}_{p - 2}(\alpha) g_{\epsilon p + 1, 1}(U, V) + c_2 \mathcal{Y}_{p - 4}(\alpha) g_{\epsilon p + 2, 0}(U, V) \label{block-ansatz}
\end{align}
which has the same form as the combination of exchange Witten diagrams \eqref{multipletS}. Previously, we used the single zero in the MRV limit to fix the coefficients of the Witten diagrams. Since conformal blocks and Witten diagrams have the same polar part, this immediately predicts that the superconformal block should be given by \eqref{block-ansatz} with
\begin{align}
\begin{split}
c_1 &= \frac{\epsilon (p^2 - k_{12}^2)(p^2 - k_{34}^2)}{8p^2(p - 1)(\epsilon p + 1)} \\
c_2 &= \frac{\epsilon^2 (p^2 - k_{12}^2)\big((p - 2)^2 - k_{12}^2\big) (p^2 - k_{34}^2)\big((p - 2)^2 - k_{34}^2\big)}{256p(p - 1)^2(p - 2)^2(p - 3)(\epsilon p + 1)(\epsilon p - \epsilon + 1)}.
\end{split}
\end{align}
Note that here we normalize the conformal blocks as 
\begin{align}
g_{\Delta, \ell}(\rho, \bar{\rho}) \sim [4|\rho|]^\Delta C^{(\epsilon)}_\ell(\mathrm{cos} \theta), \quad \rho = |\rho| e^{i\theta} = \frac{z}{(1 + \sqrt{1 - z})^2} \label{block-norm}
\end{align}
which is different from the normalization used in Appendix \ref{app:exchangeWittenMellin}. This gives rise to an extra factor of $2$ in $c_1$ compared to \eqref{sollambda}. On the other hand, we can independently reproduce this answer by imposing superconformal Ward identities on (\ref{block-ansatz}), as these Ward identities must hold for individual superconformal blocks. More concretely, we use the recursive procedure in \cite{kps14} to compute corrections to \eqref{block-norm} and solve the superconformal Ward identities order-by-order in $|\rho|$. Re-expressing \eqref{scfWardida} in the radial coordinates leads to
\begin{align}
\left [ \frac{e^{-i\theta}}{8} \frac{(1 + \rho)^3}{1 - \rho} \left ( \frac{\partial}{\partial |\rho|} + \frac{i\mathrm{sin} \theta}{|\rho|} \frac{\partial}{\partial \mathrm{cos}\theta} \right ) - \epsilon \alpha \frac{\partial}{\partial \alpha} \right ] \mathcal{G}(\rho, \bar{\rho}; \alpha) \biggl |_{\alpha = \frac{(1 + \rho)^2}{4\rho}} = 0\;. \label{radial-ward}
\end{align}
The expansion coefficients of this equation at low-lying orders already fix $c_1$ and $c_2$ to be the values given above. We have also gone up to $\mathcal{O}(|\rho|^{10})$ to check that (\ref{radial-ward}) vanishes at higher orders. Let us now turn to the case of 3d $\mathcal{N}=3$ with $\epsilon=\frac{1}{2}$. To show  the extra scalar operators are absent, we now only need to show that their contributions cannot satisfy the superconformal Ward identities. The scalar with $SU(2)_R$ spin $\frac{p-2}{2}$ can be ruled out because it does not obey the Bose symmetry in the s-channel. We then only need to consider the combinations 
\begin{equation}
\mathcal{G}_p'(U, V; \alpha) = U^{\epsilon\mathcal{E} - \tfrac{\epsilon}{2}(k_1 + k_2)} V^{\tfrac{\epsilon}{2}(\Sigma - k_4) - \epsilon\mathcal{E}} \big(d_1\mathcal{Y}_{p}(\alpha) +d_2\mathcal{Y}_{p-4}(\alpha) \big)g_{\epsilon p + 1, 0}(U, V). \label{test-ward2}
\end{equation}
By looking at several values of $p$ and the external weights $k_i$, and using the above $\rho$-expansion, we find that the only solution is the trivial one with both $d_1$ and $d_2$ zero. This concludes that $\mathfrak{osp}(3|4)$ and $\mathfrak{osp}(4|4)$ have the same superconformal block exchanging this short super multiplet.

Note that the above strategy of computing superconformal blocks can be also applied to exchanging other types of super multiplets such as the long multiplets. In particular, looking at long multiplets allows us to see the differences between the two superconformal algebras. For the $\mathfrak{osp}(3|4)$ blocks, it is easy to find values of $\Delta$ which force {\it all} bosonic components of the long multiplet to appear with a non-zero coefficient. Conversely, repeating this exercise for $\mathfrak{osp}(4|4)$ shows a surprising amount of structure in the blocks for long multiplets. To illustrate, we have been able to satisfy the Ward identity for an $\mathfrak{osp}(4|4)$ long multiplet with the ansatz
\begin{align}
\mathcal{G}^{\mathcal{N} = 4}_{\Delta, \ell, p}(U, V; \alpha) &= \mathcal{Y}_p(\alpha) g_{\Delta, \ell}(U, V) + \sum_{i, j = \pm 1} c^{(1)}_{ij} \mathcal{Y}_{p + 2i}(\alpha) g_{\Delta + 1, \ell + j}(U, V) \nonumber \\
&+ \sum_{i = 0, \pm 2} c^{(2)}_{i0} \mathcal{Y}_{p + 2i}(\alpha) g_{\Delta + 2, \ell}(U, V) + \sum_{j = \pm 2} c^{(2)}_{0j} \mathcal{Y}_p(\alpha) g_{\Delta + 2, \ell + j}(U, V) \label{long-version} \\
&+ \sum_{i, j = \pm 1} c^{(3)}_{ij} \mathcal{Y}_{p + 2i}(\alpha) g_{\Delta + 3, \ell + j}(U, V) + c^{(4)} \mathcal{Y}_p(\alpha) g_{\Delta + 4, \ell}(U, V). \nonumber
\end{align}
Note that this ansatz is much more restrictive than an expression which includes all of the operator content indicated by \cite{Cordova:2016emh}. As with \eqref{block-ansatz}, it has the property that all exchanged super descendants have the same conformal twist as the primary mod 2.

The parity selection rule evident in \eqref{long-version} is reminiscent of the ``bonus symmetry'' of 4d $\mathcal{N} = 4$ SYM \cite{i98,is99}. More recently, \cite{Chester:2014fya,acp19} found evidence of a bonus parity by computing 3d $\mathcal{N} = 8$ superconformal blocks. Our results suggest that this structure might be a feature of all 3d SCFTs with $\mathcal{N} \geq 4$ superconformal symmetry. It would be interesting to prove this conjectured bonus selection rule using the superspace approach to superconformal correlators \cite{p99}.

\section{Chiral algebra and twisted correlators}
\label{sec:chiralalgebra}
 
The holographic theories we studied in four dimensions with $\mathcal{N} = 2$ superconformal symmetry admit a chiral algebra structure through the construction of \cite{bllprv13}. For a special class of protected operators $\mathcal{O}^{\alpha_1 \dots \alpha_{2j}}(z, \bar{z})$ restricted to the plane parameterized by $(z, \bar{z})$, there exists a $\bar{z}$-dependent slice of R-symmetry space which defines operators with meromorphic OPEs. Translated into the language of correlators, four-point functions of these operators become meromorphic after the twist $\alpha = 1 / \bar{z}$, as indicated by the superconformal Ward identity \eqref{scfWardida} with $\epsilon=1$. The special operators with this property were classified in \cite{bllprv13}, and from this list, we can see that the super primary of the flavor current multiplet and its higher-weight cousins generate a chiral algebra. Thanks to the rigid structure of meromorphic OPEs, we do not need a microscopic definition of this chiral algebra to compute four-point functions of arbitrary generators to the order $1 / C_{\mathcal{J}}$ considered in this paper. Taking the selection rule \eqref{selection-p} as the one input from holography, we can fix the twisted correlators in terms of their singularities including the OPE coefficients once we impose crossing symmetry in the chiral algebra. This amounts to an independent derivation of the three-point functions \eqref{ope-sol-4d}.\footnote{The same procedure applied to the 6d $\mathcal{N} = (2, 0)$ theory would still leave ambiguities in the twisted correlators. However, in this case there \textit{is} a microscopic definition of the conjectured chiral algebra in terms of the $\mathcal{W}_N$ family of algebras \cite{brv14}.}

In this appendix, we present the details of computing meromorphic correlators in the world volume theories of D3-branes near F-theory singularities. We will first present the calculation from the field theory side, where it follows straightforwardly from the chiral algebra OPE. We then present the holographic calculation using the Mellin amplitudes obtained in Section \ref{sec:MRV}. Note that the cancellation of the non-meromorphic piece in the twisted four-point function requires all multiplets in all three channels. However, as we will explain, meromorphic correlators can be most conveniently extracted from the MRV limit where only two of the three exchange channels are present. Similar calculations for the maximally superconformal case were done in \cite{Behan:2021pzk}.

\vspace{0.5cm}

\noindent{\bf A crossing symmetric solution}

\vspace{0.2cm}

Let us start by discussing the kinematics using conventions that are more standard in two dimensional CFT. The external operators $\mathcal{O}^I_k(x,v,\bar{v})$ descend to currents $J^I_{(k)}(z,\bar{v})$ of spin $h = \frac{k}{2}$ after twisting.\footnote{We have parenthesized $k$ to avoid confusing $J^I_{(k)}$ with the Laurent modes of some current $J^I$.} Their weights with respect to $G_F$ and $SU(2)_L$ are unchanged, while there is no $SU(2)_R$ dependence anymore because of the superconformal twist. We can define the chiral algebra correlator to be
\begin{align}
& \left < J^{I_1}_{(k_1)}(z_1,\bar{v}_1) J^{I_2}_{(k_2)}(z_2,\bar{v}_2) J^{I_3}_{(k_3)}(z_3,\bar{v}_3) J^{I_4}_{(k_4)}(z_4,\bar{v}_4) \right > = \left [ \frac{z_{24} (\bar{v}_1 \cdot \bar{v}_4)}{z_{14} (\bar{v}_2 \cdot \bar{v}_4)} \right ]^{\frac{k_{12}}{2}} \left [ \frac{z_{14} (\bar{v}_1 \cdot \bar{v}_3)}{z_{13} (\bar{v}_1 \cdot \bar{v}_4)} \right ]^{\frac{k_{34}}{2}} \nonumber \\
& \hspace{3cm} \left [ \frac{\bar{v}_1 \cdot \bar{v}_2}{z_{12}} \right ]^{\frac{k_1 + k_2 - 4}{2}} \left [ \frac{\bar{v}_3 \cdot \bar{v}_4}{z_{34}} \right ]^{\frac{k_3 + k_4 - 4}{2}} \frac{\mathcal{F}^{I_1I_2I_3I_4}_{1234}(z; \beta)}{z_{12}^2 z_{34}^2} \label{2d-prefactor}
\end{align}
where $z_{ij}=z_i-z_j$ and $k_{ij}=k_i-k_j$. The dynamical part $\mathcal{F}^{I_1I_2I_3I_4}_{1234}(z; \beta)$ is a meromorphic function of the conformal cross ratio $z$ which is related to the holomorphic coordinates $z_i$ on the plane via 
\begin{equation}
z=\frac{z_{12}z_{34}}{z_{13}z_{24}}\;.
\end{equation}
It admits an expansion in $\mathfrak{sl}(2)$ blocks which are given by
\begin{equation}
g^{h_{12}, h_{34}}_h(z) = z^h {}_2F_1(h - h_{12}, h + h_{34}; 2h; z)
\end{equation}
and must obey the following crossing symmetry conditions
\begin{subequations}\label{2d-crossing}
\begin{align}
\mathcal{F}^{I_1I_2I_3I_4}_{1234}(z; \beta) &= \frac{z^{\frac{k_1 + k_2}{2}}}{(1 - z)^{\frac{k_2 + k_3}{2}}} \frac{(\beta - 1)^{\frac{k_2 + k_3 - 4}{2}}}{\beta^{\frac{k_3 - k_1}{2}}} \mathcal{F}^{I_3I_2I_1I_4}_{3214} \left ( 1 - z; \frac{\beta}{\beta - 1} \right ) \label{2d-crossing1} \\
\mathcal{F}^{I_1I_2I_3I_4}_{1234}(z; \beta) &= \frac{(\beta z)^{\frac{k_1 + k_4}{2}}}{\beta^2} \mathcal{F}^{I_4I_2I_3I_1}_{4231} \left ( \frac{1}{z}; \frac{1}{\beta} \right ). \label{2d-crossing2}
\end{align}
\end{subequations}
Importantly, only the invariance under \eqref{2d-crossing1} will be manifest in the expressions we find for $\mathcal{F}^{I_1I_2I_3I_4}_{1234}(z; \beta)$.

Unlike the $k_1\leq k_2\leq k_3\leq k_4$ ordering assumed in (\ref{GandcalG}), here we will assume without loss of generality a different ordering $k_2\leq k_1\leq k_3\leq k_4$, which guarantees 
\begin{align}
k_1 + k_2 \leq k_3 + k_4\;, \quad\quad \;\; k_2 + k_3 \leq k_1 + k_4\;. \label{2d-ineq}
\end{align}
As was explained in \cite{Behan:2021pzk}, this choice is more convenient for computing chiral algebra correlators because the singularities encountered when $z_2 \rightarrow z_1$ and $z_2 \rightarrow z_3$ determine those of the full correlator. On the other hand, the most general four-point functions with arbitrary orderings are obtained by permuting the external operators and applying the crossing relations above. With this chosen ordering, the two cases we need to distinguish are no longer \eqref{twocases} but
\begin{equation}\label{2d-twocases}
k_2+k_4 > k_1+k_3 \;\; \text{(case I$'$)}\;,\quad\quad\;\; k_2+k_4 \leq k_1+k_3 \;\; \text{(case II$'$)}\;.
\end{equation}
Since crossing symmetry in the chiral algebra will be one of our main tools, we will be most interested in case II$'$. In this case the set inequalities of \eqref{2d-ineq} and the case II$'$ condition are preserved under $1 \leftrightarrow 4$, thereby allowing us to impose \eqref{2d-crossing2} instead of just using it to define new weight arrangements. To be specific, we will choose $\mathtt{c}_s$ and $\mathtt{c}_t$ to be the independent flavor structures and derive four-point functions of the form
\begin{equation}
\mathcal{F}_{1234}(z; \beta) = \mathtt{c}_s \mathcal{F}^{(s)}_{1234}(z; \beta) - \mathtt{c}_t \mathcal{F}^{(t)}_{1234}(z; \beta)\;, \label{2d-st}
\end{equation}
where
\begin{equation}
\mathcal{F}^{(t)}_{1234}(z; \beta) = \frac{z^{\frac{k_1 + k_2}{2}}}{(1 - z)^{\frac{k_2 + k_3}{2}}} \frac{(\beta - 1)^{\frac{k_2 + k_3 - 4}{2}}}{\beta^{\frac{k_3 - k_1}{2}}} \mathcal{F}^{(s)}_{3214} \left ( 1 - z; \frac{\beta}{\beta - 1} \right ). \label{2d-st2}
\end{equation}
This guarantees invariance under \eqref{2d-crossing1}. On the other hand, \eqref{2d-crossing2} can be split into
\begin{subequations}\label{2d-crossing-explicit}
\begin{align}
\mathcal{F}^{(s)}_{1234}(z; \beta) &= \frac{(\beta z)^{\frac{k_1 + k_4}{2}}}{\beta^2} \mathcal{F}^{(s)}_{4231} \left ( \frac{1}{z}; \frac{1}{\beta} \right )\;, \label{s-only} \\
\mathcal{F}^{(t)}_{1234}(z; \beta) &= -\frac{(\beta z)^{\frac{k_1 + k_4}{2}}}{\beta^2} \left [ \mathcal{F}^{(s)}_{4231} \left ( \frac{1}{z}; \frac{1}{\beta} \right ) + \mathcal{F}^{(t)}_{4231} \left ( \frac{1}{z}; \frac{1}{\beta} \right ) \right ]\;,
\end{align}
\end{subequations}
with the help of the Jacobi identity \eqref{jacobi-c}. As we will see, these equations are consistent with \eqref{ope-sol-4d}, which is one of the main results in the body of the paper. However, we emphasize that crossing symmetry of the chiral algebra and that of the higher dimensional theory do not follow from each other in a simple way.

As discussed at the beginning of this appendix, we assume the following OPE at tree level for arbitrary $C_{k_1,k_2,p}$
\begin{align}
J^{I_1}_{(k_1)} & (z_1,\bar{v}_1) J^{I_2}_{(k_2)}(z_2,\bar{v}_2) = f^{I_1 I_2}_{\;\;\;\;\;\;\;I} \sum_{p = |k_{12}| + 2}^{k_1 + k_2 - 2} C_{k_1,k_2,p} \sum_{m = 0}^{\tfrac{k_1 + k_2 - p - 2}{2}} \frac{\left ( \frac{k_{12} + p}{2} \right )_m}{m!(p)_m} \frac{(\bar{v}_1 \cdot \bar{v}_2)^{\frac{k_1 + k_2 - p - 2}{2}}}{z_{12}^{\frac{k_1 + k_2 - p}{2} - m}} \nonumber \\
& \frac{\partial^m}{(p - 2)!} \sum_{\sigma \in S_{p - 2}} J^I_{(p) ; \bar{\alpha}_{\sigma(1)} \dots \bar{\alpha}_{\sigma(p - 2)}}(z_2) \bar{v}_1^{\bar{\alpha}_1} \dots \bar{v}_1^{\bar{\alpha}_{(p + k_{12} - 2)/2}} \bar{v}_2^{\bar{\alpha}_{(p + k_{12})/2}} \dots \bar{v}_2^{\bar{\alpha}_{p - 2}}. \label{2d-ope}
\end{align}
This assumption of the chiral algebra OPE follows from the holographic setup given by the brane construction, and is valid in the $N\to\infty$ limit. The chiral algebra correlator is fixed by demanding that the singularities implied by \eqref{2d-ope} are reproduced in each channel. Following \cite{Behan:2021pzk}, this indeed produces an expression of the form \eqref{2d-st} where
\begin{align}
&\mathcal{F}^{(s)}_{1234}(z; \beta) = z^{\frac{k_1 + k_2}{2}} \sum_{p = |k_{12}| + 2}^{k_1 + k_2 - 2} C_{k_1,k_2,p}C_{k_3,k_4,p} g_{1 - \frac{p}{2}}^{\frac{k_{21}}{2}, \frac{k_{43}}{2}}(\beta^{-1}) \sum_{m = 0}^{\tfrac{k_1 + k_2 - p - 2}{2}} \frac{\left ( \frac{k_{21} + p}{2} \right )_m \left ( \frac{k_{34} + p}{2} \right )_m}{m!(p)_m z^{\frac{k_1 + k_2 - p}{2} - m}}. \label{chiral-corr}
\end{align}
The OPE coefficients must be chosen such that \eqref{2d-crossing-explicit} are satisfied for all external weights belonging to case II$'$, which is non-trivial. While it is reassuring that this happens for $C_{k_1,k_2,p} = C_{2,2,2}$, we can in fact say more. Imposing these nontrivial crossing equations on a large number of $\left < kkqq \right >$ examples shows that in fact $C_{k_1,k_2,p} = C_{2,2,2}$ is the \textit{unique solution}. Upon using this solution, we also observe that the chiral algebra correlators for case II$'$ can be recast into the remarkably compact form
\begin{align}
\mathcal{F}^{(s)}_{1234}(z;\beta) = (C_{2,2,2})^2 \sum_{n=1+\tfrac{|k_{12}|}{2}}^{\tfrac{k_1+k_2-2}{2}}\beta^{n-1}\,z^n \label{simple-corr}
\end{align}
which shows that \eqref{s-only} is manifest.
This is very reminiscent of a simplification previously seen in the $\mathcal{N}=4$ results of \cite{Behan:2021pzk}.

\vspace{0.5cm}

\noindent{\bf Matching with holography}

\vspace{0.2cm}

According to the chiral algebra construction, our main result \eqref{chiral-corr} should also be reproduced from the 4d result after applying the twist $\alpha = \bar{z}^{-1}$. As commented earlier, showing the meromorphy of the twisted correlator requires all multiplets in all channels.\footnote{This can be explicitly checked in many examples.} However, our focus here is to extract the meromorphic correlators and compare them with the field theory result. To this end, it is most efficient to go to a special kinematic limit using the independence of the twisted correlator on $\bar{z}$. This limit turns out to be the same as the MRV limit where only two channels are present. In this limit the matching of the meromorphic correlators can be seen at the level of each multiplet.

The starting point of this calculation is the inverse Mellin integral. Taking into account the difference in the extra factors in \eqref{GandcalG} and \eqref{2d-prefactor}, and different orderings of the weights, the function we should twist is
\begin{align}
\mathcal{F}^{I_1I_2I_3I_4}_{1234}(z, \bar{z}; \alpha, \beta) &= (\alpha \beta U)^{\frac{1}{2}(k_1 + k_2) - \mathcal{E}} \mathcal{G}^{I_2I_1I_3I_4}_{2134} \left ( \frac{U}{V}, \frac{1}{V}; 1 - \alpha, 1 - \beta \right ) \nonumber \\
&= \int_{-i\infty}^{i\infty} \frac{dsdt}{(4\pi i)^2} U^{\frac{s}{2}} V^{-\frac{s}{2} - \frac{t}{2} + \frac{1}{2}(k_1 + k_4)} (\alpha\beta)^{\frac{1}{2}(k_1 + k_2) - \mathcal{E}} \label{2134-mellin} \\
& \hspace{2.5cm} \mathcal{M}^{I_2I_1I_3I_4}_{2134}(s, t; 1 - \alpha, 1 - \beta) \, \Gamma_{k_2k_1k_3k_4}. \nonumber
\end{align}
Clearly, only one of the two terms in \eqref{2d-st} needs to be computed thanks to the crossing relation (\ref{2d-st2}). In \cite{Behan:2021pzk}, the function $\mathcal{F}^{(t)}$ was computed by extracting negative powers of $1 - z$ from an analogous integral. In this work, we will make the opposite choice and compute $\mathcal{F}^{(s)}$ instead. Recalling that \eqref{2d-prefactor} becomes proportional to $z^{-\frac{k_1 + k_2}{2}} \mathcal{F}^{I_1I_2I_3I_4}(z; \beta)$ in the $(z_1, z_2, z_3, z_4) = (0, z, 1, \infty)$ kinematics, $\mathcal{F}^{(s)}$ requires us to extract the powers of $z$ up to $z^{\frac{k_1 + k_2}{2} - 1}$. These powers can only arise from single-particle poles in the s-channel. All of these are simple poles since we are dealing with Mellin amplitudes at tree level. Evaluating the $t$-integral is more difficult because its contour encircles infinitely many double-particle poles from the Gamma function factor. However, we can circumvent this difficulty by exploiting the fact that twisted correlators are independent of $\bar{z}$. This means there is no loss of generality in setting $\bar{z}$ to a special value if $\alpha = \bar{z}^{-1}$. It will be most convenient to take $\bar{z} = 1$ so that we do not have to worry about the region $t > k_1 + k_4 - s$ and localize the $t$-integral to just a single pole at $t = k_1 + k_4 - s$.\footnote{In restricting the outer integral to $s = p + 2m$ and the inner integral to $t = k_1 + k_4 - s$, we have relied on the left and right inequalities of \eqref{2d-ineq} respectively.} Note that the $\alpha=1$ limit is also the t-channel MRV limit. We now implement this strategy, and start by restoring the $SU(2)_L$ polynomial in the s-channel residue \eqref{Sp} and switching $1 \leftrightarrow 2$. This leads to
\begin{align}
\frac{\mathcal{Y}_{p-2}(\beta)}{s - p - 2m} \sum_{i = 0}^{\mathcal{E}} \mathcal{R}^{p,m; i}_{1234}(t, u) (1 - \alpha)^i &\rightarrow \frac{\mathcal{Y}_{p-2}(1 - \beta)}{s - p - 2m} \sum_{i = 0}^{\mathcal{E}} \mathcal{R}^{p,m; i}_{2134}(t, u) \alpha^i \\
&= \frac{(-1)^{\frac{2p - 4 - \kappa_t - \kappa_u}{4}}}{s - p - 2m} \sum_{i = 0}^{\mathcal{E}} \mathcal{R}^{p,m; i}_{2134}(t, u) \alpha^i \beta^{\mathcal{E} - \frac{k_1 + k_2}{2}} g_{1 - \frac{p}{2}}^{\frac{k_{21}}{2}, \frac{k_{43}}{2}} (\beta^{-1}). \nonumber
\end{align}
We then apply the chiral algebra twist and extract the poles discussed above to get
\begin{align}
\mathcal{F}^{(s)}(z; \beta) &= z^{\frac{k_1 + k_2}{2}} \frac{\Gamma \left [ \frac{\kappa_t}{2} \right ]}{2} \sum_{p = \mathrm{max} \{ |k_{12}|, |k_{34}| \} + 2}^{k_1 + k_2 - 2} (-1)^{\frac{2p - 4 - \kappa_t - \kappa_u}{4}} g_{1 - \frac{p}{2}}^{\frac{k_{21}}{2}, \frac{k_{43}}{2}}(\beta^{-1}) \nonumber \\
&\sum_{m = 0}^{\tfrac{k_1 + k_2 - p - 2}{2}} \frac{\Gamma \left [ \frac{k_{21} + 2m + p}{2} \right ] \Gamma \left [ \frac{k_{34} + 2m + p}{2} \right ]}{m! (p)_m z^{\frac{k_1 + k_2 - p}{2} - m}} \sum_{i = 0}^{\mathcal{E}} (-1)^i K^i_p  E^i_p \label{ke-nob}
\end{align}
where we have used $K^i_p$ as a shorthand for $K^i_p(k_1 + k_4 - 2m - p, k_2 + k_3)$. In other words,
\begin{align}
\begin{split}
K^i_p &= 2i(\kappa_t - 2)\left ( \frac{1}{2} \kappa_t + \frac{1}{2} \kappa_u - 2m - p \right ) \\
&- \frac{1}{4}\left ( \frac{1}{2} \kappa_t - \frac{1}{2} \kappa_u - 2i - 2m - p \right )(2p - \kappa_t - \kappa_u)(2p + \kappa_t + \kappa_u - 4).
\end{split}
\end{align}
The remaining task is to compute the inner sum of \eqref{ke-nob}. Since this sum is hypergeometric, it is a simple matter of using a Gamma function identity and $\{ \kappa_t + \kappa_u, \kappa_t - \kappa_u \} = \{ -2k_{21}, -2k_{34} \}$ to recover \eqref{chiral-corr}.
This shows the desired matching between the multiplets labelled by $p$ individually.

Let us now return to \eqref{simple-corr} which gives the simplest expression for the chiral algebra correlators. Although the exchange amplitudes in Section \ref{sec:MRV} naturally led to chiral algebra four-point functions that took the original form \eqref{chiral-corr} under the superconformal twist, the simplification \eqref{simple-corr} is instead best derived by using the reduced Mellin amplitudes. As with the $\mathcal{N} = 4$ SYM case, the $\mathcal{N}=2$ reduced amplitudes show that the complexity of the rational function that accompanies each R-symmetry monomial is bounded with respect to the external weights. Indeed, it is difficult to imagine how the hidden conformal symmetry observed in Section \ref{hiddenconformal} would have been possible were this not the case. From Section \ref{hiddenconformal}, we know that the full Mellin amplitudes are related to the reduced Mellin amplitudes by
\begin{align}
\mathcal{M}_{2134}(s, t; 1 - \alpha, 1 - \beta) &= \frac{k_2 + k_3 - u}{2} \frac{k_1 + k_4 - u}{2} \alpha \widetilde{\mathcal{M}}_{2134}(s, t; 1 - \alpha, 1 - \beta) \label{MfromMt} \\
&+ \frac{k_1 + k_2 - s}{2} \frac{k_3 + k_4 - s}{2} \alpha (\alpha - 1) \widetilde{\mathcal{M}}_{2134}(s - 2, t; 1 - \alpha, 1 - \beta) \nonumber \\
&+ \frac{k_2 + k_4 - t}{2} \frac{k_1 + k_3 - t}{2} (1 - \alpha) \widetilde{\mathcal{M}}_{2134}(s, t - 2; 1 - \alpha, 1 - \beta)\;, \nonumber
\end{align}
which follows from $\mathcal{M} = \widehat{R} \circ \widetilde{\mathcal{M}}$ after swapping $1 \leftrightarrow 2$. To see that this expression matches the sum over poles with contact terms absorbed from Section \ref{sec:correlators}, one will need to use a number of identities. However, since our goal is to make \eqref{simple-corr} transparent it is best \textit{not} to apply any of these identities. Let us now compute $\mathcal{F}^{(s)}$ by performing the inverse Mellin transformation for $\bar{z} = 1$ again, but using the expression given by (\ref{MfromMt}). The choice $1/\alpha=\bar{z}=1$ leaves only the first term of \eqref{MfromMt} after the twist. Moreover, note that only the $i = 0$ term in the sum \eqref{Mreducedgeneral} survives, which leads to
\begin{align}
\begin{split}
\mathcal{M}_{2134}(s, t; 0, 1 - \beta) &= (C_{2,2,2})^2 \sum_{j + k = \mathcal{E} - 2} \frac{\beta^j \Gamma \left [ \tfrac{\kappa_t + 2}{2} \right ]^{-1}}{j! k! \left ( j + \frac{\kappa_u}{2} \right )! \left ( k + \frac{\kappa_s}{2} \right )!} \label{convenient-m} \\
&\times \left [ \mathtt{c}_s \frac{u - k_1 - k_4}{s - s_M + 2k} - \mathtt{c}_t \frac{u - k_1 - k_4}{t - t_M + 2j} \right ].
\end{split}
\end{align}
The $t$-integral is further localized at $t = k_1 + k_4 - s$ as before, and kills the $1 - \bar{z}$ in \eqref{2134-mellin}. We can then extract all the single-particle s-poles in \eqref{convenient-m}. As long as the weight condition for case II$'$ is met, we will have $\mathcal{E} = k_2$, which makes the powers of $\beta$ the same as those in \eqref{simple-corr}. This completes the derivation of the simplified correlator \eqref{simple-corr}. 

\section{Contact terms and higher-derivative corrections}
\label{app:contact}

In this appendix we study a class of crossing symmetric solutions to the superconformal Ward identities, where we relax the condition of constant growth in the flat space limit. Such contributions will be relevant when considering, for instance, subleading higher-derivative corrections to our results. The exchange of the current multiplet ($p=2$) is fixed in terms of the central charge to be $1/C_{\mathcal{J}}$-exact, and is expected not to receive corrections. Hence,  in our ansatz below we allow for the exchange of multiplets with $p>2$ in addition to a contact term of degree $D$.\footnote{For simplicity we are assuming that there are no new states appearing in the exchanges. \label{footnote-no-extra}} For simplicity, we will focus on $\langle kkkk \rangle$ correlators. For $D=0$ we find no new solutions. This is of course expected since in the body of the paper we showed our solutions are unique, and all amplitudes were given in terms of multiplet exchange amplitudes with no additional contact terms. Below we will analyze in detail the cases $D=1,2$ and make a brief remark about $D \geq 3$.

\vspace{0.5cm}

\noindent{\bf Ansatz}

\vspace{0.2cm}

\noindent We start by writing down an ansatz for the higher-derivative corrections 
\begin{equation}
\mathcal{M}_{\rm H.D.}^{I_1I_2I_3I_4}=\mathcal{M}_{\rm H.D., \, ex}^{I_1I_2I_3I_4}+\mathcal{M}_{\rm H.D., \,con}^{I_1I_2I_3I_4}
\end{equation}
which consists of an exchange part $\mathcal{M}_{\rm H.D., \, ex}^{I_1I_2I_3I_4}$ and a contact part $\mathcal{M}_{\rm H.D., \,con}^{I_1I_2I_3I_4}$. In the exchange part, we include all multiplets $p$ allowed at $\mathcal{O}(1/C_{\mathcal{J}})$ with $p>2$
\begin{equation}\label{ansatzexchange}
\mathcal{M}_{\rm H.D., \, ex}^{I_1I_2I_3I_4}=\mathtt{c}_s\,\left( \sum_{p=4}^{2(k-1)}\Lambda_p\, \mathcal{S}'_p(s,t;\alpha)\,\mathcal{Y}_{p-2}(\beta)\right)\,+\,\rm{crossed}\,,
\end{equation}
where the prime symbol means that the $\lambda$ coefficient in the multiplet exchange amplitude (\ref{Sp}) is set to 1, but we explicitly include a new coefficient $\Lambda_p$ for each multiplet. The contact part is given by a sum over all irreducible representations of the flavor group $G_F$ in ${\bf adj}_{G_F}\times {\bf adj}_{G_F}$
\begin{align}
\mathcal{M}_{\rm H.D., \,con}^{I_1I_2I_3I_4}=\sum_{{\bf R}_a\in {\bf adj}_{G_F} \times {\bf adj}_{G_F}} \mathrm{P}_a^{I_1I_2I_3I_4}\,\mathcal{M}_{\rm H.D., \,con}^a\,.
\end{align}
Each amplitude $\mathcal{M}^a_{\rm H.D., \,con}$ is a polynomial in the four variables $s$, $t$, $\alpha$ and $\beta$. While the degree in the last two variables is dictated by R-symmetry considerations, the degree in $s$ and $t$ can be freely tuned, at least in principle.  We parameterize contact terms of degree $D$ in $s$ and $t$ by 
\begin{align}\label{generalcontact}
\mathcal{M}^a_{\rm H.D., \,con}=\sum_{0\le m\le k}\sum_{0\le n \le k-2}\sum_{\substack{0\le p+q \le D\\p\ge 0, \, q\ge 0}}c^a_{m,n;p,q}\alpha^m\,\beta^n\,s^p\,t^q\,,
\end{align}
and we look for crossing symmetric solutions to the superconformal Ward identities for increasing values of $D$. Note that the exchange part is already crossing symmetric. On the other hand, crossing symmetry will constrain the individual amplitudes $\mathcal{M}_{\rm H.D., \,con}^a$ in two ways. First, one has to impose the conditions \eqref{Bosesymmfcrossing}:
\begin{align}\label{crossingcontact}
\begin{split}
\mathcal{M}^{a}_{\rm H.D., \,con}&=\sum_b (-1)^{|\mathbf{R}_a|}({\rm F}_u)_a{}^b (\alpha-1)^{k}(\beta-1)^{k-2}\bigg(\mathcal{M}^{b}_{\rm H.D., \,con}\big|^{\{s,t,u\}\to\{t,u,s\}}_{\{\alpha,\beta\}\to\{\frac{1}{1-\alpha},\frac{1}{1-\beta}\}}\bigg)\;,\\
\mathcal{M}^{a}_{\rm H.D., \,con}&=\sum_b (-1)^{|\mathbf{R}_a|}({\rm F}_t)_a{}^b(-\alpha)^{k}(-\beta)^{k-2}\bigg(\mathcal{M}^{b}_{\rm H.D., \,con}\big|^{\{s,t,u\}\to\{u,s,t\}}_{\{\alpha,\beta\}\to\{\frac{\alpha-1}{\alpha},\frac{\beta-1}{\beta}\}}\bigg)\;.
\end{split}
\end{align}
Moreover,  Bose symmetry requires that the amplitude is also symmetric under the exchange of the first two operators.\footnote{Together with \eqref{crossingcontact},  this generates the whole crossing symmetry group.  As discussed at the end of Section \ref{sec:MRV},  this symmetry is built-in for exchange amplitudes,  so it must only be enforced on contact terms.} For the $\langle kkkk\rangle$ correlators we are discussing here,  this condition translates to
\begin{align}
\mathcal{M}_{\rm H.D., \,con}^{I_1I_2I_3I_4}(s,t;\alpha,\beta)=\mathcal{M}_{\rm H.D., \,con}^{I_2I_1I_3I_4}(s,u;1-\alpha,1-\beta)\,,\label{Bose12}
\end{align}
and given that $\mathrm{P}_a^{I_2I_1|I_3I_4}=(-1)^{|{\bf R}_a|}\mathrm{P}_a^{I_1I_2|I_3I_4}$ it corresponds to a symmetry or antisymmetry condition on $\mathcal{M}^a_{\rm H.D., \,con}$ according to the parity of the representation ${\bf R}_a$.

\vspace{0.5cm}

\noindent\underline{{\bf $\mathbf{D=0,1}$}}

\vspace{0.2cm}

\noindent Our first finding is that the Ward identities cannot be solved for $D=0,1$, regardless of the spacetime dimension $d$ or the flavor group $G_F$. This provides a consistency check for our results from Sections \ref{sec:correlators} and \ref{sec:flavor}, where contact terms in these four-point amplitudes are found to be uniquely fixed by superconformal symmetry. The non-existence of solutions for $D=1$ has interesting implications. The coupling of the super gluons to two-derivative supergravity would in general introduce such contact terms. Our result shows that in tree-level super gluon amplitudes at subleading order, where super gravitons are exchanged, contact terms should also be uniquely fixed by the exchange contributions. 

\vspace{0.5cm}

\noindent\underline{{\bf $\mathbf{D=2}$}}

\vspace{0.2cm}

\noindent Nontrivial solutions start to appear at $D=2$ for suitable choices of the coefficients $c^a_{m,n;p,q}$ and $\Lambda_p$ in \eqref{generalcontact} and (\ref{ansatzexchange}).  In the remainder of this appendix we shall discuss such solutions, which interestingly show rather different behaviors in $d=4$ and in $d=3,5,6$. We will discuss the two cases separately.

\vspace{0.2cm}

\noindent $\bf{d=4}$.  In four dimensions we find that all $\Lambda_p$ are zero. The solutions are given by polynomials of degree 2 in $s,t$ with the following structure
\begin{align}\label{contact4d}
\mathcal{M}_{\rm H.D., \,con}^a=P(s,t;\alpha)\,Q^a(\alpha,\beta)\,,
\end{align}
where $P(s,t;\alpha)$ is completely fixed 
\begin{align}\label{4dcontact_s,t}
P(s,t;\alpha)=(s+t-s\,\alpha  )^2-4 k \left(-k\,\alpha ^2 +k\,\alpha -k+\alpha ^2 s-2 \alpha  s+s+t\right)\,,
\end{align}
and has the following properties under crossing symmetry
\begin{align}
P(s,t;\alpha)=(\alpha-1)^2\,P\left(t,u;\frac{1}{1-\alpha}\right)=(-\alpha)^2\,P\left(u,s;\frac{\alpha-1}{\alpha}\right)=P(s,u;1-\alpha)\,,
\end{align}
recalling that $u=4k-s-t$. $Q^a(\alpha,\beta)$ are polynomials of degree $k-2$ in $\alpha$ and $\beta$ separately, thus containing in principle $(k-1)^2$ free parameters for each representation ${\bf R}_a$. We find that the Ward identities do not impose further constraints. Crossing symmetry, however, gives the following relations on the polynomials $Q^a(\alpha,\beta)$
\begin{align}
\begin{split}
Q^a(\alpha,\beta)&=(\alpha-1)^{k-2}\,(\beta-1)^{k-2}\,\sum_b (-1)^{|\mathbf{R}_a|}( \mathrm{F}_u)_a^{\,\,\,b}\,Q^b\left(\frac{1}{1-\alpha},\,\frac{1}{1-\beta}\right)\,,\\
Q^a(\alpha,\beta)&=(-\alpha)^{k-2}\,(-\beta)^{k-2}\,\sum_b (-1)^{|\mathbf{R}_a|}( \mathrm{F}_t)_a^{\,\,\,b}\,Q^b\left(\frac{\alpha-1}{\alpha},\,\frac{\beta-1}{\beta}\right)\,,\\
Q^a(\alpha,\beta)&=(-1)^{|{\bf R}_a|}\,Q^a(1-\alpha,1-\beta)\,.
\end{split}
\end{align}
The flat space limit of \eqref{contact4d} is simply
\begin{align}
\left.\mathcal{M}^a_{\rm H.D., \,con}\right|_{s,t\to \infty}=(s+t-s\,\alpha  )^2\,Q^a(\alpha,\beta)\,.\label{contact4d-flat}
\end{align}
which contains the familiar prefactor $(s+t-s\,\alpha  )^2$ already discussed in Section \ref{sec:flatspace}.

The vanishing of $\Lambda_p$ for higher-derivative corrections in four dimensions is consistent with the results of Appendix \ref{sec:chiralalgebra}. Under the same assumption as footnote \ref{footnote-no-extra}, the OPE coefficients of the exchanged multiplets, which survive the chiral algebra twist, were found to be completely fixed by crossing symmetry. 

Let us finish our discussion of the $d=4$ case with the following interesting remark. In terms of the reduced Mellin amplitude (\ref{Mtildedef}), this solution simply corresponds to a constant in the Mellin variables. This allows us to generalize the result for degree-two contact terms in four dimensions to general weights: simply by acting with $\widehat{R}$ on a constant we find that all that one needs to change in \eqref{contact4d} is the form of the polynomial $P(s,t; \alpha)$,  which now reads
\begin{align}
\begin{split}
P(s,t; \alpha)=&-(k_1+k_2-s)\,(k_3+k_4-s)\,\alpha\,(1-\alpha)+(k_1+k_4-t)\,(k_2+k_3-t)\,\alpha\\
&+(k_1+k_3-u)\,(k_2+k_4-u)\,(1-\alpha)\,.
\end{split}
\end{align}
Note that the flat space limit of $P(s,t; \alpha)$ does not depend on the weights. 

\vspace{0.2cm}

\noindent $\bf{d=3,5,6}$. In all dimensions other than four, the situation is more interesting. We find there are both pure contact term solutions  as in the $d=4$ case, with $\Lambda_p=0$, and new solutions with non-vanishing exchange coefficients $\Lambda_p$. 

Let us first describe the pure contact solutions. We find that the Ward identities admit polynomial solutions of degree two in $s$, $t$  only for {\it even} values of $k$.  The solutions can be written as
\begin{align}
\mathcal{M}^a_{\rm H.D., \,con}=R(s,t;\alpha)\,S^a(\beta)\,,
\end{align}
where $S^a(\beta)$ is an arbitrary polynomial of degree $k-2$ in $\beta$, constrained only by crossing.  In analogy with the case of $d=4$,  $R(s,t;\alpha)$ is a completely fixed polynomial for each value of $k$, of degree two in $s,t$ and degree $k$ in $\alpha$. We will give its expression in the flat space limit. As for the polynomial $P$ in $d=4$,  $R(s,t;\alpha)$ has simple properties under crossing
\begin{align}
R(s,t;\alpha)=(\alpha-1)^k\,R\left(t,u;\frac{1}{1-\alpha}\right)=(-\alpha)^k\,R\left(u,s;\frac{\alpha-1}{\alpha}\right)=R(s,u; 1-\alpha)\,,
\end{align}
where $u=4\,\epsilon\,k-s-t$.  The flavor dependence is completely factorized in $S^a(\beta)$ and crossing imposes the following constraints
\begin{align}
\begin{split}
S^a(\beta)&=(\beta-1)^{k-2}\,\sum_b (-1)^{|\mathbf{R}_a|}( \mathrm{F}_u)_a^{\,\,\,b}\,S^b\left(\frac{1}{1-\beta}\right)\,,\\
S^a(\beta)&=(-\beta)^{k-2}\,\sum_b (-1)^{|\mathbf{R}_a|}( \mathrm{F}_t)_a^{\,\,\,b}\,S^b\left(\frac{\beta-1}{\beta}\right)\,,\\
S^a(\beta)&=(-1)^{|{\bf R}_a|}\,S^a(1-\beta)\,. \label{all-d-flavor}
\end{split}
\end{align}
In the flat space limit, the expressions greatly simplify. We find
\begin{align}
\left. \mathcal{M}^a_{\rm H.D., \,con} \right|_{s,t\to \infty}=(s+t-s\,\alpha)^2\,\mathtt{r}(\alpha)\,S^a(\beta)\,,
\end{align}
with again the expected prefactor and $\mathtt{r}(\alpha)$ a polynomial in $\alpha$ of degree $k-2$,  satisfying 
\begin{align}\label{crossing_r_alpha}
\mathtt{r}(\alpha)=\mathtt{r}(1-\alpha)=(\alpha-1)^{k-2}\,\mathtt{r}\left(\frac{1}{1-\alpha}\right)=(-\alpha)^{k-2}\,\mathtt{r}\left(\frac{\alpha-1}{\alpha}\right)\,.
\end{align}
$\mathtt{r}(\alpha)$ can be written explicitly in terms of the R-symmetry polynomials for $\langle kkkk\rangle$,  given in \eqref{Rsymm_kkkk}:
\begin{align}
\begin{split}
\mathtt{r}(\alpha)&=\sum_{p=0}^{\tfrac{1}{2}(k-2)}\,C_{p;k}\mathcal{Y}_{4p}(\alpha)\,,\\
C_{p;k}&=-\frac{2^{7-4(k-p)} \Gamma\left[\tfrac{1}{2}(3-k)\right]^2\,\Gamma\left[\tfrac{k}{2}\right]^2\,\Gamma\left[2(k-1)\right]\,\Gamma\left[\tfrac{1}{2}(1-k-2p)\right]^2}{\sqrt{\pi}\,(p!)^2\,\Gamma\left[-\tfrac{1}{2}(1+4p)\right]\,\Gamma\left[\tfrac{1}{2}(k-2p)\right]^2\,\Gamma\left[\tfrac{1}{2}(1-2p)\right]^2}\,.
\end{split}
\end{align}
Let us now consider solutions with $\Lambda_p \neq 0$. We now find that there are $k-2$ (for $k>2$) independent extra solutions to the superconformal Ward identities for both even and odd $k$. These solutions with nonzero $\Lambda_p$ also require specific contact coefficients $c^{a}_{m,n;p,q}$ in the ansatz (\ref{generalcontact}). Taking the flat space limit, we find these solutions become
\begin{align}
\left.\mathcal{M}_{\rm H.D.}^{I_1I_2I_3I_4}\right|_{s,t\to \infty}=(s+t-s\,\alpha)^2\,\left(\mathtt{c}_s\,P_{su}(\alpha,\beta)-\mathtt{c}_t\,P_{tu}(\alpha,\beta)\right)\,,
\end{align}
where we have expressed the results in terms of the two independent flavor structures. Here $P_{su}(\alpha,\beta)$ and $P_{tu}(\alpha,\beta)$ are polynomials in $\alpha$ and $\beta$ of degree $k-2$.  The two are not independent of each other,  rather they are related by crossing via
\begin{align}
P_{tu}(\alpha,\beta)=\left(\alpha\,\beta\right)^{k-2}\,P_{su}\left(\frac{\alpha-1}{\alpha},\, \frac{\beta-1}{\beta}\right)\,,
\end{align}
with $P_{tu}(\alpha,\beta)$ symmetric under the exchange of the first two operators,  namely
\begin{align}
P_{tu}(\alpha,\beta)=P_{tu}(1-\alpha,1-\beta)\,.
\end{align}
Note that for both even and odd $k$ the solutions we found behave universally in the flat space limit, as expected. This would not be the case if we only considered polynomial solutions. Something very similar happens when considering  M-theory corrections to $AdS_7\times S^4$ super graviton four-point functions, which was studied in detail in \cite{Chester:2018dga}. There it was also found that it is necessary to include exchange contributions in order to find nontrivial solutions to the superconformal Ward identities. These solutions with non-vanishing $\Lambda_p$ were called ``meromorphic'' in \cite{Chester:2018dga}.

\vspace{0.5cm}

\noindent\underline{{\bf Higher $\mathbf{D}$}}

\vspace{0.2cm}

We will leave a careful investigation for solutions with $D\geq 3$ for future work. Here we only mention that for $d=4$, based on checking the first few values of $k$ and contact terms up to degree $D=5$, there appears to be no solution to the superconformal Ward identities with nonzero exchange contributions. It extends our previous observation at $D=2$, and seems to indicate that  all higher-derivative corrections for $d=4$  correspond to polynomial solutions. This is also consistent with our expectation from the chiral algebra.

\bibliography{refhalfsusy} 
\bibliographystyle{utphys}
\end{document}